\documentclass[12pt]{iopart}
\usepackage{epsfig}

\begin{document}
% Journal identifier can be put here if required, e.g.
%\jl{14}
%\input epsf.tex    %<-If you need EPS figures to be
                     %  called in {figure} environment for PC
%\input epsf.def   %<-If you need EPS figures to be
                     %  called in {figure} environment for Macintosh

% Definitions
\def\beq{\begin{equation}}
\def\eeq{\end{equation}\noindent}
\def\nuc#1#2{${}^{#1}$#2}
\def\mee{$\langle m_{\beta\beta} \rangle$}
\def\mnu{$\langle m_{\nu} \rangle$}
\def\gnu{$\langle g_{\nu,\chi}\rangle$}
\def\mmod{$\| \langle m_{\beta\beta} \rangle \|$}
\def\mb{$\langle m_{\beta} \rangle$}
\def\BBz{$\beta\beta(0\nu)$}
\def\BBm{$\beta\beta(0\nu,\chi)$}
\def\BBt{$\beta\beta(2\nu)$}
\def\bb{beta-beam}
\def\Mz{$|M_{0\nu}|$}
\def\Mt{$|M_{2\nu}|$}
\def\Tz{$T^{0\nu}_{1/2}$}
\def\Tt{$T^{2\nu}_{1/2}$}
\def\Tc{$T^{0\nu\,\chi}_{1/2}$}
\def\ms{$\delta m_{\rm sol}^{2}$}
\def\ma{$\delta m_{\rm atm}^{2}$}
\def\ts{$\theta_{\rm sol}$}
\def\ta{$\theta_{\rm atm}$}
\def\tot{$\theta_{13}$}
\def\be{\begin{equation}}
\def\ee{\end{equation}}
\def\today{\space\number\day\space\ifcase\month\or January\or February\or
    March\or April\or May\or June\or July\or August\or September\or October\or
    November\or December\fi\space\number\year}
% end definitions

% commands
\newcommand{\simgt}{\ \raisebox{-.25ex}{$\stackrel{>}{\scriptstyle \sim}$}\ }
\newcommand{\simlt}{\ \raisebox{-.25ex}{$\stackrel{<}{\scriptstyle \sim}$}\ }
% end commands

\title[Beta-beams]{Beta-beams}

\author{Cristina Volpe$\dag$}

\address{\dag\ Institut de Physique Nucl\'eaire Orsay, F-91406 Orsay cedex}

\begin{abstract}
Neutrino physics is traversing an exciting period, after the
important discovery that neutrinos are massive particles, that has
implications from high-energy physics to cosmology. A new method for
the production of intense and pure neutrino beams has been proposed 
recently: the ``beta-beam''. It exploits boosted
radioactive ions decaying through beta-decay. This novel concept has been 
the starting point for a new possible future facility. Its main goal is
to address the crucial issue of the existence of CP violation in the lepton
sector. Here we review the status and the recent developments
with beta-beams. We discuss the original, the medium and high-energy scenarios
as well as mono-chromatic neutrino beams produced through ion electron-capture.
The issue of the degeneracies is mentioned. An overview of low energy beta-beams 
is also presented. These beams can be used to perform experiments of interest for 
nuclear structure, for the study of fundamental interactions and for nuclear
astrophysics.
\end{abstract}


\tableofcontents

\maketitle

\section{Introduction}
The observations made by the Super-Kamiokande (Fukuda \etal 1998),
the K2K (Ahn \etal 2003), the SNO (Ahmad \etal 2001) and the KAMLAND (Eguchi \etal 2003)
experiments have brought a breakthrough in the field of neutrino physics.
The longstanding puzzles of the {\it solar neutrino deficit}, observed
by Davis' pioneering measurement (Davis 1964) and
of the atmospheric anomaly
have been clarified. We know now that
the
expected fluxes are reduced due to the neutrino oscillation phenomenon:
the change in flavour that neutrinos undergo while traveling, first
proposed by Pontecorvo (Pontecorvo 1957). 
An impressive progress has been achieved in our knowledge of neutrino
properties, thanks to these experimental advances. Most of the parameters
of the Maki-Nakagawa-Sakata-Pontecorvo (MNSP) unitary matrix (Maki \etal 1962), relating the
neutrino flavor to
the mass basis, are nowadays determined, except
the third neutrino mixing angle, usually called $\theta_{13}$.
However, this matrix might be complex, meaning
there might be additional phases -- one Dirac and two Majorana
phases, if neutrinos are Majorana particles. A non-zero Dirac phase
introduces a difference between neutrino and anti-neutrino oscillations and
implies the breaking of the $\cal{CP}$ symmetry in the lepton sector.
On the other hand, the $\cal{CP}$ violating Majorana phases can manifest themselves 
in other processes like e.g. 
neutrinoless double-beta decay.
The Majorana or Dirac nature of neutrinos,
the value of the neutrino magnetic moment, the existence of sterile
neutrinos, the neutrino mass scale (and hierarchy)
and the possible existence of $\cal{CP}$ violation are among the
crucial issues that remain open. 

Neutrino oscillations imply that neutrinos are
massive particles, contrary to what has been believed so far, and represent
the first direct experimental evidence for physics beyond the Standard Model of
elementary particles and interactions. Understanding the mechanism for
generating the neutrino masses and their small values is clearly a fundamental
question, that needs to be understood. On the other hand, the presently known
(as well as unknown) neutrino properties have important implications for other
domains of physics as well, among which astrophysics, e.g. for our
comprehension of processes like the nucleosynthesis of heavy elements,
and cosmology, for example for unraveling the asymmetry
between matter and anti-matter in the Universe. Stars and
the Universe constitute a precious laboratory for the study of the yet unknown neutrino
properties, of fundamental interactions and for the search of new physics (Raffelt 1996).  
The present limits on the neutrino magnetic moment, 
on sterile neutrinos, or on the neutrino
energy density coming from the cosmic microwave background and large
scale structures (Bilenky \etal 2003) are 
talkative examples of the strong constraints that can be extracted from
the field of neutrino astronomy. 

If the study of astrophysical and cosmological phenomena offers a unique
way to progress on the open questions, performing terrestrial -
accelerator and non-accelerator - experiments represent an irreplaceable
procedure to obtain direct (model-independent) information.
Many experiments, running, planned or under study, will bring invaluable
results in the near future. In particular, the measurement of the third,
less known, neutrino mixing angle, as well as of the Dirac phase need the
availability of very intense neutrino beams, and might require new methods
besides conventional beams (coming from the
decay of pions and muons)  pushed to their ultimate intensities,
or super-beams. The idea of a neutrino factory was first put forward,
which is based on the production,
collection, acceleration
and storage of muons to obtain muon and electron (anti)neutrino beams. 
The neutrino energies are of several
tens of GeV and the needed baseline of the order of 3000 Km. The aim is to
address in one step the leptonic $\cal{CP}$ study, the neutrino
hierarchy and the precise measurement
of the $\theta_{23}$, $\Delta m^2_{23}$ and $\theta_{13}$ oscillation
 parameters (Albright \etal 2004).

Recently Zucchelli proposed the idea of producing electron (anti)neutrino
beams using the beta-decay of boosted radioactive ions: the ``beta-beam''
(Zucchelli 2002).
Such a novel concept has three main advantages: well-known fluxes, purity
(in flavour) and collimation. This simple and
attractive idea exploits major developments in the field of nuclear
physics, where radioactive ion beam facilities, either planned or
under study, are expected to reach ion intensities of $10^{11-13}$ per second.
The main goals are to perform studies in the fields of 
nuclear structure, of fundamental
interactions, in nuclear astrophysics, solid state physics, 
as well as for medical applications. 
Such intensities can render neutrino accelerator experiments using ions 
feasible as well, thanks to the beta-beam concept.  

In the original scenario proposed by Zucchelli, the ions are produced,
collected and accelerated first at several hundred MeV per nucleon, then
at several tens GeV energies -- after injection in the Proton Synchrotron (PS)
and Super Proton Synchrotron (SPS) accelerators\footnote{Zucchelli considers
CERN as a possible site for a beta-beam facility (Zucchelli 2002).} --
and finally stored in a storage ring of 7.5 km total length
and of 2.5 km straight sections.
This facility is based on reasonable extrapolation of existing technologies
and exploits already existing accelerator infrastructure to reduce cost.
The neutrino beam produced by the ions decaying along the
straight sections point to a gigantic \v{C}erenkov detector,
located at the (upgraded) Fr\'ejus Underground Laboratory, in order
to study $\cal{CP}$ violation, through a comparison of $\nu_e \rightarrow \nu_{\mu}$ and
$\bar{\nu}_e \rightarrow \bar{\nu}_{\mu}$ oscillations.
A first feasibility study is performed in (Autin \etal 2003) and discussed in
(Lindroos 2003, Benedikt \etal 2004).
The discovery potential is now actively investigated
(Mezzetto 2003, Bouchez \etal 2003, Mezzetto 2005, Mezzetto 2006, Burguet-Castell \etal 2004,
Burguet-Castell \etal 2005, Campagne \etal 2006) 
and is also discussed in (Guglielmi \etal 2005). Other technologies are also being considered 
for the detector, like 
liquid Argon Time Projection Chambers (Rubbia C 1977, Rubbia A 2004, Ereditato and Rubbia 2004 and
2006) and iron calorimeters.

The interest of the beta-beam concept for establishing a facility producing
low-energy neutrino beams is soon recognized (Volpe 2004).
Here the ions are accelerated at $\gamma=5-14$ ($\gamma$ is the Lorentz ion boost) 
to produce neutrino
beams in the 100 MeV energy range. This idea opens new axis of research
compared to the original scenario. Such beams can in fact be used for the study of
nuclear structure, of fundamental interactions, and of nuclear astrophysics. 
In particular, the physics potential covers 
experiments on spin-isospin and isospin nuclear excitations and neutrino-nucleus interactions
(Volpe 2004, Serreau and Volpe 2004, McLaughlin 2004, Volpe 2005a), measurements of the Weinberg angle
(Balantekin \etal 2006a), of the neutrino magnetic moment (McLaughlin and Volpe 2004)
and CVC tests (Balantekin \etal 2006b), or 
core-collapse Supernova physics (Volpe 2004, Jachowitz and McLaughlin 2005 and 2006). 
A small devoted storage ring appears as more appropriate for these low energy
applications (Serreau and Volpe 2004). Its feasibility study is now ongoing
(Chanc\'e and Payet 2005).

Other scenarios for the study of $\cal{CP}$ violation
have been proposed afterwards, where the energy of the
ions is much higher, the $\gamma$ ranging from
150 (Burguet-Castell {\it et al} 2005) to several hundreds (medium energy)
to thousands (high energy) (Burguet-Castell {\it et al} 2004). 
The value of 150 GeV per nucleon comes from the maximum
acceleration that can be attained in the SPS. 
The baseline scenario in this case is the same as the original one.
On the contrary, the medium and high energy options
require major changes in the accelerator infrastructure, such as a 
refurbished SPS (or even the LHC) at CERN, as well as
bigger storage rings. An increased
distance between the source and the detector is also necessary 
to match the same oscillation frequencies. Therefore such scenarios imply
further locations for the far detector, such as the Canfranc or the Gran
Sasso Underground Laboratories. 
The physics potential is being actively investigated 
(Terranova \etal 2004, Huber \etal 2006,
Agarwalla \etal 2005, Donini \etal 2005a and 2006) and is discussed in (Terranova 2005,
Gomez-Cadenas 2005, Migliozzi 2005). 
It covers the third neutrino
mixing angle, the $\cal{CP}$ violating phase, as well as the neutrino mass
hierarchy. Some reduction of the degeneracy problem is also expected.
This enlarged discovery potential 
is due to the information that can be extracted on the energy of the events; 
while in the original scenario
the experiment is a counting measurement only.
However, even if from the physics point of view pushing to higher energies
appears as attractive, a specific and extensive feasibility study is needed,
to determine in particular the ion intensities, that drastically influence the
sensitivities, as well as the characteristics of the storage ring.
From this point of view, the medium energy option appears as more promising,
whereas the high energy one keeps at a much more speculative level.

The presence of degeneracies constitute an important open issue in the analysis 
of future accelerator neutrino experiments. This consists in the appearance of several
disconnected regions in the oscillation parameter space, besides the true solutions.
The origin of such regions is both the inherent three flavour structure of the oscillation
probabilities, and some still unknown neutrino properties, namely the
sign of $\Delta m^2_{23}$ and the octant of $\theta_{23}$. The "fake" regions are
usually referred to in the literature as the sign, the octant and the mixed degeneracies
(or clones). They might render the identification of the true values of 
$\delta$ and $\theta_{13}$ more complex, depending on our knowledge of these
unknowns at the time a beta-beam experiment takes place.
Several possible strategies can be followed to resolve the problem of the clone solutions.
An extensive analysis in the context of beta-beams is performed in
(Donini \etal 2004, Donini \etal 2005b, Donini \etal 2005c, Campagne \etal 2006, Donini 
and Fernandez-Martinez 2006, Fernandez-Martinez 2006).
An interesting and promising possibility is pointed out in (Huber \etal 2005b), i.e.
to combine accelerator experiments with atmospheric data in the same detector.

Finally, monochromatic neutrino beams produced by boosted 
ions decaying through electron capture
are proposed in (Bernabeu {\it et al} 2005a) and further
discussed in (Bernabeu {\it et al} 2005b, 2005c and 2006).
The baseline envisaged here is the same as for the original beta-beam.
In order to search for $\cal{CP}$ violation, instead of comparing neutrino versus
anti-neutrino oscillations, a comparison of $\nu_e \rightarrow \nu_{\mu}$ 
oscillations (only $\nu_e$ are available) at different neutrino energies is necessary.
Again, a specific feasibility study is needed in particular to state the ion rates at production
and in the storage ring, since such a configuration requires the acceleration and storage 
of not fully stripped ions.

An attractive aspect of a beta-beam facility is that
it offers strong synergies between different
communities, to realize a multidisciplinary program covering
hot open issues in various domains. In particular, the synergy 
between the nuclear and the
neutrino physics communities has been quickly recognized.
This has merged in a common feasibility study within the EURISOL
Design Study, financed by
the European Community in the FP6 Program (2005-2009), 
that is now ongoing. 
On the other hand the far detector can
be used for other purposes, in particular, to measure neutrinos
from a core-collapse Supernova explosion (in or outside our Galaxy), and
to improve the present sensitivity on the proton decay.
Finally, it is interesting to note that beta-beams can produce 
neutron beams as well, thanks to the beta-decay of boosted
radioactive ions which emit delayed neutrons.
This interesting possibility deserves further investigation.

Clearly the beta-beam concept has a broad physics potential,
which explains why this field has been developing very fast since its very
beginning, bringing many new ideas and scenarios. 
In this review we summarize the work done and try to give
an overall view of the field. The review is meant not only for experts, but also as
an introduction for non-experts. For this reason, many technical details
are skipped, unless strictly necessary. 
From now on, unless stated differently, with the terminology "beta-beams" we refer
to the original scenario. 
In Section I we recall the phenomenon of neutrino oscillations, the
recent discoveries in neutrino physics, as well as how neutrino oscillations
can be used to address the breaking of the $\cal{CP},\cal{T}$ and $\cal{CPT}$
symmetries in the lepton sector. Section II presents the original beta-beam
concept and baseline scenario, as proposed by Zucchelli, as well as
the following sensitivity studies.
In Section III low energy beta-beams and the physics
potential discussed so far are presented. 
Section IV is devoted to the medium and high energy scenarios.
The issue of the degeneracies is mentioned in Section V.
Mono-chromatic neutrino beams based on electron capture are discussed
in Section VI. Conclusions and perspectives are made in Section VII.

\subsection{The neutrino oscillation phenomenon}
Let us briefly remind the basic ingredients of neutrino oscillations.
This phenomenon arises if neutrinos have masses and there is mixing.
The evolution equation of the mass eigenstates in vacuum is
simply $i(d/dt)|\nu_{m}\rangle=H_{m}|\nu_{m}\rangle$. This
becomes 
$i(d/dt)|\nu_{\alpha}\rangle=UH_{m}U^{\dagger}|\nu_{\alpha}\rangle$ in the flavor basis,
where $U$ is the unitary matrix which relates
the weak $\nu_{\alpha}$ ($\alpha=e,\mu,\tau$)
to the mass $\nu_{i}$ ($i=1,2,3$) eigenstates.
Each flavour eigenstate can be written as 
$| \nu_{\alpha} \rangle =U_{\alpha i}^* | \nu_i \rangle $ .

In general, a unitary $n \times n$ matrix depends on $n(n-1)/2$ angles and $n(n+1)/2$ phases.
In the Dirac case, $2n-1$ phases can be removed by a proper re-phasing of the left-handed
fields, leaving $(n-1)(n-2)/2$ physical phases. Therefore, $\cal{CP}$ violation
is only possible in the case of $n \ge 3$ generations. In the Majorana case there is less freedom
to re-phase the fields, since the phases of the neutrino fields cannot be absorbed in 
the Majorana mass terms. Only $n$ phases can be removed, leaving $n(n-1)/2$ physical 
phases. Out of these $(n-1)(n-2)/2$ are the usual Dirac phases, while $n-1$ are specific to the Majorana
case and are called Majorana phases. The latter do not lead to any observable effects in neutrino 
oscillations and are not discussed further in this review.

In the simple case of two families, this matrix
reduces to the rotation matrix:
\begin{equation}\label{e:rot2}
\left(\begin{array}{cc} \nu_e
\\  \nu_\mu   \end{array}\right) = 
\left(\begin{array}{cc}
\cos \theta & \sin\theta \\
- \sin \theta & \cos \theta
\end{array}\right)
\left(\begin{array}{cc} \nu_1
\\ \nu_2  \end{array}\right)
\end{equation}
where the angle $\theta$ is called the neutrino mixing angle, since it mixes
the mass eigenstates $\nu_1$ with mass $m_1$ and $\nu_2$ with mass $m_2$.

\begin{figure}
\begin{center}
\epsfig{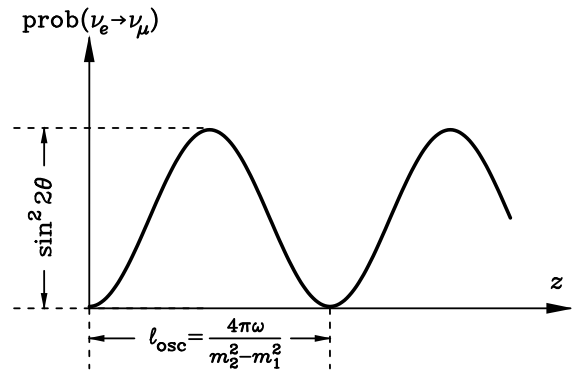}
\caption{Schematic drawing of the $\nu_e$ to $\nu_{\mu}$ appearance oscillation probability
in the case of two families
Eq.(\ref{e:2fam}). The oscillation amplitude and frequency depend on
two neutrino parameters: $\theta$ and $\Delta m^2=m_{2}^2 - m_{1}^2$, where $m_1$ and
$m_1$ are the masses of the mass eigenstates $\nu_1$and $\nu_2$ respectively.}
\label{fig:osc}
\end{center}
\end{figure}

 If one assumes that a given source is producing a neutrino flux of 
 given flavor $| \nu_{\alpha} \rangle$ at $t=x=0$, the neutrino state at a later time $t$
is then $| \nu_{\alpha}(t) \rangle = U^*_{\alpha j} e^{- i E_j t} | \nu_j \rangle$, 
 and evolves as an admixture of the mass eigenstates.
 The probability amplitude of finding the neutrino at the time $t$ in a flavour state
 $\nu_{\beta}$ is 
 \beq\label{e:amposc}
 A(\nu_{\alpha} \rightarrow \nu_{\beta}; t) = U_{\beta j} e^{- i E_j t} U^*_{\alpha j} ~.
 \eeq
 The neutrino oscillation probability, i.e. the transformation probability
 of a flavour eigenstate neutrino $| \nu_{\alpha} \rangle$ into another one $| \nu_{\beta} \rangle$
 is then 
 \beq\label{e:probosc}
 P(\nu_{\alpha} \rightarrow \nu_{\beta}; t)= |A(\nu_{\alpha} \rightarrow \nu_{\beta}; t)|^2=
 |U_{\beta j} e^{- i E_j t} U^*_{\alpha j}|^2
 \eeq
 Therefore the probability for detecting a state of a different flavor $\nu_{\beta}$, at
 a given distance $L$ from the source, is given, for relativistic particles, by:
 \begin{equation}\label{e:2fam}
 P(\nu_{\alpha} \rightarrow \nu_{\beta})\approx \sin^2 2\theta~
 \sin^2({1.27 \Delta m^2 {L\over E}} )
 \end{equation}
 where $L$ is in m (km) and $E$ in MeV (GeV).
 The mixing angle $\theta$ Eqs.(\ref{e:rot2}) and (\ref{e:2fam}) determines the
 oscillation amplitude, whereas the difference of the square of the 
 masses $\Delta m^2= m^2_1 -m_2^2$,
 gives the oscillation frequency (Figure \ref{fig:osc}). 
 The neutrino sources available (the sun, the atmosphere, reactors and accelerators)
 have different typical neutrino average energies,
 and distances between the source and the detector, and allow to probe various oscillation frequencies.
 Oscillation searches are called appearance experiments, if the appearance of a new flavour is looked for,
 or disappearance experiments, if a reduction in the neutrino flux from the source is searched for. 

Neutrino propagation through
matter can modify oscillations significantly. This is known as the 
Mikheyev-Smirnov-Wolfenstein (MSW) effect (Mikheev and Smirnov 1986, Wolfenstein 1978).
Indeed, neutrinos of all flavors couple to electrons, protons and neutrons in matter
through neutral currents. In the case of $\nu_e$ scattering through charged-currents
is also effective. 
This scattering induces a matter-induced potential acting
on the electron neutrinos, which depends on the electron 
density $n_e$:
\begin{equation}
V_e^{CC}= \sqrt{2}G_F n_e.
\end{equation}\label{e:matpot}
The neutral current contributions from electrons and protons
cancel, if the medium is electrically neutral, whereas the scattering on neutrons
induces equal potentials on all neutrinos. 
Staying in the two-flavor formalism, 
the time evolution of the flavor eigenstates is as 
follows\footnote{Terms which contribute equally to the diagonal give equal phases
and do not influence neutrino oscillations.}: 
\begin{equation}\label{e:matevo}
i{d \over dt}
\left(\begin{array}{cc} \nu_e
\\  \nu_\mu   \end{array}\right) 
= 
\left(\begin{array}{cc}
- {\Delta m^2 \over 4E}\cos2\theta +  \sqrt{2}G_F n_e & {\Delta m^2 \over 4E}\sin2\theta \\
{\Delta m^2 \over 4E}\sin2 \theta & {\Delta m^2 \over 4E}\cos2\theta
\end{array}\right)
\left(\begin{array}{cc} \nu_e
\\ \nu_{\mu}  \end{array}\right).
\end{equation}
Note that in the three-flavor formalism, 
the matter-induced potential modifies both $\nu_e$ versus $\nu_{\mu}$
and $\nu_{\mu}$ to $\nu_{\tau}$  oscillations, since these are coupled.
Matter effects can enhance neutrino oscillations
when the MSW resonance condition is met:
\begin{equation}\label{e:reson} 
{\Delta m^2 \over 2E} \cos2\theta =  \sqrt{2}G_F n_e ~.
\end{equation}
This means, that independently on the value of the vacuum mixing angle $\theta$,
mixing in matter is maximal if the MSW condition Eq.(\ref{e:reson}) is satisfied.

\begin{figure}
\begin{center}
\epsfig{file=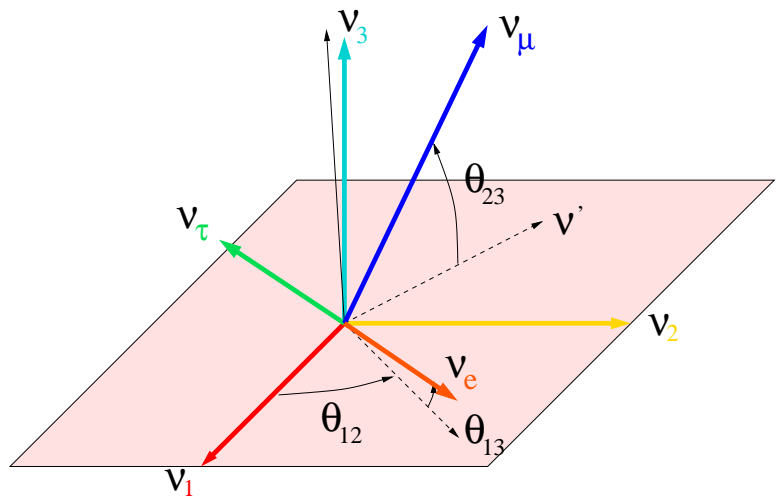,scale=0.5}
\caption{Representation of the 3-d ``rotation'' relating the neutrino
flavour and mass basis Eqs.(\ref{e:5}-\ref{e:mnsp}).}
\label{fig:eulerangles}
\end{center}
\end{figure}

In the case of three families, the neutrino flavour and mass eigenstates are related through
\begin{equation}\label{e:5}
\left(\begin{array}{cc} \nu_e
\\  \nu_\mu  \\ \nu_\tau \end{array}\right)
=
\left(\begin{array}{ccc}
U_{e1} & U_{e2}  & U_{e3} \\
U_{\mu1} & U_{\mu2}  & U_{\mu3} \\
U_{\tau1} & U_{\tau2}  & U_{\tau3}
\end{array}\right)
\left(\begin{array}{cc} \nu_1
\\ \nu_2 \\  \nu_3 \end{array}\right)
\end{equation}
This matrix is called the Maki-Nakagawa-Sakata-Pontecorvo (MNSP) matrix (Maki \etal 1962),
the analogue of the Cabibbo-Kobayashi-Maskawa (CKM) matrix, and is usually
parametrized as:
\begin{equation}\label{e:mnsp} 
U=
\left(\begin{array}{ccc}
c_{12} c_{13} & s_{12} c_{13} & s_{13} e^{-i \delta}
 \\
 -s_{12} c_{23} - c_{12} s_{23} s_{13} e^{i \delta} &
 c_{12} c_{23} - s_{12} s_{23} s_{13} e^{ i \delta} &
 s_{23} c_{13}
 \\
 s_{12} s_{23} - c_{12} c_{23} s_{13} e^{i \delta} &
 -c_{12} s_{23} - s_{12} c_{23} s_{13} e^{i \delta} & c_{23} c_{13}
\end{array}\right)
\end{equation}
where $c_{ij}=\cos \theta_{ij}$ and $s_{ij}=\sin \theta_{ij}$,
$\theta_{ij}$ are the rotation angles (Figure \ref{fig:eulerangles}) and $\delta$ is a
$\cal{CP}$ Dirac violating phase.
If neutrinos are Majorana particles, the matrix U is multiplied by an additional matrix 
having only non-zero diagonal terms, i.e. $diag(1,e^{i\alpha_1},e^{i\alpha_2})$, including
two extra $\cal{CP}$ violating phases. As far as the oscillation frequencies are 
concerned, since $\sum_i \Delta m^2_{i}=0$, only two frequencies are independent.
Note that the oscillation phenomenon gives information on the $\Delta m^2_{i}$ only, while
the neutrino mass scale needs to be obtained from other measurements, in particular
from the endpoint spectrum of nuclear beta-decay - like the tritium based experiment KATRIN 
(Angrik \etal 2004) -
or from the possible observation of the neutrinoless double-beta decay in nuclei, where
the latter is also sensitive to the neutrino nature and to the Majorana $\cal{CP}$ 
violating phases (see e.g. Elliott and Engel 2004).  

\begin{figure}
\begin{center}
\epsfig{file=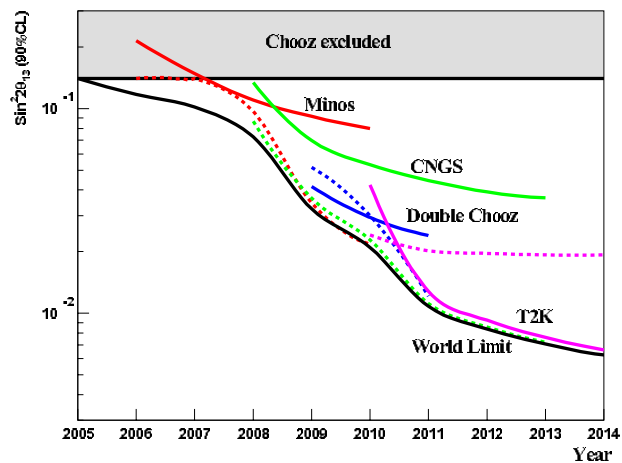,scale=0.5}
\end{center}
\caption{Sensitivity on $\sin^2 2\theta_{13}$ as a function of time.
For each experiment, its sensitivity is shown (solid line), as well as
the world sensitivity without the experiment (dashed line). The overall
world sensitivity as a function of time is also shown (Guglielmi \etal 2005). }
\label{fig:theta13vstime}
\end{figure}

In the last years 
much progress has been obtained in our knowledge of neutrino properties 
(for a detailed description see e.g. Bilenky \etal 2003). 
Two $\Delta m^2$ and mixing angles are now known,
i.e. $\Delta m^2_{12}=7.9~10^{-5}~$eV$^2,~\sin^2 \theta_{12}=0.3$, 
 and 
$\Delta m^2_{23}=2.4~10^{-3}~$eV$^2,~\sin^2 \theta_{23} =0.5$, 
often referred to as the solar
and the atmospheric oscillation parameters 
respectively,
extracted from the SNO (Ahmad \etal 2001) and KAMLAND (Eguchi \etal 2003) experiments on one hand, and 
from the Super-Kamiokande (Fukuda \etal 1998) and K2K (Ahn \etal 2003) 
experiments on the other hand.  
The sign of the solar oscillation parameter $\Delta m^2_{12}$ is known, while the
atmospheric one $\Delta m^2_{23}$ remains unknown. 
This introduces degeneracies in the determination of the other unknown parameters
(Section V). Note that the LSND collaboration has found a positive oscillation signal at 
$\Delta m^2=1~$eV$^2$ as well (Athanassopoulos \etal 1998 and 1996). 
The running miniBOONE and the future BOONE experiments will confirm/refute these observations.
Waiting for these results, in this review, we will not consider the third oscillation frequency and will
stay within the three flavour formalism. 

The precise value 
of the $\theta_{13}$  angle is not yet known, the present upper limit $\sin^2 2 \theta_{13} < 0.1$ 
is extracted from a reactor experiment called CHOOZ (Apollonio \etal 1999). 
Future experiments
will improve the present sensitivities on $\sin^2 2 \theta_{13}$ and search first
in the $10^{-1}-10^{-2}$ range (Figure \ref{fig:theta13vstime}). 
The $10^{-3}-10^{-4}$ range will only be reached
with beta-beams while values smaller than $10^{-4}$  can be explored with a neutrino factory. 
It is important to remind that the knowledge of $\theta_{13}$ is crucial for
$\cal{CP}$ violation searches, since if by any chance nature
has decided that $\theta_{13}$ is strictly zero, there cannot be any $\cal{CP}$
violation through the Dirac $\delta$ phase.

\subsection{$\cal{CP}, \cal{T}, \cal{CPT}$ violation searches}
The study of the weak interaction and of neutrino properties has brought
some of the building blocks of the Standard Model.
Since 1964, we know that $\cal{CP}$ is one of the symmetries broken
by the weak interaction, as first observed
in K mesons (Christenson \etal 1964), and recently in B mesons. The $\cal{CPT}$ theorem ensures
that under given conditions (Lorentz invariance, locality of the interactions)
a modern quantum field theory description of particles and their interactions
preserves the $\cal{CPT}$ symmetry. This is the case for renormalizable abelian
theories. However some variants of string theories predict a breaking of
Lorentz invariance and therefore of the $\cal{CPT}$ symmetry. Note that, if 
$\cal{CPT}$ is conserved, an obvious consequence
of $\cal{CP}$ violation is the violation of the $\cal{T}$ symmetry. 

Neutrinos can be used to study these symmetries and their breaking through the comparison of
neutrino oscillations. If $\cal{CP}$ is conserved, the
oscillation probability between particles and the corresponding antiparticles
coincide:
\begin{eqnarray}\label{e:CPgeneral}
{\cal{CP}}~~:~~~~~P(\nu_{\alpha} \rightarrow \nu_{\beta}; t)=P(\bar{\nu}_{\alpha} \rightarrow
\bar{\nu}_{\beta};
t) .
\end{eqnarray}
The action of the particle-antiparticle conjugation on the lepton mixing matrix $U$ Eq.(\ref{e:mnsp})
amounts to the change $U \rightarrow U^*$, meaning that $\cal{CP}$ is only conserved if the mixing
matrix $U$ is real or can be made real by re-phasing of the lepton fields.
The search for the possible existence of $\cal{CP}$ violation in the lepton sector 
and of a non-zero the Dirac phase $\delta$ Eq.(\ref{e:mnsp}) consists in
searching for possible differences between neutrino versus anti-neutrino oscillations.

Time reversal interchanges the initial and the final states, so if $\cal{T}$ is
conserved one has:
\begin{eqnarray}\label{e:T}
{\cal{T}}~~~~:~~~~~P(\nu_{\alpha} \rightarrow \nu_{\beta}; t)=P({\nu}_{\beta} \rightarrow \nu_{\alpha};
t) .
\end{eqnarray}
Finally, the oscillation probabilities are invariant under the combined action 
of $\cal{CP}$ and $\cal{T}$, if the following equality holds:
\begin{eqnarray}\label{e:CPT}
{\cal{CPT}}~:~~~~~P(\nu_{\alpha} \rightarrow \nu_{\beta}; t)=P(\bar{\nu}_{\beta} \rightarrow 
\bar{\nu}_{\alpha};
t) .
\end{eqnarray}
  
The study of these symmetries requires in particular appearance oscillation
experiments, since disappearance experiments like $P(\nu_e \rightarrow
\nu_e)$ are time-reversal invariant, 
and clearly insensitive to $\cal{CP}$
breaking effects. 
For example, by using Eqs.(\ref{e:probosc}) and (\ref{e:5}-\ref{e:mnsp}),
the $\nu_{\mu} \rightarrow \nu_e$ appearance probability can be rewritten as (Richter 2000) :

\begin{eqnarray}\label{e:probCP}
      P (\nu_{\mu}  \rightarrow \nu_e) =
 4 c^2_{13} { s^2_{13}} s^2_{23} \sin^2{\frac{\Delta m^2_{13}L}{4E_\nu}}
\times \left[ 1 + \frac{2 a}{\Delta m^2_{13}} (1 - 2 s^2_{13}) \right]
\\
 +8c^2_{13}s_{12} s_{13} s_{23} (c_{12}c_{23} { \cos \delta}-s_{12}s_{13}s_{23})
   \cos{\frac{\Delta m^2_{23} L}{4E_\nu}} \sin{\frac{\Delta m^2_{13}
L}{4E_\nu}} \sin{\frac{\Delta m^2_{12} L}{4E_\nu}}  \nonumber \\
 - 8c^2_{13} c_{12} c_{23} s_{12} s_{13} s_{23} {\sin \delta}
   \sin{\frac{\Delta m^2_{23} L}{4E_\nu}}\sin{\frac{\Delta m^2_{13}
L}{4E_\nu}}\sin{\frac{\Delta m^2_{12} L}{4E_\nu}} \nonumber \\
 +4 {s^2_{12}} c^2_{13} ({c^2_{13}
c^2_{23}+s^2_{12}s^2_{23}s^2_{13}-2c_{12}c_{23}s_{12}s_{23}s_{13} \cos
\delta})
 \sin{\frac{\Delta m^2_{12} L}{4E_\nu}} \nonumber \\
  - 8c^2_{12}s^2_{13}s^2_{23}
   \cos{\frac{\Delta m^2_{23} L}{4E_\nu}}\sin{\frac{\Delta m^2_{13}
L}{4E_\nu}}{\frac{aL}{4E_\nu}}(1-2s^2_{13}). \nonumber
\end{eqnarray}

\begin{figure}
\begin{center}
\epsfig{file=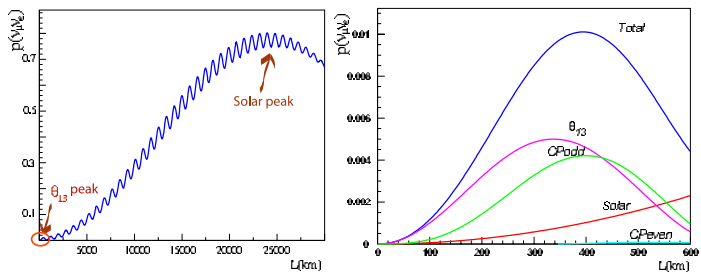,scale=0.5}
\caption{Drawing of the $P(\nu_{\mu} \rightarrow \nu_e)$ oscillations
as a function of the distance, assuming a monochromatic beam of 1 GeV
neutrinos, and matching the solar oscillation frequency for $\delta=0$ (left
figure) and the atmospheric oscillation frequency for $\delta=-\pi/2$
(right figure). In the latter case, the different terms contributing to the
oscillation probability Eq.(\ref{e:probCP}) are shown (Guglielmi {\it et al} 2005).} 
\label{fig:CPterms}
\end{center}
\end{figure}

\begin{figure}[t]
\begin{center}
\epsfig{file=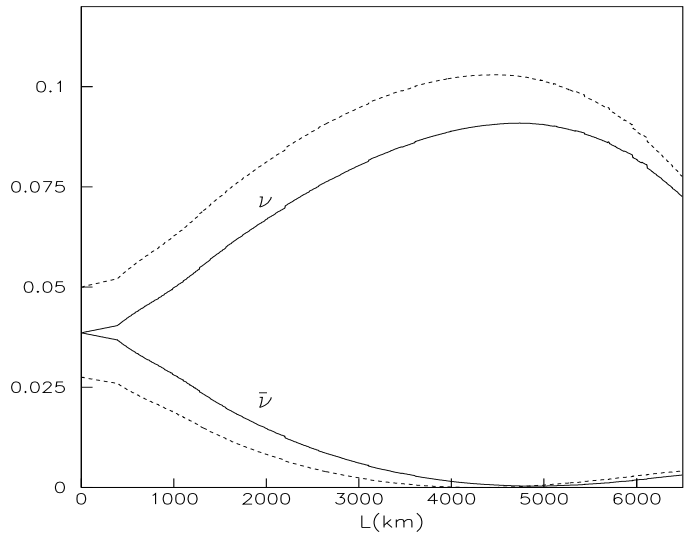,width=10cm,height=8cm}
\caption[a]{Comparison of the $\nu_{e} \rightarrow \nu_{\mu}$ and
the $\bar{\nu}_{e} \rightarrow \bar{\nu}_{\mu}$ appearance oscillation probability, as a
function of the baseline distance $L$ between the source and the detector,
evaluated at the first atmospheric oscillation maximum
$E/L=|\Delta m_{23}^2|/2 \pi$. From about $L=O(1000)$~km, matter effects
become significant and can be exploited to extract the sign of 
$\Delta m_{23}^2$.
The solid and dashed curves show the
true $\cal{CP}$ violation effects for $\delta=0^\circ$ and 90$^\circ$.}
\label{f:mattereffects}
\end{center}
\end{figure}

The first line contains terms driven by $\theta_{13}$, the second and 
third lines contain $\cal{CP}$ even and odd terms respectively, while the forth line
is driven by the solar parameters (Figure \ref{fig:CPterms}). Matter effects developed at the first
order are included in the last line, where 
$a [$eV$^2] = \pm 2\sqrt2 G_F n_e E_{\nu} = 7.6 10^{-5} \rho [$g/cm$^3] E_{\nu}
[$GeV$] $. 
The $\cal{CP}$ odd term as well as matter effects change
sign by replacing neutrinos with anti-neutrinos. The occurrence of
matter effects, at baselines of about  $O(1000)$ km (Figure \ref{f:mattereffects}),
allows to extract information on the mass hierarchy,
since their sign depends on the sign of $\Delta m_{23}^2$. These effects can be
distinguished from the true $\cal{CP}$ violating ones, because they have a
different energy dependence. 

Clearly, the knowledge of the Dirac $\delta$ phase depends on the knowledge
of all other oscillation parameters at the time an experiment searching for
$\cal{CP}$ violation takes place. 
As mentioned above, 
a particularly important piece of information 
is the value of the angle $\theta_{13}$. 
Note that 
if $\theta_{13} \ne 0$, it might be difficult from one single experiment to
unambiguously extract information about $\theta_{13}$ and $\delta$ 
because of the correlations and the degeneracies (Section V).
Anyway, 
pure, intense and collimated neutrino beams are needed to determine very small $\theta_{13}$ values
and a non-zero phase, such as those produced with beta-beams.

\section{The original beta-beam scenario}

\subsection{The beta-beam concept and neutrino fluxes}
When a nucleus undergoes a beta-decay, it emits either an electron neutrino, or an
electron anti-neutrino. In the rest frame of the nucleus, such neutrinos are
emitted isotropically. However, if the ions are accelerated, the momentum 
parallel to the ion beam gets boosted, while the perpendicular one keeps
unchanged. As a consequence, the neutrino beam gets collimated with the
emittance of the neutrino flux inversely proportional to the Lorentz
gamma factor. A neutrino beam produced with this procedure is what we call,
a ``\bb'' (Zucchelli 2002). The neutrino flux is well known, since the ion
intensity can be precisely determined and the flux is pure in flavour.
Its expression can be determined analytically as follows 
(Serreau and Volpe 2004).

The decay rate of a nucleus in the rest ($cm$) frame can be written as:
\begin{equation}
\label{e:1}
\frac{dW}{dt}\Big|_{cm}=\Phi_{cm} (E_\nu)\,dE_\nu\,\frac{d^2\Omega}{4\pi}\,,
\end{equation}
where $E_{\nu}$ and $\Omega$ denote respectively the energy and the solid
angle of the emitted (anti)neutrino, and where the neutrino flux $\Phi_{cm} (E_\nu)$ 
is given by the well-known formula:
\beq
\label{e:2}
\Phi_{cm} (E_\nu)=b\,E_\nu^2\,E_e\,
\sqrt{E_e^2-m_e^2}\, F(\pm Z,E_e)\,\Theta(E_e-m_e)\,.
\eeq
where the constant $b=\ln 2/m_e^5 ft_{1/2}$, with $m_e$ the electron mass
and $ft_{1/2}$ the ft-value. The quantities appearing in the above expression
are the energy $E_e=Q_{\beta}-E_{\nu}$ of the emitted lepton (electron or positron),
$Q_{\beta}$ is the $Q_{\beta}$--value of the decay. The Fermi function $F(\pm Z,E_e)$
accounts for the Coulomb modification of the spectrum.

In the laboratory frame, where the boosted nucleus has a velocity $v=\beta c$,
the decay rate reads:
\beq\label{e:events}
\frac{dW}{dt}\Big|_{lab}=\frac{1}{\gamma}\,\Phi_{lab} (E_\nu,\theta)\,
dE_\nu\,\frac{d^2\Omega}{4\pi}\,,
\eeq
where $\gamma=1/\sqrt{1-\beta^2}$ is the time dilation factor and where $E_{\nu}$ 
and $\Omega\equiv(\theta,\varphi)$ now denote the energy and solid angle of the 
emitted (anti)neutrino in the laboratory ($lab$) frame, $\theta$
being the angle of emission with respect to the beam axis. The boosted flux
$\Phi_{lab}(E_\nu,\theta)$ is given by:
\beq
\label{bflux}
\Phi_{lab}(E_\nu,\theta) = 
\frac{\Phi_{cm}(E_\nu\gamma[1-\beta\cos\theta])}{\gamma[1-\beta\cos\theta]} \, .
\eeq

\noindent
The ions are stored in a storage ring of total length $S$ with straight sections of
length $D$. In the stationary regime the mean number of ions
in the storage ring is $\gamma\tau g$, where $\tau=t_{1/2}/\ln 2$ is the
lifetime of the parent nuclei and $g$ is the number of injected ions per unit time.
The total number of neutrinos emitted per unit time from a portion $d\ell$ of the
decay ring is therefore
\beq
\frac{dN_\nu}{dt}=\gamma\tau\,g\times\frac{dW}{dt}\Big|_{lab}
\times\frac{d\ell}{S}\,.
\eeq

\noindent
Let us consider, for example, a cylindrical detector of radius $R$ and depth
$h$, aligned with one of the the storage ring straight sections, and placed at
a distance $d$ from the latter. After integration over the useful decay path and over
the volume of the detector, the total number of events per unit time is:
\beq
\label{dNevdt}
\frac{dN_{ev}}{dt}=g\tau nh\times
\int_0^\infty dE_\nu\,\Phi_{tot}(E_\nu)\,\sigma(E_\nu)\,,
\eeq
where $n$ is the number of target nuclei per unit volume, $\sigma(E_\nu)$ is
the relevant interaction cross section with the target, and the total neutrino flux is
\beq
\label{phitot}
\Phi_{tot}(E_\nu)=\int_0^D \frac{d\ell}{S}\,\int_0^h \frac{dz}{h}\,
\int_0^{\bar{\theta}(\ell,z)} \frac{\sin\theta
d\theta}{2}\,\Phi_{lab}(E_\nu,\theta)\,,
\eeq
with
\beq
\label{theta} \tan\bar{\theta}(\ell,z)=\frac{R}{d+\ell+z}\,.
\eeq
\begin{figure*}[!htp]
%\vspace{-3.cm}
\epsfig{file=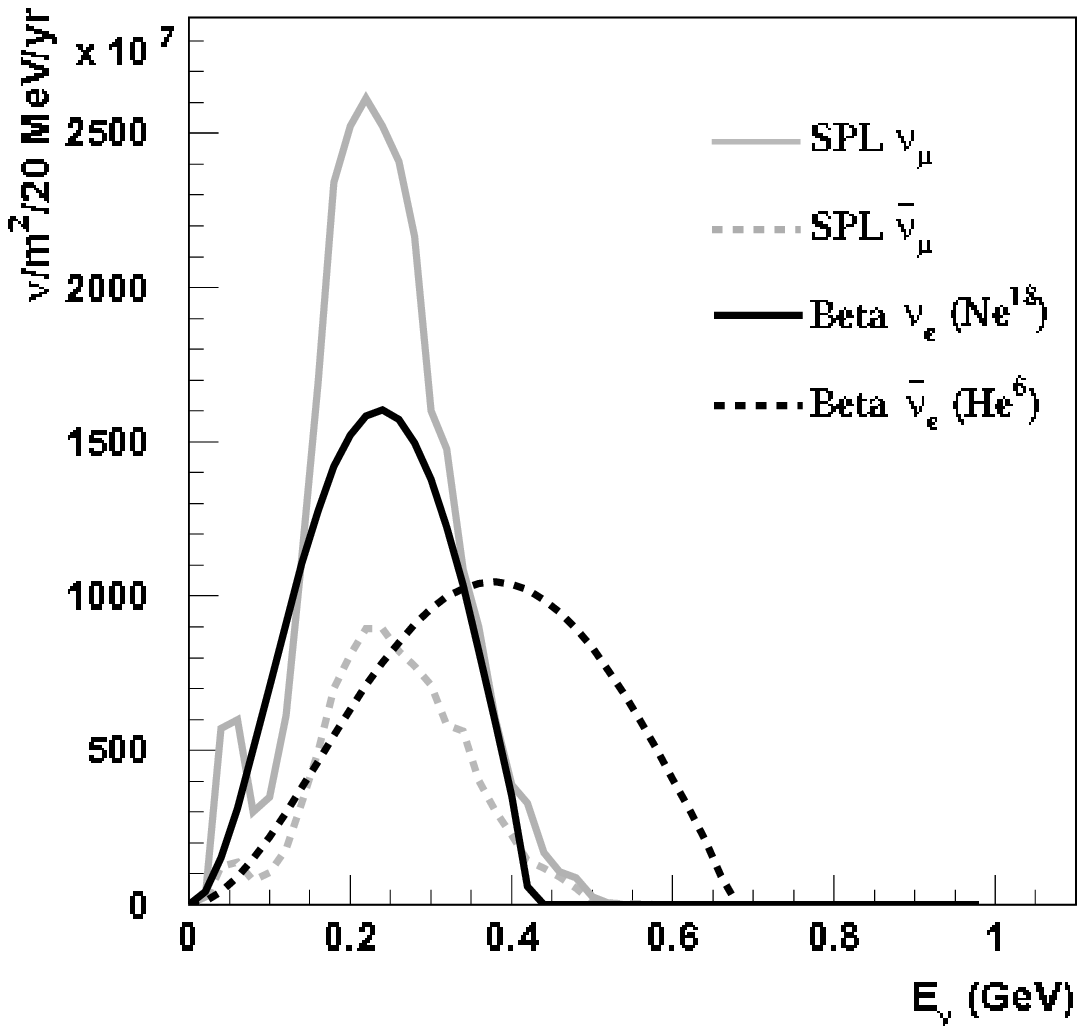,width=0.45\textwidth}
\epsfig{file=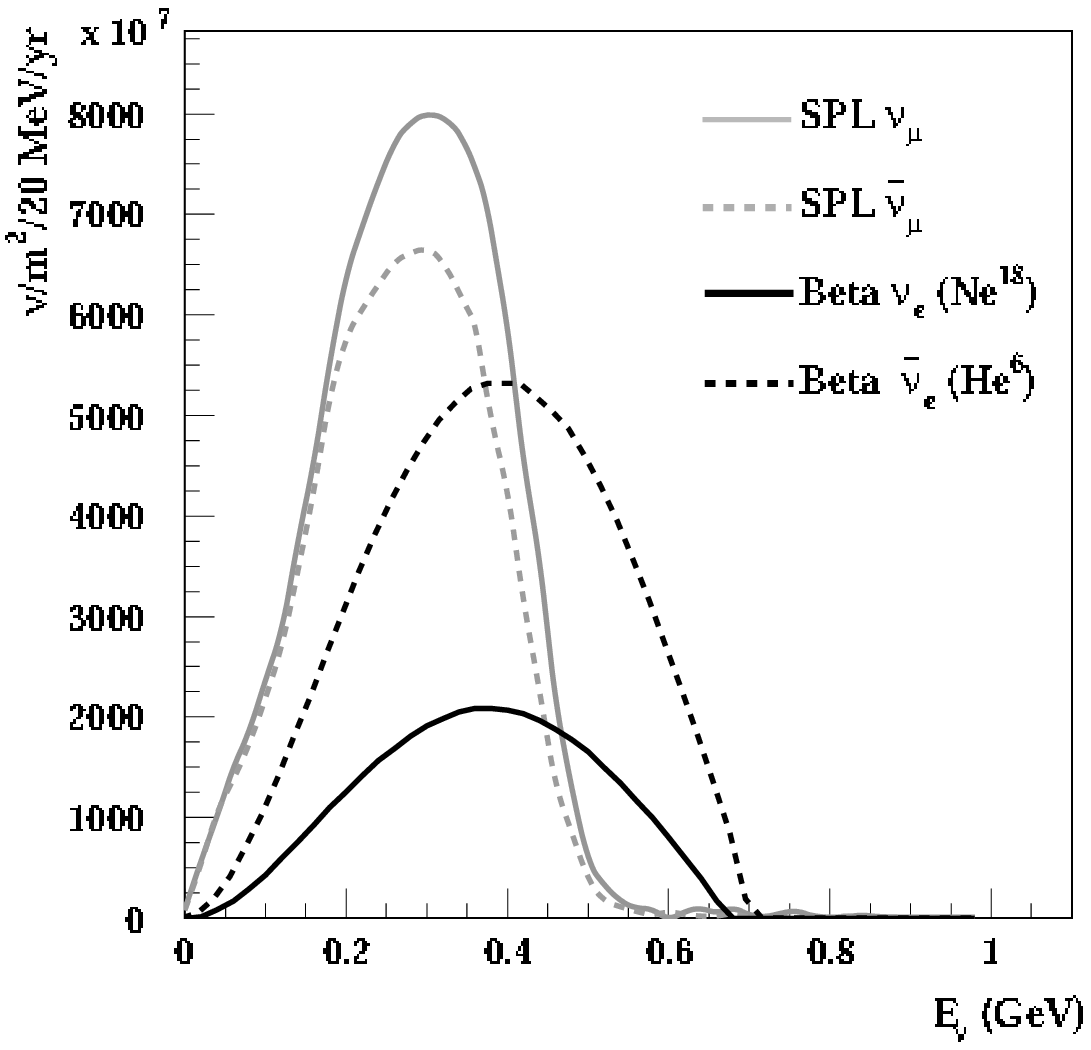,width=0.45\textwidth}
\caption{Comparison of neutrino fluxes from a super-beam (SPL) and a beta-beam. 
The neutrino beams are produced at CERN and sent to the Fr\'ejus Underground Laboratory,
130 km from CERN. 
Two options for the beta-beam are shown here. Left: The ions circulate together
in the storage ring, with $\gamma=60$ (100) for $^{6}$He ($^{18}$Ne)
(Mezzetto 2005). Right:
The ions circulate at the same $\gamma=100$, independently, in the storage ring.
Note that the average neutrino energies are related to the ion boost through
$E_{\nu} \approx 2 \gamma Q_{\beta}$ (Guglielmi \etal 2005).}
\label{fig:fluxes}
\end{figure*}

\begin{figure}
\begin{center}
\epsfig{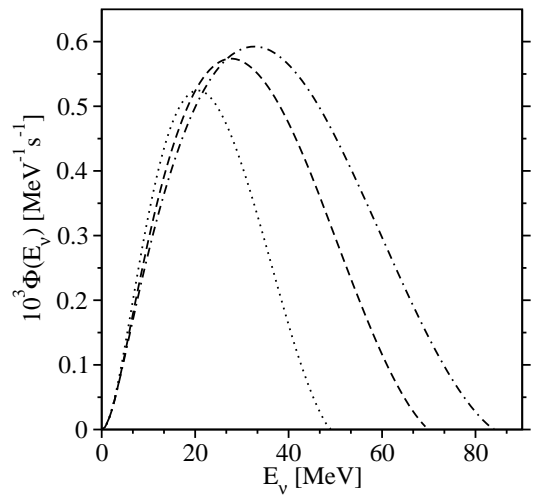}
\vskip 0.15cm
\caption{ Anti-neutrino fluxes from low energy beta-beams.
The curves correspond to $\gamma=7$ (dotted line), $\gamma=10$
(broken line) and $\gamma=12$ (dash-dotted line) for the boosted
helium-6 ions. Similar curves are obtained for neutrinos from neon-18 ions.
The detector is located
at 10 m from a small devoted storage ring (see Section III) (Balantekin
\etal 2006a). }
\label{fig:lowbbfluxes}
\end{center}
\end{figure}

Figures \ref{fig:fluxes} and \ref{fig:lowbbfluxes} show examples of the neutrino fluxes from
beta-beams where the ions are boosted at $\gamma$ of several hundreds or from
low energy beta-beams with $\gamma=7-14$, respectively.    

\subsection{The baseline}
\begin{figure}
\begin{center}
\epsfig{file=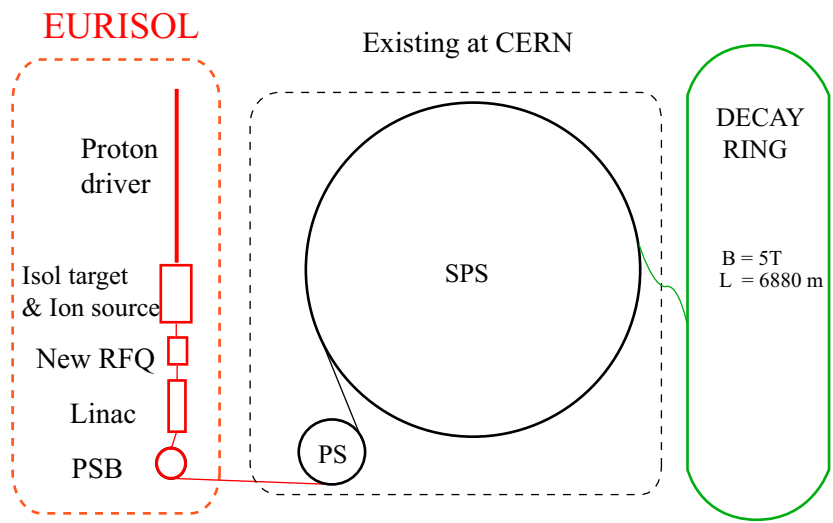,scale=0.5}
\caption{Schematic drawing of a beta-beam facility complex at CERN.
The production, collection and acceleration part shown on the left
is very close to the EURISOL project (see EURISOL), at present under study, which aims
at producing very intense radioactive ion beams. The ions are
then injected in the Proton-Synchrotron (PS) and Super Proton-Synchrotron
(SPS) accelerator infrastructures (central part) already existing at CERN,
to achieve about a hundred GeV per nucleon ($\gamma=100$).
Finally the ions are stored to decay in an hippodrome-like storage ring
(right part), that needs to be built. One of the long straight sections
points to a far detector (Zucchelli 2002, Autin \etal 2003).}
\label{fig:baseline}
\end{center}
\end{figure}

The original conceptual design for a beta-beam beam facility 
includes
three main steps (Zucchelli 2002, Autin {\it et al} 2002, Bouchez
{\it et al} 2003, Benedikt {\it et al} 2004, Lindroos and Volpe 2004, Benedikt 2005).
In the first stage protons delivered
by a Superconducting Proton Linac (SPL) 
impinge on a target to produce radioactive
species of interest, using an ISOL target.
Such an intense proton driver (at present under study) 
would deliver 2mA of 2.2 GeV (kinetic energy) protons, while the ISOL target
would need 100 $\mu$A, that is 5 \% of the proton intensity only.
After ionization and bunching,
the ions are accelerated to low energy, i.e. 300 MeV/A. 
This part of the facility is very close to the conceptual design of
the EURISOL project (see EURISOL), aiming to the production and acceleration of
very intense radioactive ion beams through the ISOL technique.
In the second stage
energies of about 100 GeV/A are attained using e.g. CERN accelerator
infrastructures, like the PS and SPS. Finally the ions are stacked
longitudinally in a new decay ring. Each of these steps require specific
feasibility studies, in particular the bunching at low energy, space-charge
limitations in the PS and SPS (keeping CERN in mind as a possible site),
decay losses in the accelerator chain and, at high energy, the stacking
process in the decay ring. 
Both the decay ring, the proton source and the low energy
accelerators do not exist at CERN today. Nevertheless, the use of existing
accelerator infrastructure represents an
important saving. 
The beta-beam complex is shown schematically in 
Figure \ref{fig:baseline}.

The choice of the isotopes is based on several criteria. The half-life
should be such as to minimize the decay losses in the accelerator
chain. Another constraint comes from the required collimation and energy of
the neutrinos at the detector, which depends on the energy of the decaying
ions through $E_{\nu} \approx 2 \gamma Q_{\beta}$. 
These (and other) considerations point at present at two isotopes of
particular interest, namely $^{6}$He and $^{18}$Ne as anti-neutrino and
neutrino emitters respectively. Their lifetimes are 0.8 s and 1.6 s, while their $Q_{\beta}$-values are of about 4 MeV. 
Both these ions can be produced in large quantities by the so-called ISOL
method. These two noble gases have the advantage that they can easily diffuse
from the target. For helium, it consists either of a water cooled tungsten
core or of a liquid lead core which works as a proton to neutron converter
surrounded by a beryllium oxide, aiming for $10^{15}$ fissions per second.
This yields a quasi-continuous beam, so the ions have to be bunched
prior to further acceleration.
Efficient bunching and full stripping of a high-intensity beam can be achieved
using a high frequency ECR source (Sortais 2003). While such system does not
exist today, theoretical calculations show that it can be constructed.
Note that very intense $^{8}$Li and $^{8}$B beams can be achieved with a  
new ``ionisation cooling technique'' (Rubbia \etal 2006).

16 bunches (consisting for example of 2.5 $10^{12}$ ions each for helium)
are first accumulated in the PS, reduced to 8 bunches during the acceleration
to intermediate energies, injected in the SPS and then
in the storage ring in batches of four 10 ns bunches. This time
structure is required to efficiently reduce the atmospheric background.
Note that the SPS can accelerate $^6$He and $^{18}$Ne ions at a maximum $\gamma$
value of 150 and of 250 respectively (Lindroos 2005). 
Due to the relativistic
time dilatation, the ion lifetimes reach several minutes, so that
stacking the ions in the decay ring is mandatory to get enough
decays and hence high neutrino fluxes.
The ring has the shape of an hippodrome, with a total length of 6880 m
and straight sections of 2500 m each (36 \%). 
The average number of ions in the storage ring between two stacking
cycles is given by $N_{average}=N_{bunch} \gamma \tau /T$ where 
$N_{bunch}$ is the number of particles with mean-life $\tau$
having Lorentz boost $\gamma$, injected into the ring every $T$ seconds. 
\noindent

The ion injection and merging 
with existing high density bunches in the decay ring represent
an important 
challenge and require a new scheme. Stacking
means a regular top-up of the stored beam before the existing particles cease
to be useful. Many stacking methods involve some form of a beam cooling to
increase phase space density and make room for more particles within a given
transverse and longitudinal acceptance. For light ions at high energy, the
classical methods of electron and stochastic cooling are excluded. Instead, a
new scheme has been proposed combining bunch rotation and asymmetric
bunch pair merging. The new bunches are off momentum and are injected in a
high dispersion region on a matched dispersion trajectory. Subsequently, 
each injected bunch rotates a quarter turn in longitudinal phase space until
the initial conditions for bunch pair merging are met. In the final step,
each small fresh bunch is moved into the core of a large stored one and
then phase space mixing occurs. The fact that only the central part of the
existing bunch is combined with an incoming dense one results in a net
increase in the core intensity of the resultant stored beam. The surrounding
older ions are pushed out toward the bucket separatrix, where eventually
the oldest are lost. This new method, called asymmetric bunch pair merging, 
has recently been demonstrated at the CERN PS.

A principal difference between the acceleration to high energies of
stable and radioactive ions is the additional losses due to the radioactive
decays. The predicted intensities for $^{6}$He and $^{18}$Ne are shown in Table 
\ref{tab:intensities}.
The conventional losses through the accelerator chain are expected 
to be less than 50 \% (based on operational experience at CERN). 
The isotopes proposed for the beta-beam conceptual
design have been chosen such that no long-lived activity is left to
contaminate the low energy machines. A first
simulation of losses in the decay ring yields a dose rate in the arcs of 2.5
mSv/h after 30 days of operation and one day of cooldown (Magistris and Silari 2003).
Furthermore, the induced radioactivity in the ground water will have no
impact on public safety. It has also been confirmed that the total loss in
the accelerator chain will be beyond the nominal 1 W/m that permits hands-on
maintenance (except for the PS).

\begin{table}
\caption{\label{tab:intensities} Intensities 
for the $^{6}$He  and $^{18}$Ne ion beams along the acceleration chain. 
Only beta-decay losses are taken into
account. The number of ions in the decay ring, including conventional losses, is 
$0.5 N_{tot}$ (Autin {\it et al} 2002, Lindroos 2003, Benedikt \etal 2004, Benedikt 2005).}
\begin{tabular}{@{}lccc}
\\ \hline
Machine & $^6$He ions extracted & $^{18}$Ne ions extracted & Batches 
\\ \hline \hline
Source & 2$\times 10^{13}$ & 8$\times 10^{11}$ ions/s & dc \\
Rapid Cycling Synchrotron & 1.00$\times 10^{12}$ & 4.09$\times 10^{10}$ & 16 \\
PS & 1.01$\times 10^{13}$ & 5.19$\times 10^{11}$ & 1  \\
SPS & 0.95$\times 10^{13}$  & 4.90$\times 10^{11}$ & $\infty$  \\
Decay Ring ($N_{tot}$) & 2.02$\times 10^{14}$ & 9.11$\times 10^{12}$ & - \\
\hline
\end{tabular}
\end{table}

\subsection{Sensitivity to $\cal{CP}$ violation}
The major goal of the beta-beam facility proposed by (Zucchelli 2002) is
to measure $\cal{CP}$ violation in the lepton sector.
This requires a comparison of neutrino versus anti-neutrino oscillations
in an appearance experiment Eq.(\ref{e:CPgeneral}) (note that disappearance experiments like
$\nu_e \rightarrow \nu_e$ are not sensitive to the $\cal{CP}$ violating phase 
$\delta$). Since beta-beams
produce pure electron neutrino (or anti-neutrino) beams, such experiment should
search for differences in $\nu_{e} \rightarrow \nu_{\mu}$
versus $\bar{\nu}_e \rightarrow \bar{\nu}_{\mu}$ oscillations
Eq.(\ref{e:probCP}). The scenario considered is to produce the neutrino
beams at CERN (Figure \ref{fig:baseline})
and fire them to a water \v{C}erenkov detector of 440 kton
fiducial volume -- like MEMPHYS (Mosca 2005a, Campagne \etal 2006),
UNO (Jung 2000), or HYPER-K (Itow \etal 2001) --
located at the (upgraded) Fr\'ejus Underground Laboratory, 130 km from CERN.
If Lorentz ion boosts around 60-100 are chosen (Fig.\ref{fig:fluxes}), this
distance magically fits the  $\nu_{e} \rightarrow \nu_{\mu}$ first
oscillation maximum at the atmospheric scale (Fig.\ref{fig:CPterms}).
The detector is of the same kind of Super-Kamiokande (Fukuda \etal 1998),
but has a fiducial mass 20 times larger, requiring a cavern of about 1 million cubic meters. 
The ion intensity considered is 2.9$\times 10^{18}$ $^6$He and 1.1$\times 10^{18}$ $^{18}$Ne per year
(Autin \etal 2003).

The physics potential of the original baseline scenario is being actively
investigated (Mezzetto 2003, Bouchez \etal 2003, Mezzetto 2005, Mezzetto 2006, Campagne \etal 2006)
and is discussed in (Guglielmi \etal 2005, Mosca 2005b, Bouchez 2005).
The signal corresponds to the muons produced by
$\nu_{\mu}$ charged-current events in water, mainly via quasi-elastic
interactions at these energies. Such events are selected by requiring a
single-ring event, with the same identification algorithms used by the
Super-Kamiokande experiment (Fukuda \etal 1998), and by the detection
of the electron from the muon decay.
At such energies the energy resolution is very poor due
to the Fermi motion and other nuclear effects. For these reasons, a $\cal{CP}$
violation search with $\gamma=60-100$ is based on a counting experiment only.
Energy cuts can be exploited when going to higher $\gamma$, as first 
suggested in (Burguet-Castell 2004) (see Section IV) 
e.g. to resolve some of the degeneracies and get information on the neutrino
hierarchy.

The beta-beam has no intrinsic
backgrounds, contrary to conventional sources. However, inefficiencies in
particle identification, such as single-pion production in neutral-current
$\nu_e$ ($\bar{\nu}_e$) interactions, electrons (positrons) misidentified
as muons, as well as external sources, like atmospheric neutrino interactions,
can produce backgrounds. The background coming from single pion production
has a threshold at about 450 MeV, therefore giving no contribution for
$\gamma < 55$. Standard algorithms for particle identification
in water \v{C}erenkov detectors are quite efficient in suppressing the fake
signal coming from electrons (positrons) misidentified as muons, especially
if the signal for the muon decaying into electrons is also required.
Concerning the atmospheric neutrino interactions, 
estimated to be of about 50/kton/yr, this important background is reduced to
1 event/440 kton/yr by requiring a time bunch length for the ions of 10 ns.

The choice of the ion acceleration is a compromise between having the
$\gamma$ factor as high as possible, to profit of larger cross sections
and better focusing of the beam on one hand, and keeping it as low as
possible to minimize the single pion background and better match the
$\cal{CP}$ odd terms Eq.(\ref{e:probCP}) on the other hand. If one requires the ions to circulate
together in the decay ring (Figure \ref{fig:baseline}), this constraints the ratio of the ion boosts
to be $\gamma(^{6}$He)/$\gamma(^{18}$Ne)=3/5. In this case the optimal boosts turn
out to be $\gamma(^{6}$He)=60 and $\gamma(^{18}$Ne)=100, giving
0.24 GeV and 0.36 GeV as $\bar{\nu}_e$ and $\nu_e$ average energies
respectively (Figure \ref{fig:fluxes} left). 
This choice is assumed in (Bouchez \etal 2003, Mezzetto 2005).

\begin{figure}[htb]
\centerline{\epsfig{file=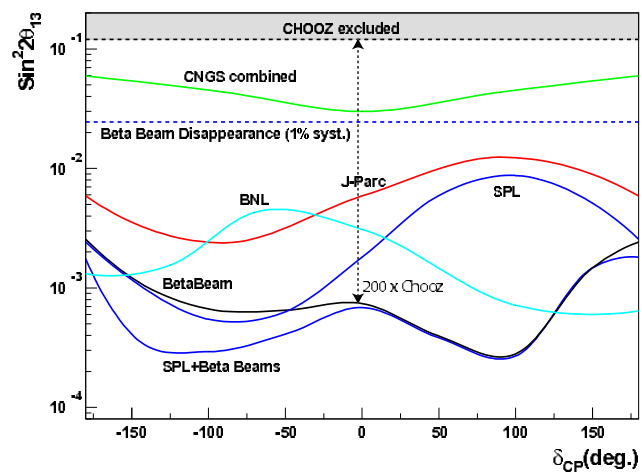,scale=0.5}  }
\caption{Sensitivity on $\theta_{13}$ (90 \% C.L.) as a function of $\delta$
for a beta-beam, a super-beam (SPL), and their combination.
The present CHOOZ limit (Apollonio \etal 1999) is shown, as well as the sensitivities of
CNGS and JPARC (Migliozzi and Terranova 2003) and BNL (Diwan \etal 2003). The appearance results
are calculated considering 5 years measuring time (Mezzetto 2005).}
\label{fig:th13-60-100}
\end{figure}

Table \ref{t:events} shows the expected events with and without oscillations.
One can see that the beam related background is zero, while some
background events are expected especially for the case of neon-18, due to the
single-pion misidentified events.
The expected sensitivity on $\theta_{13}$ is shown in Figure
\ref{fig:th13-60-100}  after 5 years run, in comparison with planned or proposed
experiments. 
Note that the value of this angle can be explored both with $\bar{\nu}_e$ and $\nu_e$
disappearance and with $\bar{\nu}_{\mu}$ and $\nu_{\mu}$ appearance experiments.
Clearly, a beta-beam alone as well as a combination of a super-beam and a beta-beam sent to the same
detector are powerful and allow to reach sensitivities on $\sin^2 2\theta_{13}$
in the range of $10^{-3}-10^{-4}$, improving the present CHOOZ upper limit 
(Apollonio \etal 1999) by a factor of about 200.
\begin{table}
\caption{\label{t:events}
Number of events expected after 10 years, 
for a beta-beam produced at CERN (Figure \ref{fig:baseline}) and sent to 
a 440 kton water \v{C}erenkov detector located at an (upgraded) Fr\'ejus Underground Laboratory, at 130 km 
distance. The results correspond to $\bar{\nu}_e$ (left)  and  $\nu_e$ (right). 
The different $\gamma$
values are chosen to make the ions circulate together in the ring (Mezzetto 2005).}
\begin{tabular}{@{}lrr}
&  $^6$He               &   $^{18}$Ne                \\
&       ($\gamma=60$)  &  ($\gamma=100$)  \\
\hline
CC events (no oscillation) & 19710   & 144784  \\
Oscillated ($\sin^2 2\theta_{13}=0.12$, $\delta=0$) & 612 & 5130   \\
Oscillated ($\delta=90^\circ$,$\theta_{13}=3^\circ$)&  44     &  529   \\
Beam background            &  0      &  0\\
Detector backgrounds       &   1     &  397\\
\end{tabular}
\end{table}
				 
\begin{figure}[ht]
\centerline{\epsfig{file=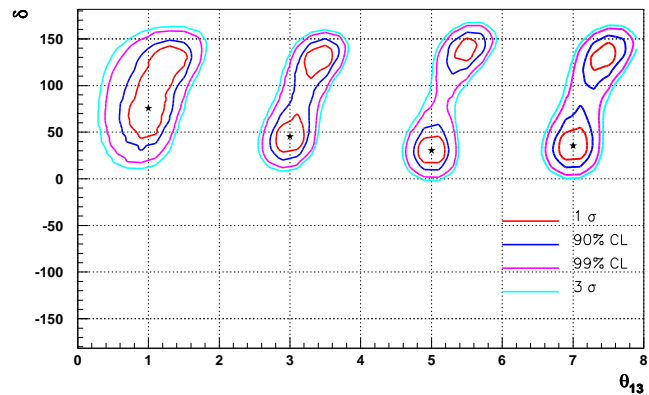,width=0.55\textwidth}  }
\caption{Fits to $\theta_{13}$ and $\delta$ after ten years measurement time,
computed with the events shown in Table \ref{t:events}.
The stars indicate the parameters' values, while the different curves show the
various confidence levels (Mezzetto 2005).}
\label{fig:many_plots}
\end{figure}

The sensitivity in the ($\theta_{13}$,$\delta$) plane is shown in Figure
\ref{fig:many_plots}, having fixed the other oscillation parameters to the values
 $\sin^2 2\theta_{23}=1,~\Delta m^2_{23}=2.5~10^{-3}$eV$^2$, sign($\Delta m^2_{23}$)=+1, 
  $\sin^2 2\theta_{12}=0.8,~\Delta m^2_{12}=7.1~10^{-5}$eV$^2$. 
  The assumed corresponding errors are of 5 \% and of 1\% on $\Delta m^2_{23}$
  and on $\sin^2 2\theta_{23}$ respectively, as expected from the T2K experiment (Itow \etal 2001),
  while 10 \% error is taken for the solar oscillation parameters.
  The curves are determined
  by fitting the number of muon-like events in the $\bar{\nu}_{\mu}$ and $\nu_{\mu}$
  oscillation experiments simultaneously (Table \ref{t:events}). 
  Correlations between $\theta_{13}$ and $\delta$ are fully accounted for and 
  have a negligible effect (Figure \ref{fig:many_plots}, see section V for a discussion on the intrinsic
  degeneracy). 
  The ambiguity on the sign of $\Delta m^2_{23}$
  is translated in a direct ambiguity on the sign of $\delta$, since matter effects at 130 km are
  negligible Eq.(\ref{e:probCP}) (Figure \ref{f:mattereffects}). The ambiguity coming from
  the choice $\theta_{23}$ or $\pi/2 - \theta_{23}$ is formally taken into account but has no
  effects in the results shown here, since the choice $\theta_{23}=45 ^\circ$ is made.
  A detailed discussion of the issues of degeneracies, correlations and clones is made in Section V.
    
\begin{figure}[th]
\centerline{\epsfig{file=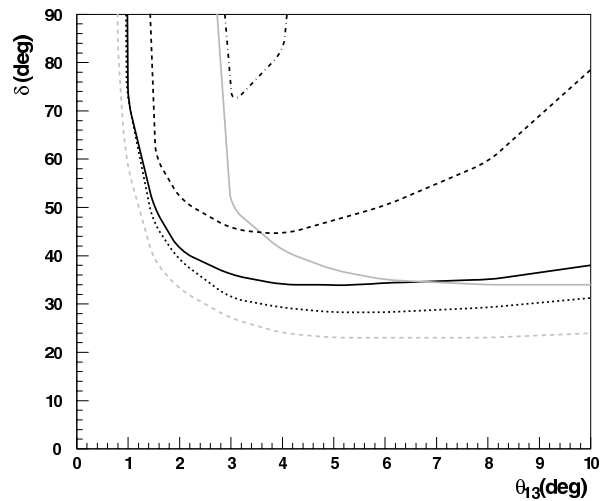,width=0.50\textwidth} }
\caption{Discovery potential for $\delta$ at 3 $\sigma$, considering 10 years
measurement time, with the two ions running at $\gamma=60,100$.
The effect of systematic errors is shown. The curves
correspond to 2 \% (solid),  5 \% (dashed),  10 \% (dot-dashed) systematic
errors. The results for a beta-beam with the two ions running at the same
$\gamma=75$ with 2 \% systematic error is also shown (dotted). The two grey
curves present the sensitivity of a super-beam (solid) and super-beam combined to a
beta-beam (dashed) with 2 \% systematic error (Mezzetto 2005).}
\label{fig:CP:delta}
\end{figure}

  In order to achieve the highest sensitivity
  it is clear that the systematic errors have to be kept at their lowest possible
  level. Such errors
come essentially from the knowledge of the cross sections -- which at the
neutrino energies of the beta-beams (several hundred MeV) are not well known --
as well as from the precision with which all the other oscillation parameters are
known, at the time such an experiment takes place.
By placing a close detector of 1 kton at least at 1 km from the decay tunnel,
one can exploit the same neutrino beams to perform precision cross section measurements (Mezzetto 2003).
Note that, at the moment, the cross section uncertainty introduces variations in the
sensitivities found in the literature (see e.g. Singh \etal 2006a for a recent calculation of the 
neutrino-oxygen cross sections with beta-beams). 
As far as the energy and the flux of the neutrino beams are concerned, one should keep
in mind that contrary to conventional beams, in the case of beta-beams these
are completely defined by the intensity of the ions and the kinematics
Eqs.(\ref{e:events}) and (\ref{bflux}). 
The flux at the far detector is also completely determined 
by the divergence of the beam which is
perfectly known. 
One can therefore hope to measure the signal and the backgrounds with a residual systematic error
of 2 \%. Figure \ref{fig:CP:delta} shows the importance of keeping the systematic errors at their lowest 
level.

\begin{figure*}[thb]
\centerline{\epsfig{file=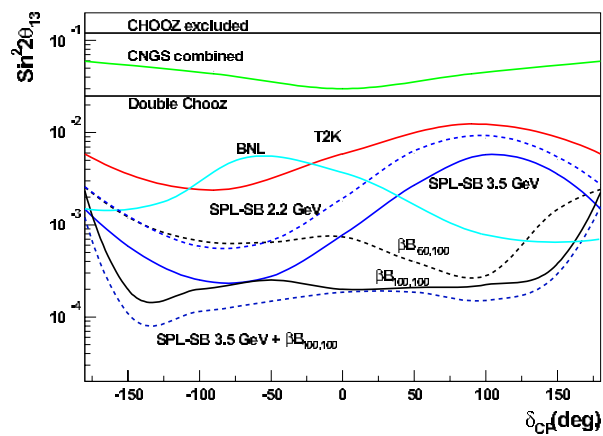,scale=0.5}  }
\caption{Sensitivity on the third neutrino mixing angle (90 \% C.L.) as a function
of the $\cal{CP}$ violating phase, with 2 \% systematic errors, as in Figure \ref{fig:th13-60-100}. 
The discovery potential of Double CHOOZ (Ardellier \etal 2004) is also presented.
Here two different
beta-beam options are compared, i.e. $\beta B_{60,100}$ and $\beta B_{100,100}$ where
$\gamma(^{6}$He)=60 and $\gamma(^{18}$Ne)=100 and $\gamma(^{6}$He) = $\gamma(^{18}$Ne)=100 are used
respectively
(Figure \ref{fig:fluxes}). The results correspond to 5 years $\nu_{\mu}$
running time for the super-beam case (SPL-SB) with either 2.2 or 3.5 proton kinetic energy, 
and 5 years $\nu_e$ and $\bar{\nu}_e$ measurement time for the beta-beam case (Guglielmi \etal 2005).  }
\label{fig:th13-100}
\end{figure*}

\begin{figure*}[thb]
\centerline{\epsfig{file=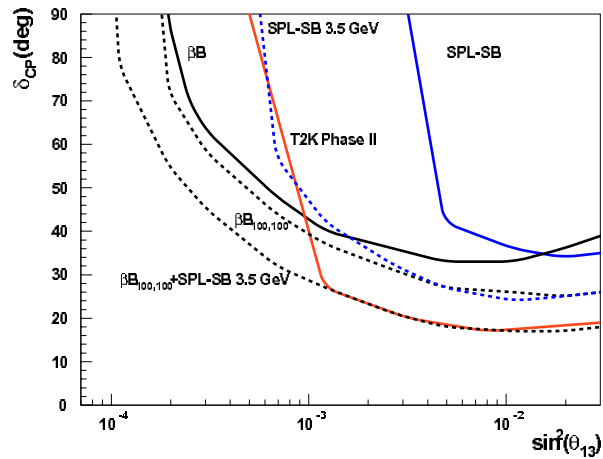,scale=0.5}  }
\caption{Same as Figure \ref{fig:th13-100}, but here all the results correspond to 10 years of data taking 
(Guglielmi \etal 2005).}
\label{fig:delta-th13-100}
\end{figure*}

A different optimization of the ion acceleration can be obtained if each type of
ion (either the $\nu_e$ or the $\bar{\nu}_e$ emitter) 
circulates separately at its optimal $\gamma$ value, as suggested in
(Mezzetto 2003, Burguet-Castell \etal 2005, Lindroos 2005).
In this case the
single fluxes can be doubled by filling all the batches with the same type of ions (instead of
alternating), which
gives the same integrated number of decays per year. 
The physics potential obtained following such a prescription is studied with
$\gamma(^{6}$He) = $\gamma(^{18}$Ne)=75 in (Mezzetto 2003) and with
$\gamma=100$ (Figure \ref{fig:fluxes}) in (Mezzetto 2006, Guglielmi \etal 2005, Campagne \etal 2006). 
These two choices 
well match the CERN-Fr\'ejus distance
of 130 km. 
The sensitivity on the third neutrino
mixing angle as well as on the $\cal{CP}$ violating phase is presented in Figures
\ref{fig:th13-100} and \ref{fig:delta-th13-100} respectively, showing an increased
discovery potential for both observables compared to the case where the
two types of ions circulate together in the storage ring. The beta-beam option
alone is clearly a better strategy than a super-beam alone, as already apparent
in Figures \ref{fig:th13-60-100} and \ref{fig:CP:delta}. Finally, the comparisons show that
the largest discovery potential can be attained by a combination of a super-beam and a beta-beam.

Since a beta-beam would exploit at most 10 \% of the SPL protons (the injector),
one could produce both a beta-beam and a super-beam and send these neutrino
beams to the same detector. This choice has several advantages. The use of the
two beams does not only increase the statistics, but also gives the possibility to reduce
the systematic errors and offers 
the necessary redundancy to firmly establish $\cal{CP}$ violating
effects, within the reach of the experiments. Besides, the combination of a beta-beam and
of a super-beam offers the largest discovery potential both with respect to the third mixing angle
and to the $\cal{CP}$ violating phase. The last important advantage 
is that also $\cal{T}$ and $\cal{CPT}$ violation studies can be performed 
through a comparison of $\bar{\nu}_e \rightarrow \bar{\nu}_{\mu}$ oscillations (from beta-beams) to
$\bar{\nu}_{\mu} \rightarrow \bar{\nu}_{e}$ Eq.(\ref{e:T}) and to
$\nu_{\mu} \rightarrow \nu_{e}$ Eq.(\ref{e:CPT}) 
oscillations (from super-beams) respectively  
(Mezzetto 2003).

An extensive study of the beta-beam option at $\gamma=100$ is made in (Campagne \etal 2006)
based on the GLoBES software (Huber \etal 2005a),
including correlations and degeneracies (Section V) and using atmospheric data in the analysis
(Huber \etal 2005b). The GloBES software is a convenient tool to simulate long-baseline experiments
and compare different facilities on the same footing. Correlations and degeneracies are
fully taken into account letting all oscillations parameters to vary in the fits.
A detailed comparison is made 
among the beta-beam, the super-beam\footnote{Note that the super-beam here has a new
optimization of the beam energy as well as of the secondary particle focusing and decay. This new optimization
has a significant impact on the physics performance (Campagne \etal 2006).} 
(with 3.5 GeV proton kinetic energy), and T2K phase II (T2HK) 
facilities. The latter corresponds to the second phase of the T2K experiment in Japan, with a 4 MW proton
driver and the Hyper-K detector located at 295 km (Itow \etal 2001). It is a competing proposal with
beta-beams with a similar size and timescale.
In this work it is pointed out that if one assumes the $\cal{CPT}$ symmetry,
a combination of a beta-beam and a super-beam sent at the same detector can be used to replace the
anti-neutrino measurement by the $\cal{T}$ conjugate one. This implies that if both beams are
available, the full information can be extracted with the neutrino beams only, reducing the 10
years measurement time down to 5. The corresponding sensitivity comes out to be very close to
the 10 years measurement time with neutrinos and anti-neutrinos from one single experiment.
 
\begin{figure}
  \centering
  \includegraphics[width=0.9\textwidth]{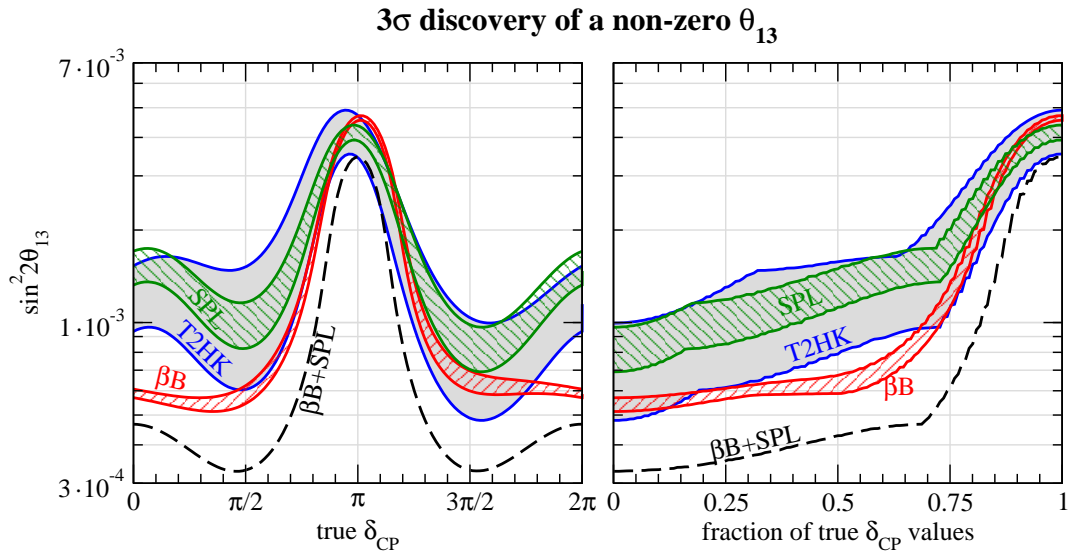}
  \protect\caption{Sensitivity on $\theta_{13}$ at 
  $3\sigma$ for
  the ($\gamma=100$) beta-beam ($\beta$B), a super-beam (SPL), 
  and T2HK (see text) as a function of the true value of $\delta_{\rm{CP}}$
  (left) and as a function of the fraction of all possible
  values of $\delta_{\rm{CP}}$ (right). The width of the bands corresponds
  to values with 2\% to 5\% systematical errors. The
  dashed curves correspond to the combination of the beta-beam and the super-beam with
  10~yrs measurement time each and 2\% systematical error (Campagne \etal 2006).\label{f:100theta}}
\end{figure}

\begin{figure}
  \centering
   \includegraphics[width=0.65\textwidth]{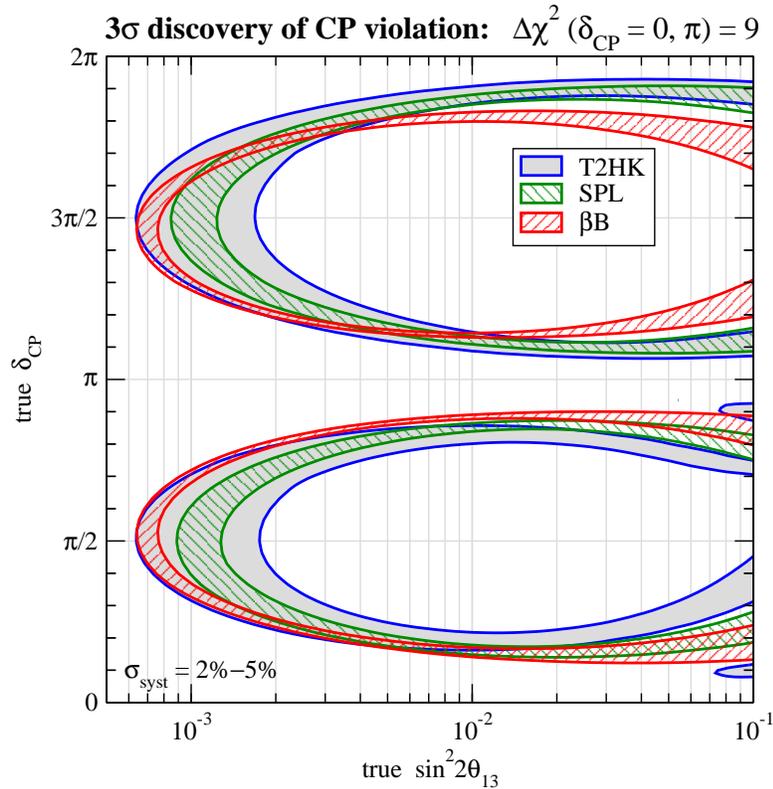}
   \protect\caption{Same as Figure \ref{f:100theta}, but for the $\cal{CP}$ 
   discovery reach. For parameter values inside the ellipse-shaped curves CP conserving
   values of $\delta_{\rm{CP}}$ can be excluded at $3\sigma$ $(\Delta\chi^2>9)$ (Campagne \etal 2006). 
   \label{f:100cp} }
\end{figure}

Figures \ref{f:100theta} and \ref{f:100cp} show the sensitivity on the third mixing angle and the 
discovery reach with respect to $\cal{CP}$ respectively.
One can see that the three facilities have quite similar performances and can clearly reach
values as small as $\sin^2 2 \theta_{13}=5~10^{-3}$, independently on $\delta$.
Besides, for a large fraction of $\delta$ values, a sensitivity of $~10^{-3}$ can be attained.
Assuming 2 \% systematical error, in the best case $5(7)~10^{-3}$ can be reached for the beta-beam
and T2HK (SPL). In a combined beta-beam and super-beam (SPL) experiment with 10 years measurement time
$4~10^{-4}$ is reached for 45 \% of $\delta$ values. Concerning $\cal{CP}$, the maximal violation
corresponding to $\delta=\pi/2$ and $3 \pi/2$ can be discovered down to 
$\sin^2 2 \theta_{13}=6(8)~10^{-4}$ (99 \% C.L.) for the beta-beam (super-beam); while the best
sensitivity is obtained for $\sin^2 2 \theta_{13} > 10^{-2}$ since at this value 
$\cal{CP}$ violation can be established for 73 \% (75 \%) for all $\delta$ values with
a beta-beam (super-beam). One can also see that the impact of systematical errors is 
crucial in the case of the T2HK experiment while the beta-beam and the SPL are much less sensitive.
The comparison with T2HK shows that the $\gamma=100$ beta-beam option has a very similar
discovery reach and offers a competitive strategy for future neutrino precision oscillation
measurements.

Most of the literature concerning beta-beams considers the water \v{C}erenkov technique for the far detector.
There are nowadays three proposal for very large water \v{C}erenkov detectors: MEMPHYS in Europe
(Mosca 2005a, Campagne \etal 2006), 
UNO in the United States (Jung 2000) and HYPER-K in Japan (Itow \etal 2001).
This mature technology has already demonstrated excellent performances at the neutrino energies
of several hundred MeV to 1 GeV. 
Other techniques are now being discussed, in particular
the liquid Argon Time Projection Chamber (Rubbia C 1977, Rubbia A 2004, Ereditato and Rubbia 2004 and
2006) and iron calorimeters.
It is important to emphasize that the large scale detector considered here can be exploited
for a multidisciplinary program covering crucial issues in astrophysics, such as the detection of 
neutrinos from a(n) (extra-)galactic core-collapse Supernova, and in high energy physics, like
proton decay.
The field of neutrino astronomy has started with Davis' pioneering experiment (Davis 1964)
which has measured solar neutrinos from the first time. Afterwards, neutrinos from a core-collapse
Supernova have been observed 
during the explosion of the SN1987A, located in the Large Magellanic Cloud,
at about 50 kpc from the Earth. For these important observations, both R.
Davis and M. Koshiba received (with R. Giacconi) the Nobel Prize in 2002.
Nowadays, neutrinos are currently used as a probe of astrophysical objects.
In particular, the measurement of the neutrino lightcurve from a future
core-collapse Supernova explosion constitute an essential ingredient 
to unravel the explosion mechanism, which keeps being a mystery: 
in the most recent tri-dimensional models the shock
wave stalls and fails to eject the mantle.
The neutrinos closely follow
the explosion, from the collapse and the accretion phases, to the cooling of
a proton-neutron star, or the formation of a black hole. 
Besides 
a recently proposed technique, that consists in adding Gadolinium to a water \v{C}erenkov detector
(Beacom and Vagins 2004), can increase the sensitivity to the diffuse Supernova neutrino background
from past Supernova explosions, for which there is at present an upper limit.
The observation of both future and past Supernova neutrinos would bring invaluable
information not only for our knowledge of Supernova physics, but also on neutrino
properties which are still poorly known (Raffelt 1996, Fogli \etal 2005).
The water \v{C}erenkov detector considered for the beta-beam project can measure about
150000 events from a Supernova at 10 kpc -- while presently running detectors can
observe at most 10000 events -- and about 50 events from Andromeda. A flux of 
250 events/10 years/500 kton from the diffuse background that can be increased by a factor
of 10 with Gadolinium.
Concerning proton decay one can attain a sensitivity up to $10^{35}$ years in the $e^+~ \pi_0$ channel
and a few $10^{34}$ years in the $K^+~ \bar{\nu}$ channel (Jung 2000, Itow \etal 2001, 
Mosca 2005a).

\section{Low energy beta-beams}
Very soon the idea of exploiting the beta-beam method to establish a facility
that produces low energy neutrino beams -- in the 100 MeV energy
range -- has been proposed, opening new axis of research in various
domains of physics (Volpe 2004). This corresponds to a range of
about 7-14 for the ion boosts $\gamma$ (Figure \ref{fig:lowbbfluxes}).
The physics potential of such a facility is now being analyzed.
Several applications have been studied
concerning nuclear structure studies and neutrino-nucleus interactions
(Volpe 2004, Serreau and Volpe 2004, McLaughlin 2004, Volpe 2005a), 
electroweak tests of the Standard Model
(McLaughlin and Volpe 2004, Balantekin \etal 2006a, Balantekin \etal 2006b)
as well as core-collapse Supernova physics (Volpe 2004, 
McLaughlin and Jachowitz 2005 and 2006).
The physics case is also discussed in (Volpe 2005b, Volpe 2005c, Volpe 2006).
A crucial outcome of (Serreau and Volpe 2004) 
is that 
a small devoted storage ring appears as more appropriate for low energy applications.
The feasibility study of this storage ring (concerning e.g. size, ion intensities, stacking method,
space charge effects) is now ongoing (Chanc\'e and Payet 2005, Benedikt \etal 2006).
Here we summarize the status on the physics potential of low energy
beta-beams\footnote{Note that, after the proposals of pushing
the ion Lorentz boosts to values higher than 60-100, 
the original scenario (Section II) is sometimes 
misleadingly referred to as "low energy" in the literature, to distinguish it from 
the medium and high-energy (Section IV).}.

\subsection{Nuclear structure, neutrino-nucleus, nuclear astrophysics applications}
\noindent
In (Volpe 2004) it has been pointed out that the availability of neutrino beams in the 100 MeV
energy range offers a unique opportunity to study spin-isospin and isospin nuclear excitations,
for which nowadays little experimental information is available.
Both  the isospin and spin-isospin
collective (and non-collective) modes are excited when a neutrino encounters a nucleus, 
due to the
vector and axial-vector
nature of the weak interaction.
The (super-)allowed Fermi
transitions  -- due to the vector current and therefore to the isospin
operator -- offer a well known example of such excitations at low momentum transfer.
A precise knowledge of these nuclear transitions is essential for determining the 
unitarity of the CKM matrix, the analog of the MNSP matrix
Eq.(\ref{e:mnsp}) in the
quark sector (see e.g. Hardy and Towner 2005). 
Another less known but still intriguing example is
furnished by the allowed Gamow-Teller transitions  -- due to the
axial-vector current and therefore to the
spin-isospin operator -- in mirror nuclei, which are a tool for the search of
the possible existence of second-class currents in the weak interaction
(Weinberg 1958, Wilkinson 2000). These terms transform in an
opposite way under the $\cal{G-}$ parity transformation\footnote{The $\cal{G-}$ parity transformation
corresponds to the product of the
charge-conjugation and of a
rotation in isospin space.} as the usual vector
and axial-vector terms, and are not present in the
Standard Model. 
Because of their importance, 
both the allowed Fermi and Gamow-Teller
transitions have been studied for a very long time
in nuclear physics through beta-decay and charge-exchange 
reactions (for the Gamow-Teller ones).
While the precision achieved in the description of the (super-)allowed Fermi
transitions is really impressive, our understanding of 
the allowed Gamow-Teller transitions still requires the use of an effective
axial-vector coupling form factor, to take into account the "quenching" of the predicted
transitions, compared to the ones measured in beta-decay or charge-exchange reactions 
(see for example Osterfeld 1992).
The still open "quenching" problem represents a
limitation in our description of the weak spin-isospin nuclear response,
in spite of the crucial role that it plays in various hot issues in 
nuclear astrophysics and in high energy physics. 

\begin{figure}
\begin{center}
\includegraphics[angle=-90.,scale=0.4]{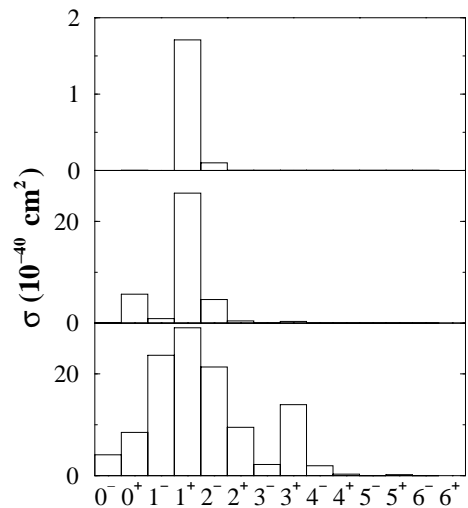}
\end{center}
\protect\caption{{\sc Nuclear structure studies with low energy beta-beams:}
Contribution of isospin and spin-isospin nuclear states excited in
the charged-current $^{208}$Pb$(\nu_{e},e^-)^{208}$Bi reaction
$(10^{-40}~$cm$^2)$
for increasing neutrino energy, i.e. $E_{\nu_e}=15~$MeV (up), $30~$MeV (middle),
$50~$MeV (bottom).
The histograms show the Isobaric Analogue State ($J^{\pi}=0^+$),
the allowed Gamow-Teller ($J^{\pi}=1^+$), the spin-dipole
($J^{\pi}=0^-,1^-,2^-$), as well as
states of higher multipolarity ($J^{\pi}=2^+,3^-,3^+,4^-,4^+$).
For the latter no experimental information is available.
Their contribution to the total cross section
becomes significant when the impinging neutrino energy increases
(Volpe 2004).\label{f:Pb}}
\end{figure}

Little or no experimental information is available for the spin-isospin and isospin
nuclear excitations such as the spin-dipole
($J^{\pi}=0^-,1^-,2^-$),
or the states of higher multipolarity (e.g. $J^{\pi}=2^+,3^-,3^+,4^-,4^+$).
These states come into play 
when a weak probe transfers a finite momentum to a nucleus, like in muon capture or in
neutrino-nucleus interactions.
Supplementary information on the corresponding weak transition amplitudes can be 
furnished by electron scattering studies, which however explore nuclear excitations
induced by the vector current only.
As an illustrative example Fig.\ref{f:Pb} shows the contribution
of spin-sospin and isospin transitions excited in 
neutrino scattering on lead, and their evolution 
when the neutrino energy increases.
A quantitative estimate of the importance of such states is also gathered
by computing the flux-averaged cross sections\footnote{Flux-averaged cross sections are
obtained by folding the cross sections
with the neutrino flux of the source.}, which
are the relevant quantities for experiments.
If one considers the 
neutrino fluxes corresponding to the decay-at-rest of muons,
the spin-dipole states ($J^{\pi}=0^-,1^-,2^-$)
contribute by about 40 \% in $^{12}$C~(Volpe {\it et al} 2000)
and $^{56}$Fe~(Kolbe and Langanke 1999), 
and by about 68 \% in $^{208}$Pb~(Volpe {\it et al} 2002).
The contribution from the states of higher multipolarity is about 5 \% and 25 \%
in iron and lead respectively; while it amounts to
about 30 \% in carbon~(Volpe {\it et al} 2000) and 60 \% in lead~(Volpe
{\it et al} 2002)
if neutrinos are produced from pion decay-in-flight.
Since low energy beta-beams have the specificity that the average neutrino
energy can be increased by increasing the Lorentz boost of the ions, they constitute a
promising tool for the study of these states, through a systematic
study on various nuclear targets and different neutrino energies.
Even though the measured cross sections are, in the majority
of cases, inclusive, experimental information on these
states can be extracted 
by changing the Lorentz ion boosts since different pieces of the nuclear
response are important at different energies (Volpe 2004, Figure \ref{f:Pb}).

\begin{figure*}[t]
\vspace*{-4cm}
\epsfig{file=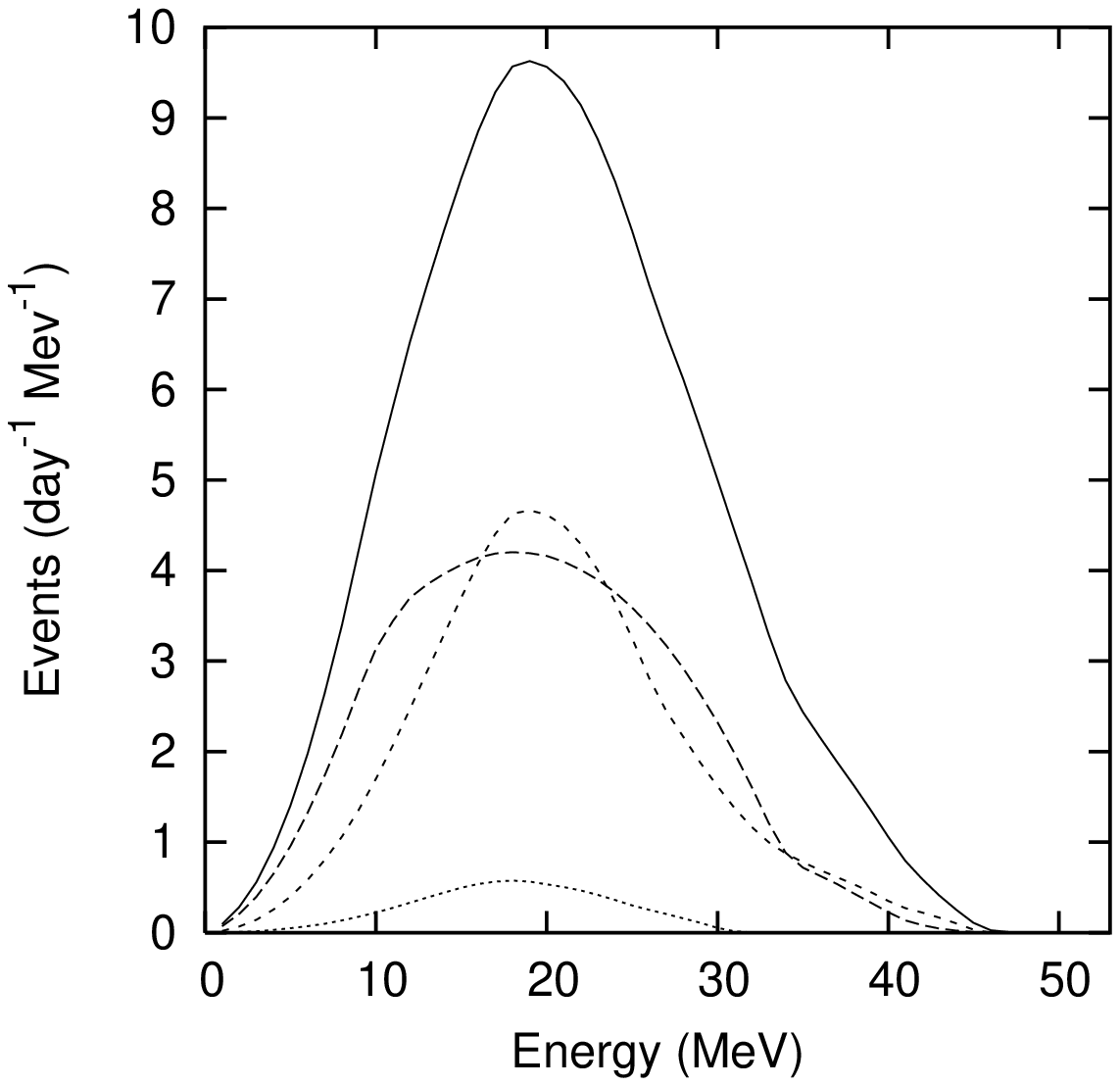,width=0.45\textwidth}
\epsfig{file=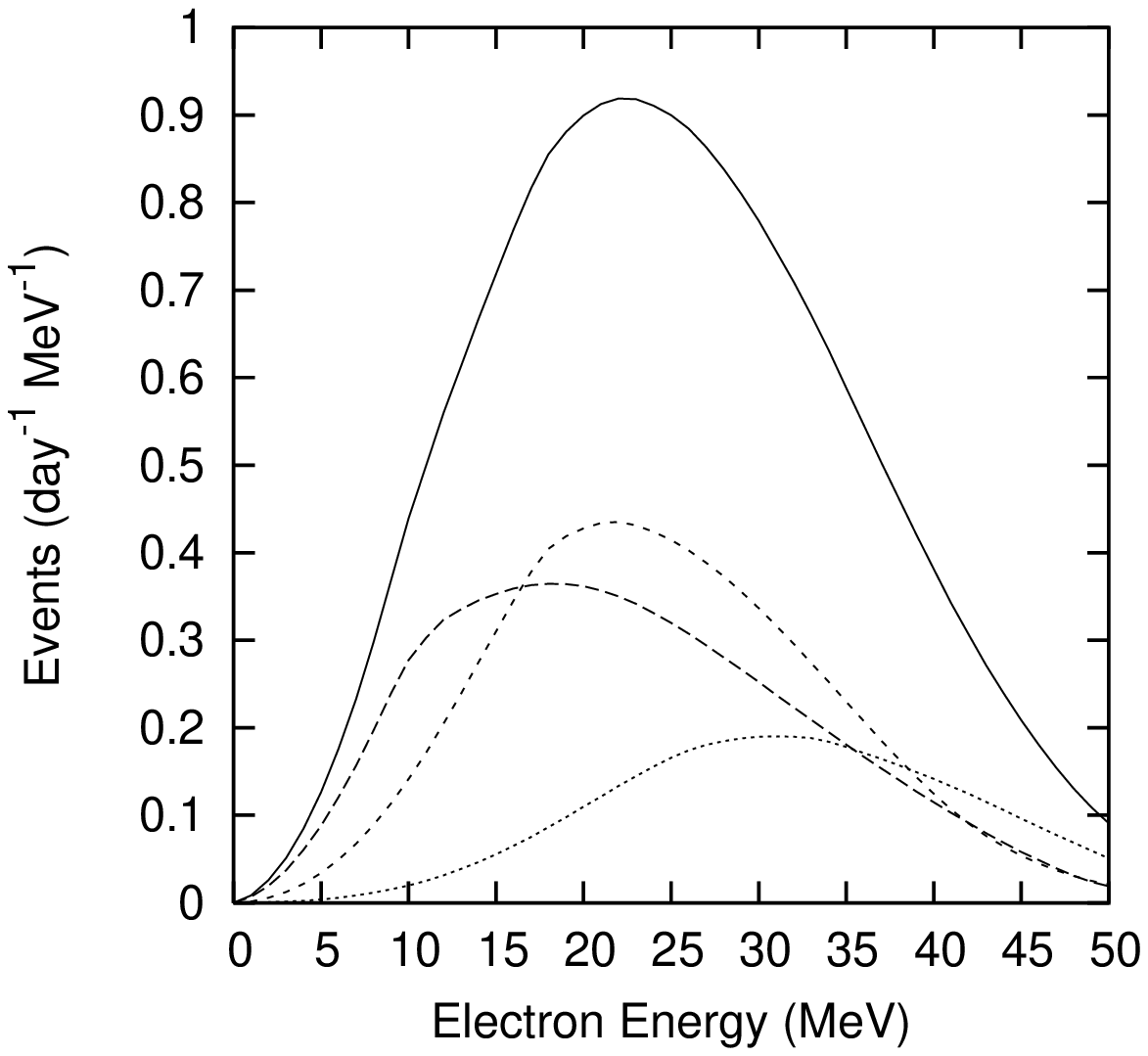,width=0.45\textwidth}
\vskip -0.5cm
\caption{Total electron spectrum coming from charged current neutrino scattering on lead (solid line),
with neutrinos coming from pion decay-at-rest (left) and from low energy beta-beams
with $\gamma=10$ (right).
The other curves present the contribution from the allowed Fermi and Gamow-Teller states (long dashed
line), from the $0^-,1^-,2^-$ and the $2^+,3^-,3^+,4^-,4^+$ states 
(dotted line) (McLaughlin 2004).}
\label{f:snslead}
\end{figure*}

A facility based on conventional sources (decay of pions and muons)
offers another possible strategy to dispose of low energy neutrinos beams.
The first feasibility studies show that the neutrino intensities
are expected to be higher with conventional sources (Avignone F T \etal 2000, Efremenko 2004) 
than with beta-beams. On the other hand,
the neutrino beams obtained with the two options present complementary 
features, both for the flavour content and for the energy. Conventional
sources provide us with neutrinos of different flavours and
a Michel spectrum (from muon decay-at-rest) peaked at about 35 MeV and with 53.8 MeV maximum energy
\footnote{See for example (Singh \etal 2006b) for a recent calculation of neutrino-nucleus cross sections
with various nuclear targets and conventional sources.}.  
Beta-beams are pure in flavour and have the 
important advantage that the neutrino energy can be tuned by varying the acceleration 
of the ions. This is a particularly important feature for extracting nuclear structure
information, as discussed above.
In (McLaughlin 2004) neutrino scattering on lead is taken as an example 
to show another procedure to extract
information on the different spin-isospin and isospin excitations, namely
through a comparison of measurements with conventional
beams and low energy beta-beams (Figure \ref{f:snslead}). In fact, the corresponding neutrino fluxes
are in the same energy range for $\gamma=7$, but their shape and average energies
are different. Besides, the measurement
of the cross section without or with (one or two) neutrons can be used for the same purpose,
in the specific
case of the lead nucleus (McLaughlin 2004). 

A better knowledge of the weak nuclear response can have a broad impact, since 
neutrino-nucleus interactions represent a topic of current
great interest for neutrino physics and nuclear astrophysics.
The motivations come from the need for a precise knowledge of neutrino detector responses,
both in neutrino experiments and in core-collapse Supernovae observatories, 
and also from open issues in core-collapse Supernovae physics, such 
as unraveling the site for the nucleosynthesis of heavy elements
(McLaughlin and Fuller 1995, Haxton \etal 1997,
Qian \etal 1997, Borzov and Goriely 2000,
Balantekin and Fuller 2003, Terasawa \etal 2004).
or from neutrino-nucleosynthesis (Woosley \etal 1990, Heger \etal 2005, Haxton 2004, 
Langanke and Pinedo 2004). 
However, the available experimental neutrino-nucleus scattering data in
the relevant energy range is limited, since deuteron, carbon and iron
are the only nuclei investigated so far. As a consequence one has to rely on 
the numerous theoretical
predictions and on extrapolations to the 
nuclei and energies of interest. These calculations exploit a variety
of approaches, among which Effective Field Theories, the Shell Model and the
Random-Phase-Approximation, and the Elementary Particle Model (Kubodera and Nozawa 1994, Kolbe \etal 2003).
The cross sections estimates agree quite well at very low energies, where the
nuclear response is dominated by the allowed Fermi and Gamow-Teller transitions.
However, important discrepancies appear at higher neutrino energies, when the other nuclear
excitations become important (Figures \ref{f:Pb} and \ref{f:snslead}).
In this energy region, the calculations are in fact largely subject to nuclear
structure uncertainties and model dependencies
(e.g. treatment of the continuum, choice of the forces, higher-order correlations).
The present discrepancies between the predicted and measured neutrino-carbon cross sections
and between the calculations in the case of lead are talkative examples (Volpe 2005c) of
the difficult theoretical task.
Systematic neutrino-nucleus interaction studies performed with low energy beta-beams
offer a perfect tool to explore the nuclear response in this energy region in
great detail, and put the theoretical predictions on a firm ground. 

\begin{table}
\caption{{\sc Neutrino-nucleus interaction rates at a low energy
beta-beam facility  ($\gamma=7-14$):}
The expected events per year ($3.2 \times 10^{7}$ s) on several target nuclei obtained 
for $\gamma=14$ with Eqs.(\ref{e:1}-\ref{phitot})
are presented. The effect of the storage ring geometry is shown
($S$ and $D$ are the total and straight section lengths respectively).
The detectors are located at 10 meters from the storage ring and have cylindrical
shapes ($R$=1.5 m and $h$=4.5 m for deuteron, iron and lead,
$R$=4.5 m and $h$= 15 m for oxygen, where $R$ is the radius and $h$ is the
depth of the detector).
Their mass is indicated in the second column (Serreau and Volpe 2004). \label{t:lowevents} }
\begin{tabular}{@{}llll}
\hline
 Reaction                        & Mass  & Small Ring & Large Ring \\
                       & (tons) & ($S$=450 m, $D$= 150 m) & ($S$=7 km, $D$=2.5 km) \\   
  \hline
  $\nu +$D            &  35   &  2363      & 180 \\
   $\bar\nu +$D       &  35   &  25779     &   1956      \\
      $\nu + ^{16}$O          &  952       &  6054  & 734       \\
	   $\bar\nu +^{16}$O         &  952  &  82645    & 9453      \\
	      $\nu +^{56}$Fe           &  250  &  20768 & 1611      \\
		       $\nu +^{208}$Pb        &  360  &  103707 &  7922      \\
			    \hline
\end{tabular}
\end{table}

The number of neutrino-nucleus scatterings expected at a low energy beta-beam
facility has been calculated in (Serreau and Volpe 2004). The effects of
the geometry of the storage ring has in particular been analyzed. Two sizes
are considered: the one from the original scenario (Figure \ref{fig:baseline}) and
a small storage ring, as the one envisaged for the future GSI facility
(see GSI). The results of exact numerical calculations using Eqs.(\ref{e:1}-\ref{phitot})
are shown in Table \ref{t:lowevents} for several target nuclei. The detectors
are those considered for the ORLAND facility (Avignone F T \etal 2000).
An analytical formula is derived to scale the rates in Table \ref{t:lowevents} to
different storage ring sizes. Note that for a close detector, due to the flux anisotropy,
the rates do not simply scale as $S/D$, as in the case of a far detector 
(like those considered in Sections II, IV and VI). 
It is clear that 
interesting neutrino-nucleus rates can be attained at such a facility.
A crucial outcome of (Serreau and Volpe 2004) is that a small devoted storage ring
is much more favorable for low energy applications. The physical reason is simple.
The flux emittance is large because of the very low $\gamma$ values:
only the ions which decay close to the detector contribute to the rates.
The feasibility study of the small storage ring is now ongoing
(Benedikt \etal 2006).

A novel procedure to determine the response of a target nucleus in a supernova neutrino detector 
directly, through the use of low energy beta-beams, is pointed out 
in (Jachowicz and McLaughlin 2005 and 2006).
It is shown that the cross sections folded with a
supernova neutrino spectrum can be well reproduced by linear combinations of
beta-beam spectra. This comparison offers a direct way to extract the main
parameters of the supernova neutrino flux. The proposed procedure appears quite
stable against uncertainties coming from the experiment, or the knowledge of the
cross section, that give rise to a "noise" in the expansion parameters. 

Finally, it has been recently pointed out (Volpe 2005a) that 
neutrino-nucleus interactions are also important for the search
of neutrinoless double-beta decay in nuclei. 
In fact, by rewriting the neutrino exchange potential in momentum space and
by using a multiple decomposition, the
two-body transition operators, involved in the
former, can be rewritten as a product of the one-body operators 
involved in neutrino-nucleus interactions
(except for the short range correlations 
as well as possible phases present in the two-body process). Neutrino-nucleus scattering data offer a
potential new constraint for the predictions on the neutrinoless double-beta
decay half-lives. At present these calculations suffer of important discrepancies for the
same candidate nucleus.  
Beta-decay (Muto \etal 1989, Aunola and Suhonen 1996), 
muon capture (Kortelainen and Suhonen 2002 and 2004), 
charge-exchange reactions (Akimune 1997, Bernabeu \etal 1988)
and double-beta decay with the emission of two neutrinos (Rodin \etal 2003) have been used
to constrain the calculations so far.  Neutrino-nucleus
measurements would have the advantage that, if both neutrinos and anti-neutrinos are
available, the nuclear matrix elements involved in the
two branches of neutrinoless double-beta
decay -- from the initial and the final nucleus to the intermediate one --
can be explored.

\subsection{Fundamental interaction studies}
\begin{figure}[t]
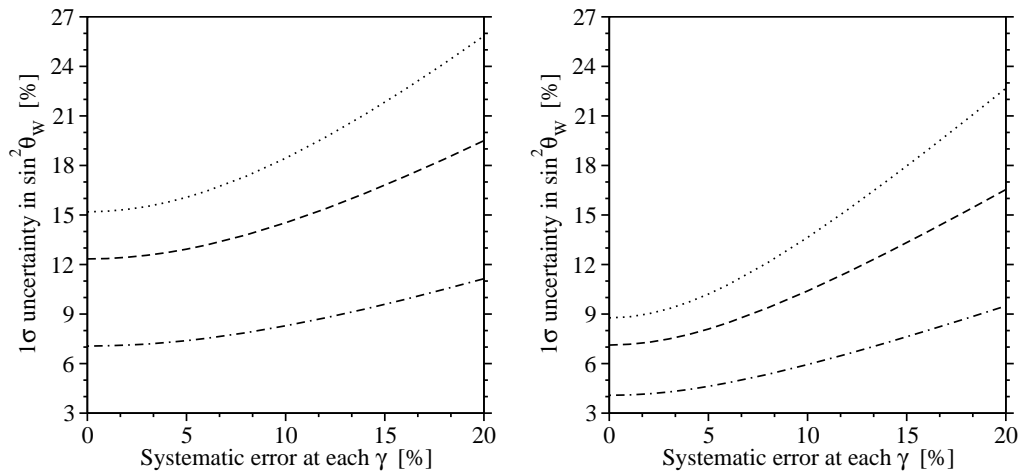

\begin{center}
\begin{minipage}{6.8cm}
\includegraphics[width=6.5cm]{systa}
\end{minipage}
\begin{minipage}{6.8cm}
\includegraphics[width=6.5cm]{systb}
\end{minipage}
\end{center}
\caption{One sigma uncertainty in the Weinberg angle as a function of the
systematic error at each $\gamma$ for $\gamma=12$ (dotted line), for
$\gamma=7,12$ (broken line),
and for $\gamma=7,8,9,10,11,12$ (dash-dotted line).  The helium-6 intensity
at the storage ring is $2.7\times 10^{12}$~ions/s~ on the left
panel (Chanc\'e and Payet 2005),
and with an increased intensity of 
$8.1\times 10^{12}$~ions/s on the right panel.  In both cases,
the measurement duration at each $\gamma$ is one year ($3\times 10^7$~s)
(Balantekin \etal 2006a).\label{fig:weinangle}}
\vskip 0.75cm
\end{figure}

Three applications for fundamental interaction studies of low energy beta-beams 
have been discussed so far: the measurement of the Weinberg angle
at low momentum transfer (Balantekin \etal 2006a),
a CVC test with neutrino beams (Balantekin \etal 2006b) and
a measurement of the neutrino magnetic moment (McLaughlin and Volpe 2004).
The measurement of the Weinberg angle represent an important test of the
electroweak theory. Several
experiments at different $Q^2$ exist,
namely the atomic parity violation
(Bennett and Wieman 1999) and Moller scattering at $Q^2 = 0.026$ GeV$^2$
(Anthony \etal 2005) which combined with the measurements
of $\sin^2\theta_W$ at the $Z^0$ pole (ALEPH \etal 2005),
are consistent with the expected
running of the weak mixing angle.
However, recent measurement of the
neutral- to charged-current ratio in muon anti-neutrino-nucleon scattering
at the NuTEV experiment disagrees with these results by about 3 $\sigma$
(Zeller \etal 2002). A number of ideas were put forward to explain the
so-called NuTEV anomaly (Davidson \etal 2002, Loinaz \etal 2004, Miller and
Thomas 2005, Giunti and Laveder 2002). However, a complete
understanding of the physics behind the NuTEV anomaly is still lacking;
probing the Weinberg angle through additional experiments with different
systematic errors would be very useful.
\begin{figure}[t]
\begin{center}
\includegraphics*[width=6.5cm]{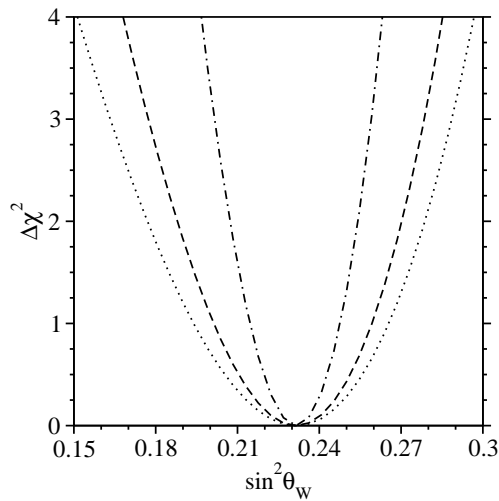}
\end{center}
\caption{{\sc A Weinberg angle measurement with low energy beta-beams:}
$\Delta\chi^2$ with $\gamma=12$ (dotted line), with $\gamma=7,12$
(broken line), and with $\gamma=7,8,9,10,11,12$ (dash-dotted line).  The
results were
obtained considering a one year ($3 \times 10^7$~s) measurement duration
at each $\gamma$, and for a helium-6 intensity at the storage ring
of $2.7\times 10^{12}$~ions/s (Chanc\'e and Payet 2005).  The count number error
was considered to be purely statistical.  The 1$\sigma$ ($\Delta \chi^2=1$)
relative uncertainty in the Weinberg angle is 15.2\% for $\gamma=12$,
12.3\% for $\gamma=7,12$, and 7.1\% for $\gamma=7,8,9,10,11,12$ (Balantekin \etal 2006a).}
\vskip 0.75cm
\label{fig:chi2}
\end{figure}
The possibility of using a low energy beta-beam facility
to carry out such a test with neutrino-electron scattering 
at low momentum transfer, i.e. $Q^2 = 10^{-4}$~GeV$^2$
is investigated in
(Balantekin \etal 2006a).
In particular it is shown that since the neutrino flux and average energy are
well known in the case of beta-beams, the number of counts is in principle
sufficient to extract information on the Weinberg angle.
 A fully efficient 1 kton \v{C}erenkov detector is located
at 10 meters from a small storage ring having 1885 m total length and 678 m straight
sections, in which the expected intensities for
$\gamma =7-14$ are 0.5$\times 10^{11}$ helium-6/s and 2.7$\times 10^{12}$ neon-18/s
(while the intensities at production are the same as in Table
\ref{tab:intensities}). These numbers are the outcome of a preliminary feasibility
study of the small storage ring (Chanc\'e and Payet 2005).
The (anti)neutrino on electron events are identified by an angular cut.
The background from neutrino-proton scattering is suppressed
by the use of Gadolinium (Beacom and
Vagins 2004). Figure \ref{fig:chi2} shows the improved precision one gets
by combining measurements with different gamma values. The results corresponds to one
(3x10$^7$ s) year measurement at each gamma and a purely statistical total
count error.
Figure \ref{fig:weinangle} shows the precision with which the
Weinberg angle can be measured after the inclusion of both statistical and
systematic errors. In particular, if the systematic error can be kept below
10 \%, a precision of 10 \% seems to be within reach at a beta-beam
facility. 

\begin{figure}[t]
\begin{center}
\includegraphics[scale=0.4]{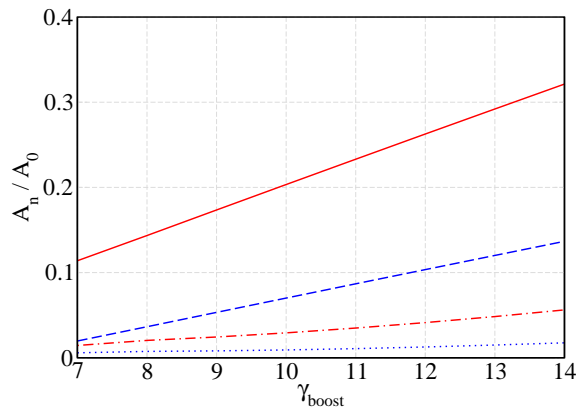}
\end{center}
\caption{{\sc A CVC test:}
The $\bar{\nu}_e + p \rightarrow e^+ + n$ reaction is used to test the weak magnetism form factor using
neutrinos from low energy beta-beams and a water \v{C}erenkov detector. The figure shows the
ratio of the first ($A_1$) and second ($A_2$) order Legendre polynomial coefficients to the zeroth 
order one ($A_0$, the number of events) as a function 
of the ion Lorentz boost. These expansion coefficients determine the angular distribution of the emitted positrons
 with (solid is for $A_1/A_0$; dash-dotted is for $A_2/A_0$)
 and without (dashed is for $A_1/A_0$; dotted is for $A_2/A_0$) the weak magnetism term. 
The overall effect of the weak magnetism term on the angular distribution is 
 considerably larger than on the total number of events
 and increases with the Lorentz ion boost.
  Note that when the weak magnetism is considered, 
 its value is set to its CVC value (Balantekin \etal 2006b). \label{fig:cvcang}}
\end{figure}

The CVC hypothesis connects weak and electromagnetic 
hadronic currents.  Several tests of CVC have been performed in the past 
(see e.g. Thomas 1996),
concerning, in particular, the vector form factor, through super-allowed nuclear 
beta-decays studies (see e.g. Hardy and Towner 2004).
Verifying that the CVC hypothesis correctly predicts tensor terms 
-- often referred to as weak magnetism -- is of fundamental 
importance~(Deutsch \etal 1977).  
So far, this contribution to the weak currents 
has been tested in an experiment involving the beta-decay of Gamow-Teller 
transitions in mirror 
nuclei in the A=12 triad~(Lee \etal 1963, Wu 1964, Lee and Wu 1965). 
A test based on neutrino-nucleon 
collisions at low momentum using low energy beta-beams is proposed in
(Balantekin \etal 2006b). This would have,
in particular, the advantage that 
there is no uncertainty coming from nuclear structure calculations.  
It is important to keep in mind that,
in addition to providing important tests of the Standard Model, 
neutrino-nucleon reactions play a crucial role in understanding the dynamics of the 
core-collapse supernovae~(Balantekin and Fuller 2003, Horowitz 2001), 
the yields 
of the r-process nucleosynthesis that could take place in such 
environments~(Meyer \etal 1998) 
and also
contribute to the energy transfer (from the accretion-disk 
neutrinos to the nucleons) in gamma-ray bursts 
models~(Ruffert \etal 1997, Kneller \etal 2004).  
Finally, understanding the subtleties of 
the neutrino-nucleon interactions are crucial to the 
terrestrial observation of neutrino signals~(Vogel and Beacom 1999, 
Beacom \etal 2002). In
(Balantekin \etal 2006b)
the sensitivity to the weak magnetism term
that can be achieved 
in the $\bar{\nu}_e + p \rightarrow e^+ + n$ reaction,
both with the total number of events and with the angular
distribution of the emitted positrons, is studied. 
For this purpose the same setup, as for the measurement of the Weinberg angle is taken, and
the interaction of anti-neutrinos on protons in a water \v{C}erenkov detector is considered.
\begin{figure}
\begin{center}
\includegraphics[scale=0.4]{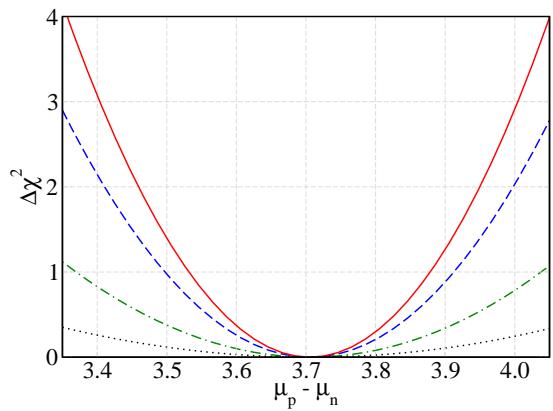}
\end{center}
\caption{ Same as for the previous 
Figure, but for $\Delta \chi^2$ obtained from the angular distribution of events (see text) in
 the cases when the total error
  is purely statistical (solid), and when one includes 
 2\% (dashed), 5\% (dash-dotted) and 10\% (dotted) systematic errors.  The 1$\sigma$ 
 ($\Delta \chi^2=1$) relative uncertainty in $\mu_p - \mu_n$ is 4.7\%, 5.6\%, 9.0\% and 
 more than 20\%, respectively.  These results were obtained considering $\gamma=12$ 
 (Balantekin \etal 2006b).}
\label{fig:cvcresults}
\end{figure}
The angular distribution can be expanded in Legendre polynomials whose coefficients are shown in
Figure \ref{fig:cvcang}, as a function of the Lorentz ion boost.
While the zeroth order coefficient is nothing but the total number of events,
the first order measures the strength of a linear term which favors forward peaked events.
The results show that, when systematic errors are taken 
into account, the angular distribution is a much better tool than the total number of events
to extract information on the weak magnetism form factor. In particular, if those
errors are kept below 5 \%, a one year measurement of the weak magnetism is possible at a 1 $\sigma$
level of 9 \%, if the ions in the storage ring are boosted to $\gamma=12$.
An even better measurement is expected if the ions are boosted to $\gamma's$ larger than 12,
because of the increasing importance of the weak magnetism contribution with the impinging neutrino energy.
This way of probing the weak magnetism form factor at low momentum transfer constitutes a new test of
the Conserved Vector Current Hypothesis. 

\begin{figure}[t]
\begin{center}
\begin{minipage}{7.5cm}
\vspace{-4.cm}
\centerline{\includegraphics[angle=0,width=7cm]{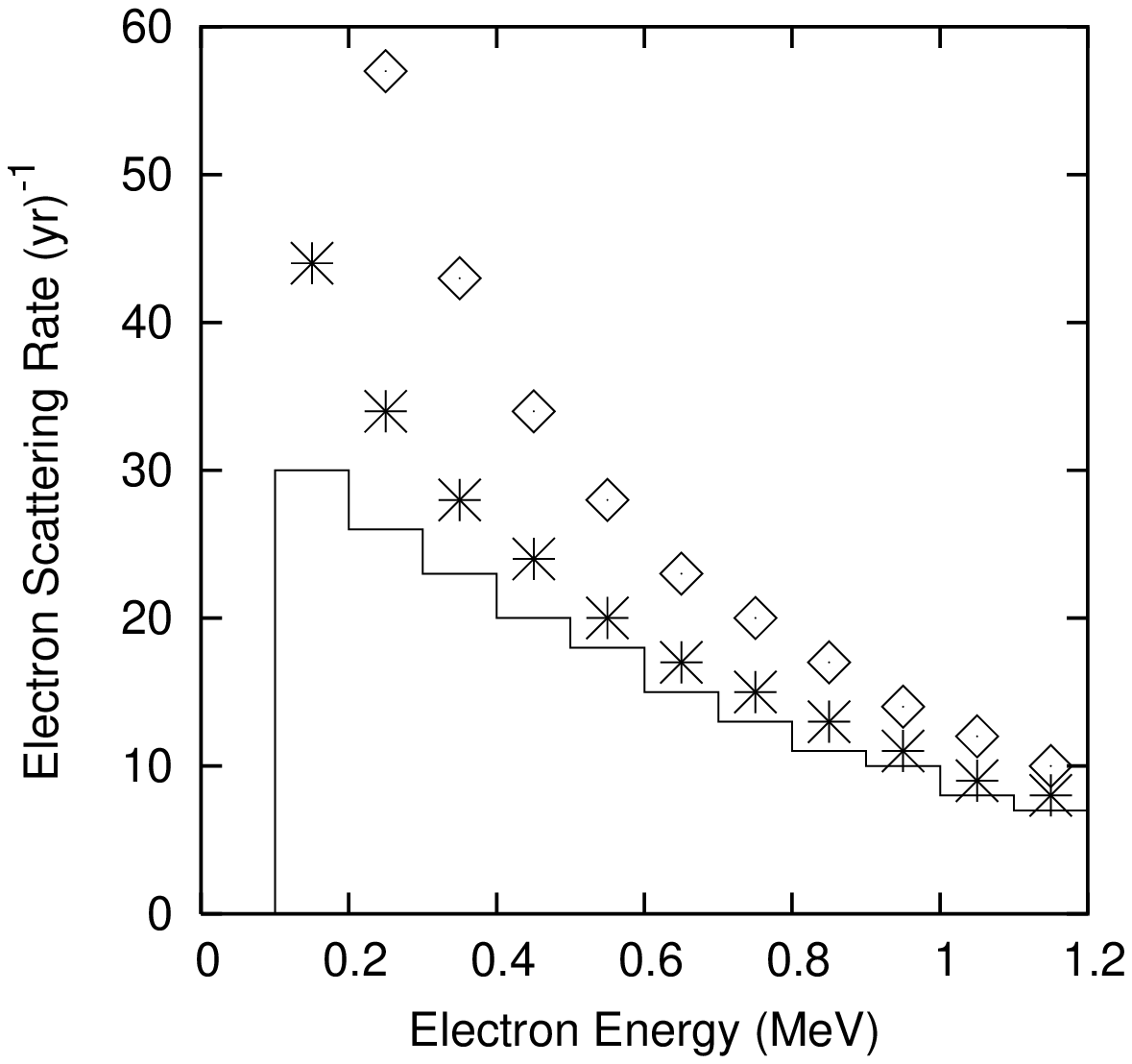}}
\end{minipage}\begin{minipage}{7.5cm}
\vspace{-4.cm}
\centerline{\includegraphics[angle=0,width=7cm]{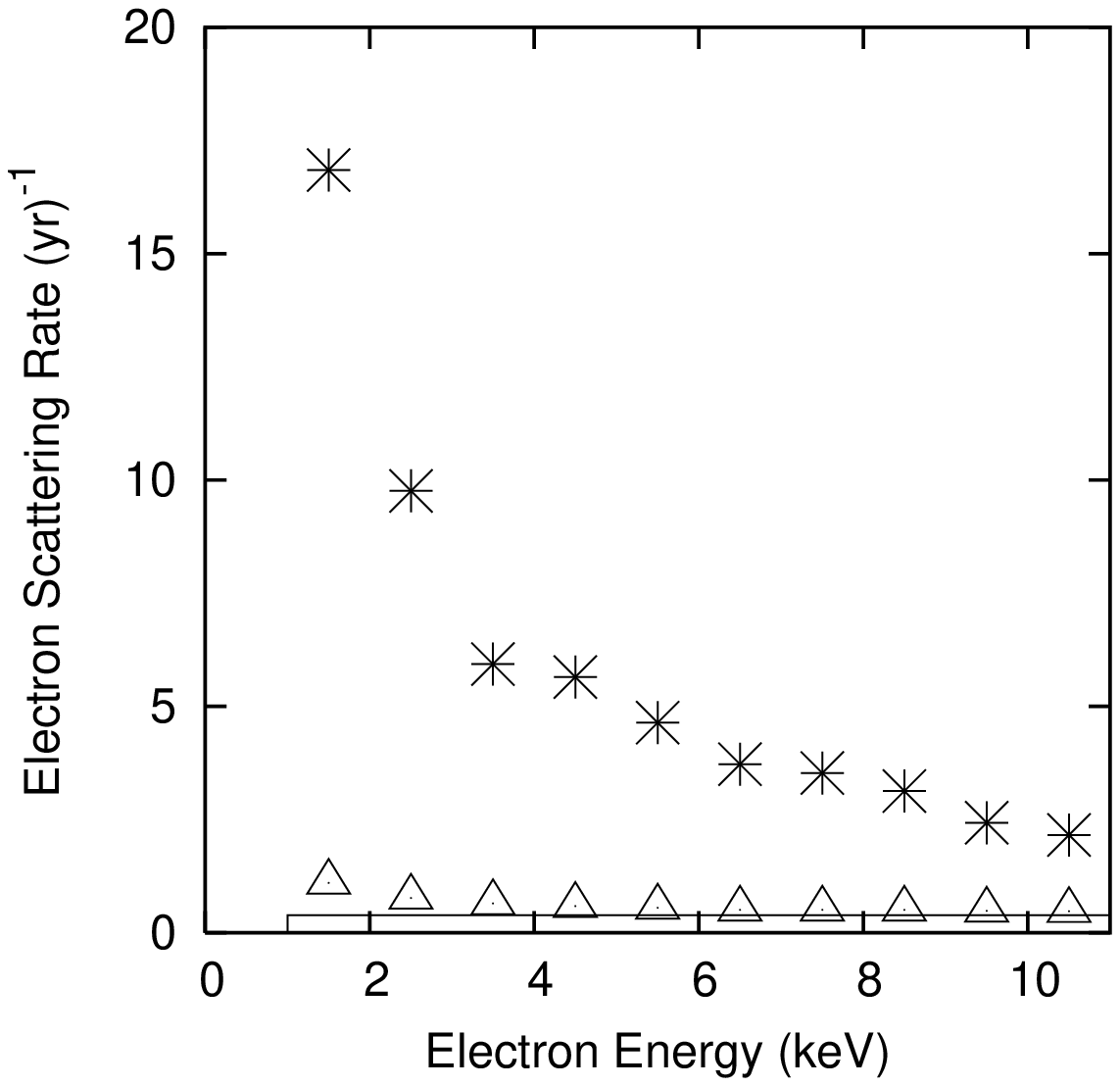}}
\end{minipage}
\end{center}
\caption{\small {\sc Measurement of the neutrino magnetic moment 
with low energy beta-beams:}
Number of neutrino-electron scattering events from 
helium-6 ions, produced at the rate $10^{15}$ per second, and sent inside a
$4 \pi$ detector of 10 m in radius. 
Similar results are obtained with neon-18 ions. 
(Note that for this application ion acceleration is not needed.)
The symbols show
the number of scatterings if the neutrino has a magnetic moment of
$\mu_\nu = 10^{-10} \mu_B$ (diamonds), 
$\mu_\nu = 5 \times 10^{-11} \mu_B$ (stars) and
 $\mu_\nu = 10^{-11} \mu_B$ (triangles).
The histogram shows the results 
for a vanishing neutrino magnetic moment (McLaughlin and Volpe 2004).
}\label{fig:he6-his-kev}
\end{figure}

It is interesting to note that one might use the ion decay at rest as an intense neutrino 
source, in order
to explore neutrino properties that are still poorly known, such as the
neutrino magnetic moment (McLaughlin and Volpe 2004). Direct measurements to
achieve improved limits are precious, since the observation of a
large magnetic moment points to physics beyond the Standard Model.
Note that such an application does not require any acceleration of the ions. Once
produced, the ions are fired to a target inside a $4 \pi$ detector. The
measurement of the (anti)neutrino interaction with the electrons, as a
function of the electron recoil, is then used to set limits on the
neutrino magnetic moment.
Current bounds come from direct
measurements using reactor neutrinos, and are in the range
$\mu_{\nu} < 0.9-4 \times 10^{-10} \mu_B$ 
(Daraktchieva \etal 2005, Li \etal 2003, 
Reines \etal 1976, Vogel and Engel 1989, Vidyakin \etal 1992, 
Derbin \etal 1993).
From solar neutrino-electron scattering at Super-Kamiokande a limit of
$ < 1.5 \times 10^{-10} \mu_B$ at 90 \% C.L. has been obtained
(Beacom and Vogel 1999). 
Upper limits
in the range $10^{-11}- 10^{-12} \mu_B$ are also inferred,
from astrophysical and cosmological considerations (Raffelt 1996),
the exact values being model dependent.
The prospects from using low energy beta-beams to improve the direct
bounds has been studied in comparison with
reactor neutrinos and a very intense tritium source (McLaughlin and Volpe 2004).
While the advantage of using beta-beams is that the neutrino flux is
perfectly known, the main limitation for this application
is clearly the intensity of the ions.
Figure \ref{fig:he6-his-kev} shows
the electron-neutrino scattering events in the
range of  0.1 MeV to 1 MeV and 1 keV to 10 keV respectively.
The results shown are obtained for an (increased) intensity of $10^{15}~\nu/s$ (i.e.
$10^{15}~$ ions/s).  If there is no magnetic moment, this intensity
produces about 170
events in the 0.1 MeV to 1 MeV range
per year and  3 events in the 1 keV to 10 keV range per year.
These numbers increase to 210 and 55 respectively in the case of a
magnetic moment of  $5 \times 10^{-11} \mu_B$.
In conclusion, by measuring
electron recoils in the keV range with a beta-beam source one
can, with a sufficiently strong source, have a very clear signature for
a neutrino magnetic moment of few $ 10^{-11} \mu_B$.
A detailed study
of the background, at these small electron recoil
energies, and a precise knowledge of the ion-induced background, are now needed.

\section{Medium and high energy scenarios}

The idea of pushing the Lorentz ion boosts to higher values than
the ones proposed in the original scenario (Zucchelli 2002) is first
proposed in (Burguet-Castell \etal 2004). In particular, the sensitivity
to $\cal{CP}$ violation of the original scenario (Section II) is compared
to two new baselines where the ions are accelerated at either $\gamma=350/580$
for helium-6/neon-18,
and sent to a distance of $L=720$~km, or at 
$\gamma=1500/2500$ and sent to $L=3000$~km.
The former case would require a refurbished SPS to reach such
accelerations at CERN, while the water \v{C}erenkov detector
detector could, for example, be located
at the Gran Sasso Laboratory. 
The latter case would necessary need something
like LHC and a baseline such as the CERN-Canary islands with e.g. a 40 kton 
tracking calorimeter. 
If the baseline is adjusted appropriately one can remain close
to the atmospheric oscillation maximum.
Working at higher boosts has essentially two advantages.
\begin{figure}[t]
\begin{center}
\includegraphics*[scale=0.5]{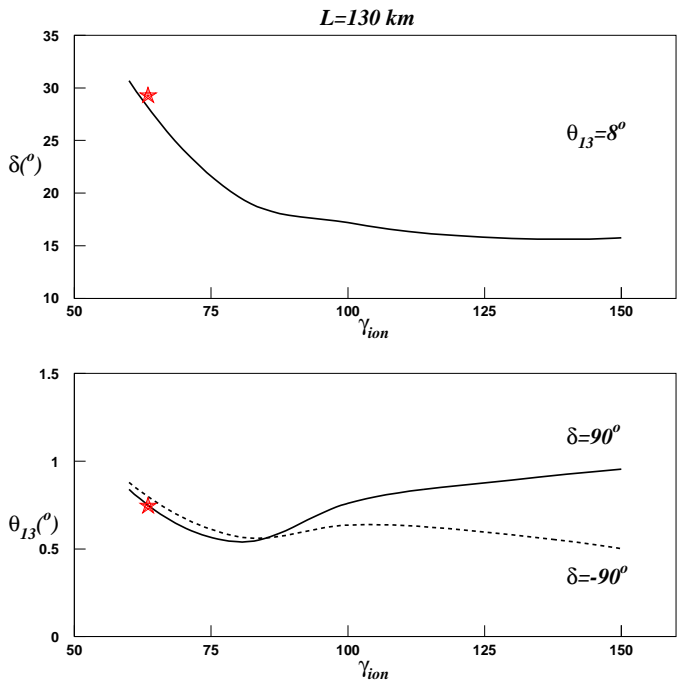}
\end{center}
\caption{Sensitivity curves (99 \% C.L.) for $\delta$ (upper) and
$\theta_{13}$ (lower), as a function of the ion boost. The
CERN-Fr\'ejus baseline with a 440 kt water \v{C}erenkov detector are
considered. The (anti)neutrino emitters run at the same gamma.
The stars present the results obtained for the original option
where the ions run at $\gamma=60,100$ (Section II) (Burguet-Castell \etal 2005).}
\label{f:g100}
\end{figure}
The appearance oscillation
signal increases linearly with gamma (and then saturates
for $\gamma \approx 400$ in a water detector), which results in a gain in
sensitivity. 
Besides, some ambiguities on the neutrino parameters due to the degeneracies
can be disentangled, both by using the neutrino energy reconstruction in
the detector, and by exploiting matter effects (that become significant
for baselines of $O(1000)$~km, Figure \ref{f:mattereffects}). 
Ref.(Burguet-Castell \etal 2004) has triggered a number of studies
(Terranova \etal 2004, Burguet-Castell \etal 2005b, Huber \etal 2006,
Agarwalla \etal 2005). The medium and high energy options have also
been discussed in (Terranova 2005,
Gomez-Cadenas 2005, Migliozzi 2005). In (Donini \etal 2005a and 2006) the medium
$\gamma$ (350-580) case is analyzed exploiting a CERN-Gran Sasso Laboratory
baseline with a non-magnetized iron detector. It is shown that
the $\cal{CP}$ violating phase can be determined down to $30^{\circ}$ 
(99 \% C.L.), if $\theta_{13} > 3^{\circ}$.
In (Agarwalla \etal 2005) the high
energy option is discussed with a baseline from CERN to Rammam or Singara where
an iron calorimeter
detector could be located.
Performances can be found in the literature for
$\gamma=150$ (Burguet-Castell \etal 2005), 200 (Huber \etal 2006),
350 (Burguet-Castell \etal 2005), 500 (Huber \etal 2006, Terranova \etal 2004),
1000 (Huber \etal 2005a), 2000 (Burguet-Castell \etal 2004) and 2488 (Terranova \etal 2004).
Precision tests of several observables such as the weak mixing angle, the
neutrino magnetic moment and potential leptonic Z' couplings,
exploiting neutrino-electron scattering and the medium energy ($\gamma=500$) beta-beam
are discussed in (De Gouvea and Jenkins 2006), in comparison with nuclear
reactors, neutrino factories and conventional beams.

\begin{figure}[t]
\begin{center}
\epsfig{file=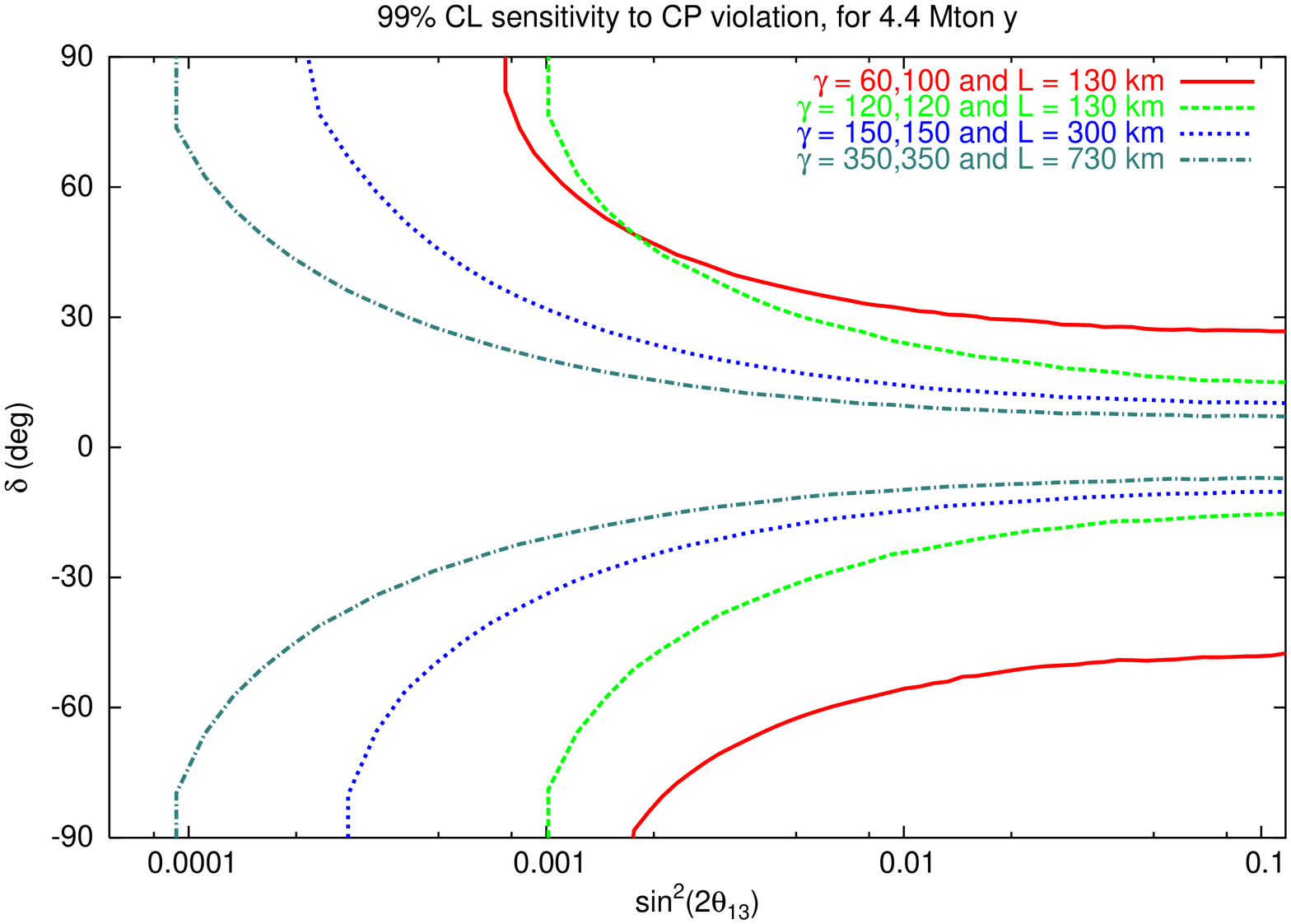,width=7cm,height=7cm,angle=-90}
\epsfig{file=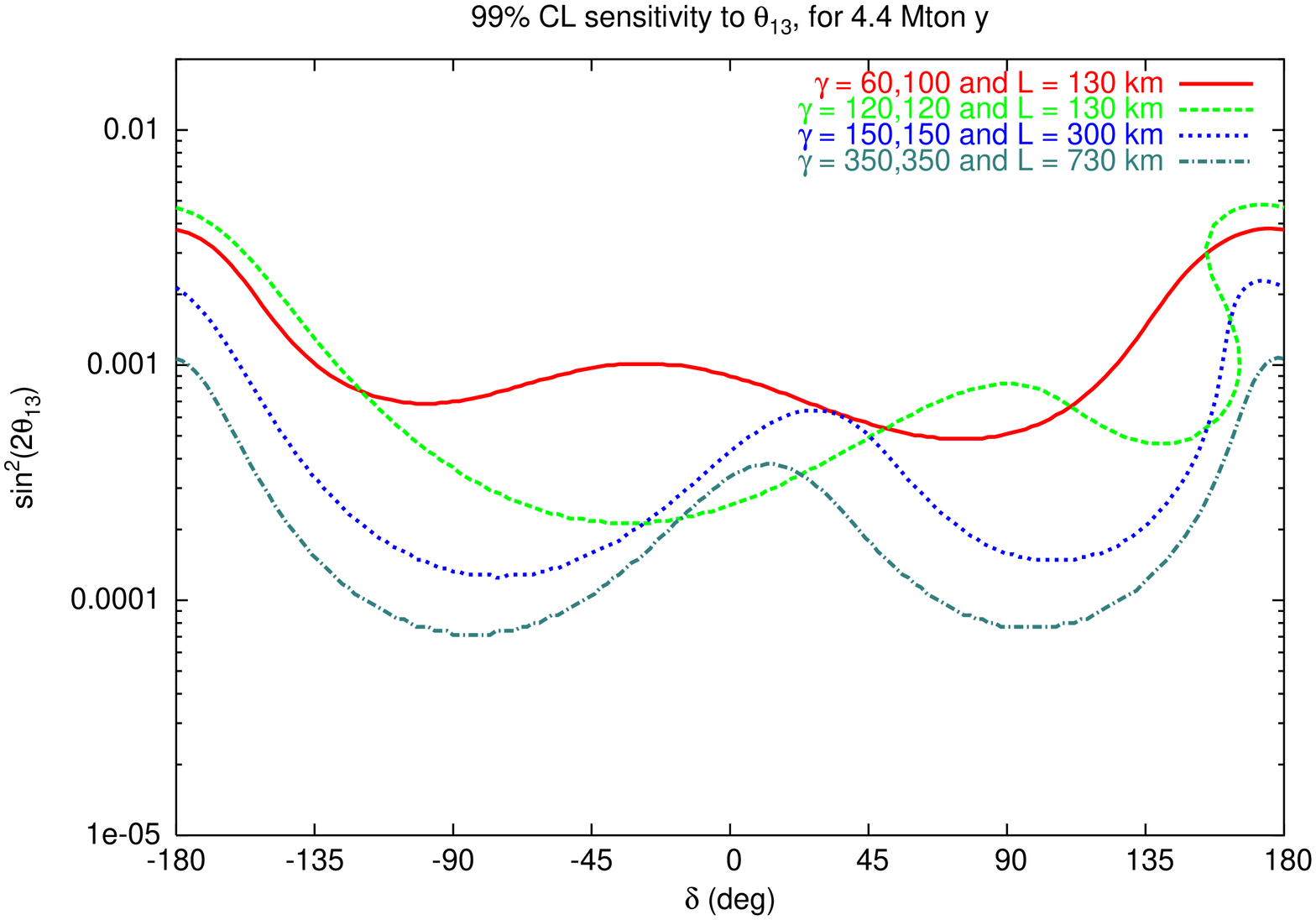,width=7cm,height=7cm,angle=-90}
\caption{Exclusion plots (99 \% C.L.) for $\delta$ (left) and
$\theta_{13}$ (right) for the original scenario in comparison with
the options $\gamma=120,150,350$ and $L=130,300,730$ km for the 
source-detector distance,
respectively. The solar and atmospheric oscillation
parameters are fixed to the present best values (Burguet-Castell \etal 2005).}
\label{f:g100-350}
\end{center}
\end{figure}

In (Burguet-Castell \etal 2005) a detailed analysis of the sensitivity
with a  Lorentz ion boost of $\gamma=150$ is made. This choice corresponds to
the maximum acceleration that
can be achieved using the existing CERN-SPS (Lindroos 2005). In particular, 
the following three issues are addressed: {\it i)} a sensitivity study of the
CERN-Fr\'ejus baseline with a megaton water \v{C}erenkov detector and 
$\gamma \ge 100$; {\it ii)} the optimal baseline at this $\gamma$ and
with such setup; {\it iii)} the (dis)advantages of using different values
of $\gamma$ for the (anti)neutrino emitters.
The intensities used are the same as in the original scenario, namely
2.9$\times 10^{18}~^{6}$He and 1.1$\times 10^{18}~^{18}$Ne, while
the measurement period is 10 years.
Note that the water \v{C}erenkov
detector considered is suitable for quasi-elastic events. Therefore the
sensitivity improves until the inelastic cross section starts to dominate
(for $\gamma \ge 400$).
The angular and energy cuts as well
as the time structure of the ion bunches help keeping the atmospheric
background at a very low level, while
the beam-induced neutral current single-pion production,
that can mimic the appearance signal, is manageable even for $\gamma=100$.

\begin{figure}[t]
\epsfig{scale=0.75,file=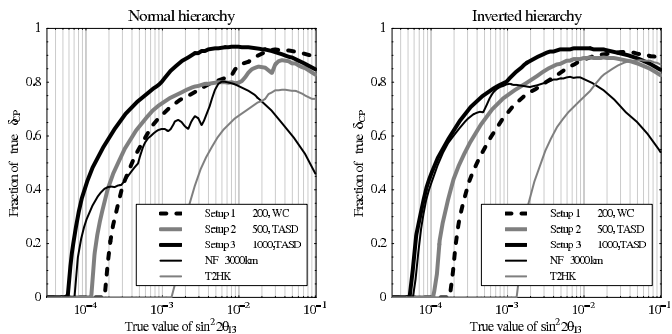}
\caption{The figure shows a comparison of the $\cal{CP}$ violation sensitivity
for the normal (left) and inverted (right) mass hierarchy as a function of the
true values for $\sin^2 \theta_{13}$ and $\delta_{CP}$ at the 3 $\sigma$ C.L.
A comparison is made between beta-beams with different ion boosts and detector
technologies (WC and TASD) 
on one hand, and neutrino factories (NF@3000 km)
and super-beams (T2HK) on the other hand.
On the y-axis is shown the values of $\delta_{CP}$ 
"stacked" to the  $\cal{CP}$ fraction 
(Huber \etal 2006).}
\label{f:senscomp}
\end{figure}
Figure \ref{f:g100} shows the sensitivity on $\delta$ and $\theta_{13}$
as a function of the ion boost for the CERN-Fr\'ejus baseline.
It is clearly seen that by using $\gamma=100$ the increased event rates
imply a definite gain
in the $\delta$ sensitivity compared to the lower gamma values first envisaged,
while its behavior keeps rather flat in the range 100-150 (since for these
cases the neutrino flux at lower energies contributing to the event rate
is practically the same).
The study of the optimal source-detector $L$ distance, for the maximum SPS
ion acceleration of 150, points to $L=300$~km, i.e.
the distance corresponding to the first atmospheric neutrino oscillation
maximum.
\begin{figure}[t]
\begin{center}
\includegraphics[width=0.72\textwidth]{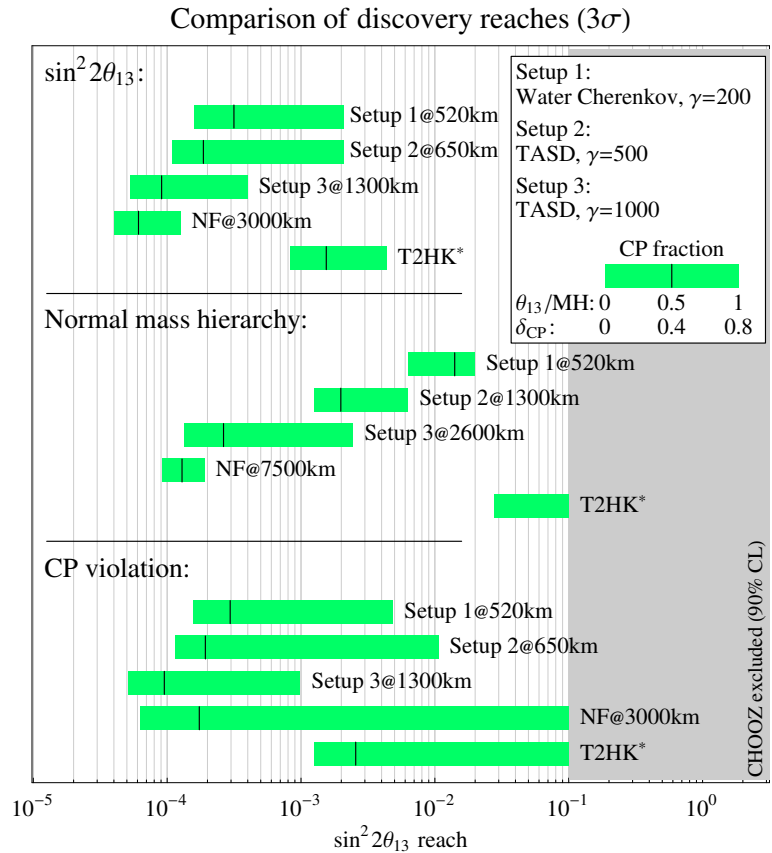}
\end{center}
\caption{A comparison of the discovery reach  at the $3\sigma$ 
confidence level. The bars represent the best case (left end), ``typical
case''  (middle line), and worst case (right end) in $\delta_{CP}$, where the 
respective sensitivities are computed with the CP fractions in the plot
legend.  Note that for CP violation, a CP fraction of one can never be 
achieved since values close to $0$ or $\pi$ cannot be distinguished from CP 
conservation. Within each category, the most competitive setups from this 
study are compared. 
For all shown sensitivities, a normal mass hierarchy is assumed (Huber \etal 2006).}
\label{f:summarize}
\end{figure}

A comparison of the $\delta$ and $\theta_{13}$ exclusion plots for the
original scenario, the $\gamma=120,150$ with $L=130,300$~km, 
and 350 with  and 730~km from (Burguet-Castell \etal 2004),
is presented in Figure \ref{f:g100-350}. 
Note that the energy dependence of the signal can be well 
reconstructed since the
neutrino energies are above the Fermi momentum of the target nuclei (this is
not the case for the original scenario). This
information can be used at $\gamma=150~(350)$ and $L=300~(700)$ to resolve
the intrinsic degeneracies. 
A major issue for such a baseline
is to find a location to house the detector, the Canfranc Underground
Laboratory being now discussed as a possible option.

An interesting and detailed analysis of the $\cal{CP}$ violation sensitivity
is made in (Huber \etal 2006), for Lorentz ion boosts values ranging from 100
to 3000, and is also discussed in (Winter 2006). Note that here the performance at high gammas is studied considering
that the intensities are reduced (and not constant as in previous studies)
for very high gammas. 
While a 500 kt water \v{C}erenkov detector is considered for a boost of 200,
a 50 kt totally active scintillator detector (TASD) is taken for higher ion accelerations.
The whole study is performed by using the GLoBES software (Huber \etal 2005a)
which allows, in particular, to include degeneracies and correlations (see Section V).
A detailed comparison with neutrino factories with a 50 kt magnetized iron detector
and very intense super-beams
(as those that will be available at T2HK) is made (Figure \ref{f:senscomp}).
Figure \ref{f:summarize} presents the main results of this work as far as the third neutrino
mixing angle, the Dirac phase and the normal mass hierarchy are concerned.
One can see the excellent performance of the $\gamma=200$ setup compared both to super-beams and
to neutrino factories. Such a discovery reach is achieved by including the second oscillation
maximum to disentangle correlations and degeneracies. This also depends on the choice of
approximately equal neutrino and anti-neutrino event rates for all setups.
On the other hand the higher gammas appear as competitive options to a 
neutrino factory.
In conclusion, the discovery reach for $\gamma > 150$ is very promising as well.
However, at least for the moment, it remains more speculative, since it is neither based
on an existing accelerator complex nor on a robust estimation of the ion intensities.
These scenarios clearly deserve further investigations.

\section{The issue of the degeneracies}
A characteristic feature in the analysis of future neutrino accelerator experiments is the presence
of the so-called "parameter degeneracies", which is the appearance of various disconnected regions in
the multi-dimensional oscillation parameter space, besides those corresponding to the true solutions.
The appearance of such degeneracies is due to both the inherent three flavour structure of the oscillation
probabilities, and some still unknown neutrino oscillation parameters. 
The presence of the latter degeneracies in the analysis of a beta-beam experiment  
depends on our knowledge of neutrino oscillation properties at the time such measurement is performed.
These degeneracies are of four different kinds, namely the intrinsic, the sign,
the octant and the mixed degeneracies (or clones).
The classification corresponds to the following origins: 
\begin{figure}
\vspace{-0.5cm}
\begin{center}
\begin{tabular}{c}
\hspace{-0.3cm} \epsfxsize10cm\epsffile{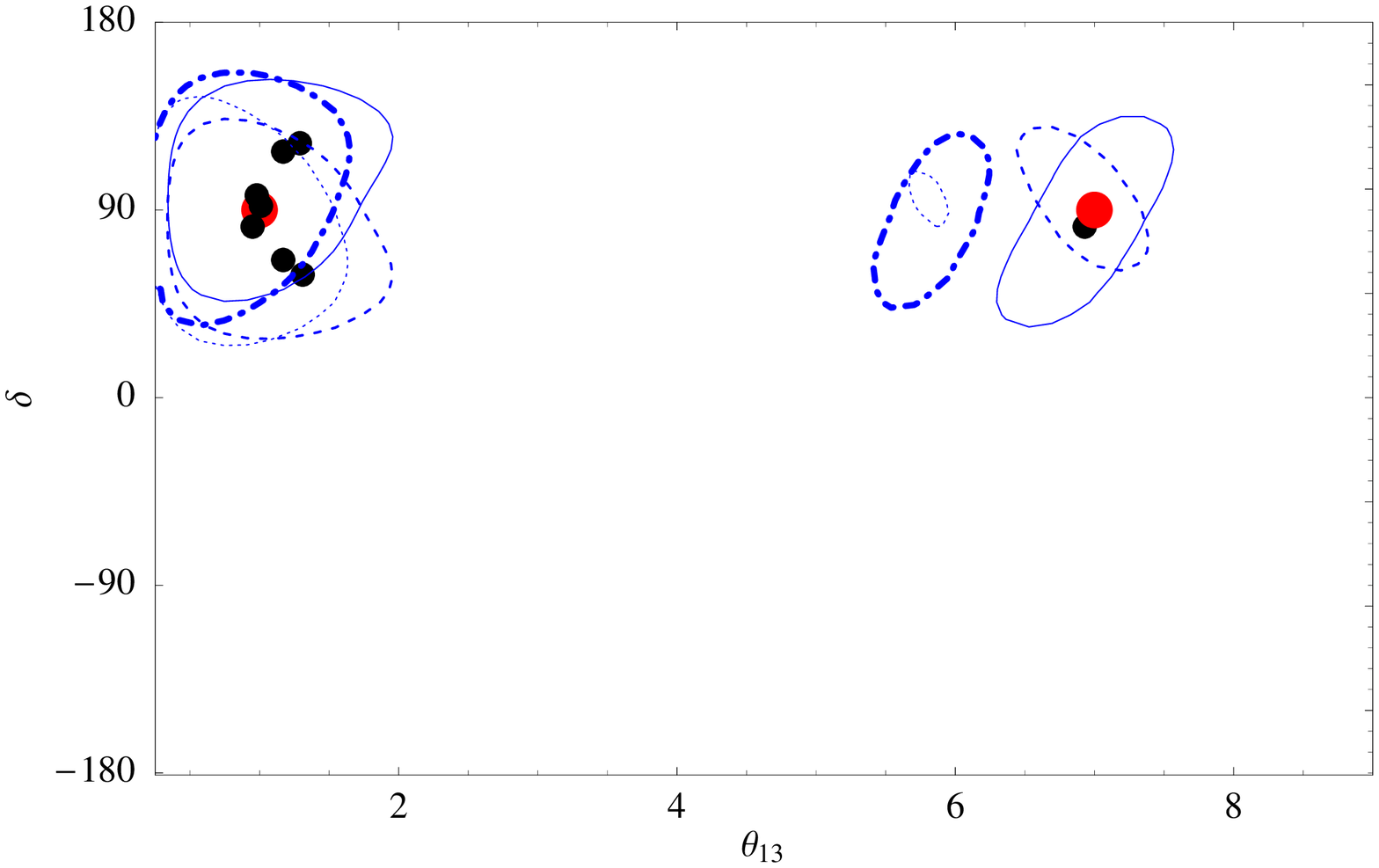} \\
\hspace{-0.3cm} \epsfxsize10cm\epsffile{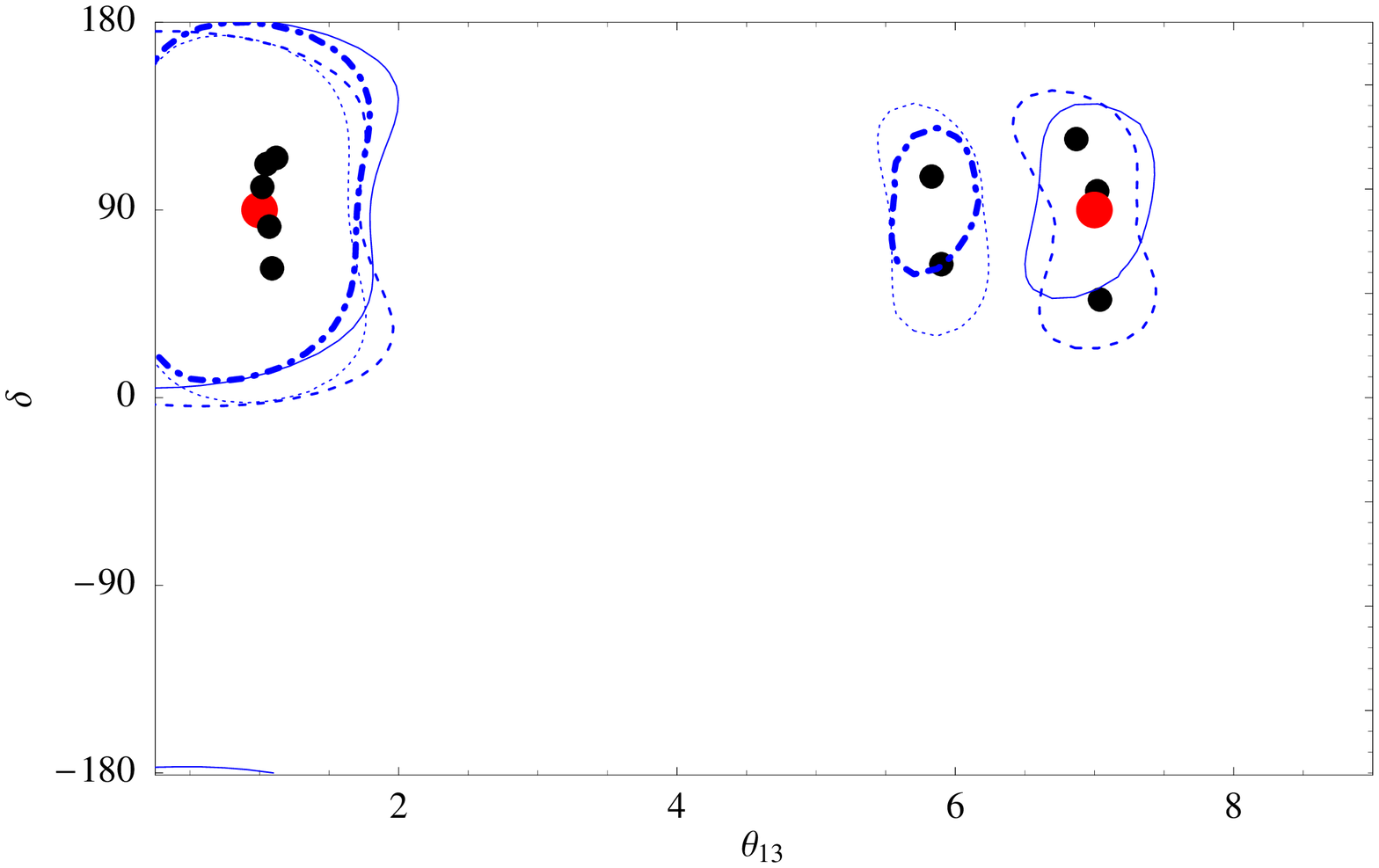} \\
\hspace{-0.3cm} \epsfxsize10cm\epsffile{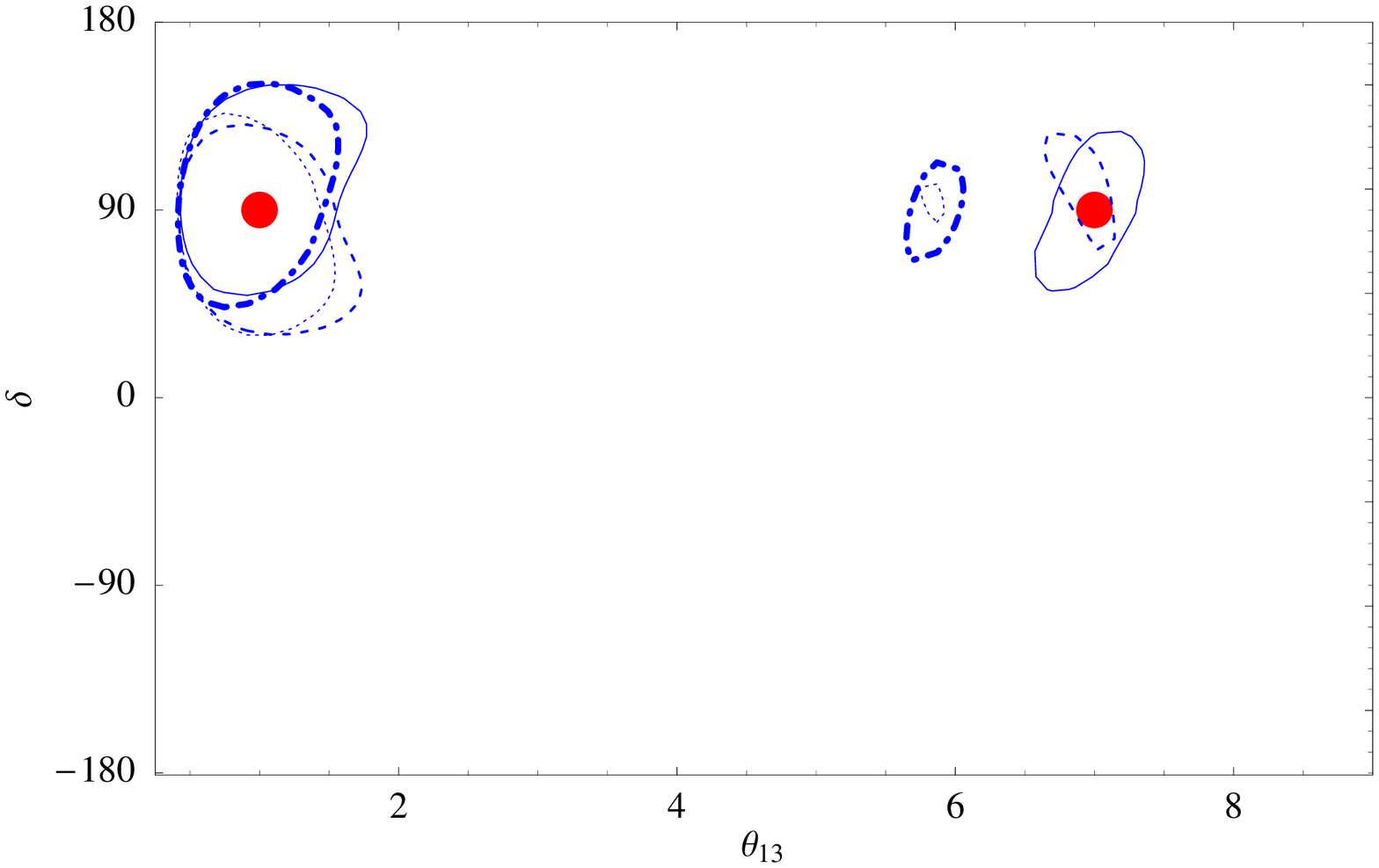} \\
\end{tabular}
\caption{The figure shows the 90\% CL contours for $\theta_{13}=1^\circ,7^\circ$
and $\delta=90^\circ$ (light red circle), after a 10 year
$\beta$-Beam (upper), a 2+8 year Super-Beam (middle) and 
a combined measurement (lower figure). The different clone regions are shown (see text):
the intrinsic (solid), the sign (dashed), the octant (dot-dashed) and  
the mixed (dotted line)  degeneracies.
The theoretical clone locations are presented as well (dark black dots) (Donini \etal 2005b).}
\label{f:clone1}
\end{center}
\end{figure} 

\begin{itemize}
\item 
for an appearance experiment the oscillation probability Eq.(\ref{e:probCP}) obtained for neutrinos at fixed baseline
and energy has a continuous number of solutions ($\delta,\theta_{13}$). If the same experiment is performed for
both neutrinos and anti-neutrinos with the same energy, only two disconnected solutions are left that satisfy 
the equation 
$P_{\alpha,\beta}^{\pm}(\delta,\theta_{13})=P_{\alpha,\beta}^{\pm}(\bar{\delta},\bar{\theta}_{13})$
(where $\pm$ refers to neutrino and anti-neutrino respectively). These two regions correspond to the true
solution and the intrinsic clone (Burguet-Castell \etal 2001).
\item The sign of $\Delta m^2_{23}$ is still unknown. The positive and negative solutions give two different
regions called the sign clones (Minakata and Nunokawa 2001).
\item The two octants $\theta_{23} > \pi/4$ and $\theta_{23} < \pi/4$ are possible, giving in general two
possible solutions for $\delta$ and $\theta_{13}$. This is called the octant degeneracy (Fogli and Lisi 1996).
\item A "wrong" sign and a "wrong" octant give rise to a new degeneracy called the mixed clone.   
\end{itemize}
The ensemble of these four regions for the true solution and the intrinsic clone gives eight 
different solutions
also called the eightfold degeneracy (Barger \etal 2002).
In (Donini \etal 2004) it is pointed out that 
the clone solutions can be significantly different if determined
from the oscillation probability, or from the number of events, showing that the degenerate 
solutions should 
be computed from the number of events, which is the physically measured quantity. 
Analytical formulae to determine the clone locations as a function of the real solutions are given.
The effect of possible 
combination of
super-beams and beta-beams  (Section II) 
as well as of a neutrino factory with super-beams and/or 
beta-beams is discussed.    
A detailed analysis of the eightfold degeneracy is performed for the first time in (Donini \etal 2005b).  
It is shown that combining a super-beam and a beta-beam does not solve the problem of 
degeneracies, 
since, at the neutrino energies involved, both are counting experiments, tuned at the
same L/E value. Some clone regions are eventually reduced (Figure \ref{f:clone1}).
\begin{figure}[t!]
\vspace{-0.5cm}
\begin{center}
\begin{tabular}{cc}
\hspace{-1.0cm} \epsfxsize8.5cm\epsffile{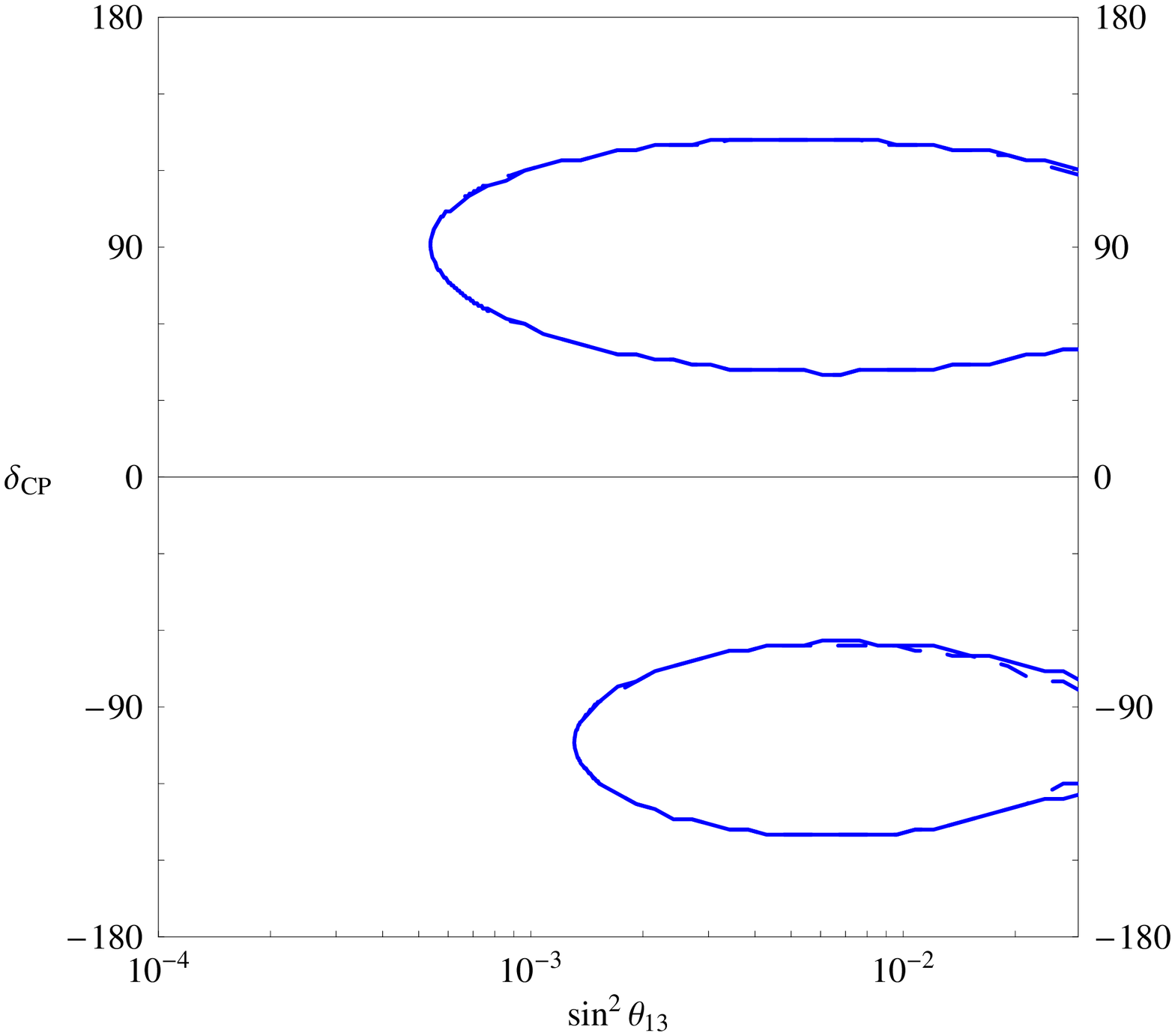} &
\hspace{-0.5cm} \epsfxsize8.5cm\epsffile{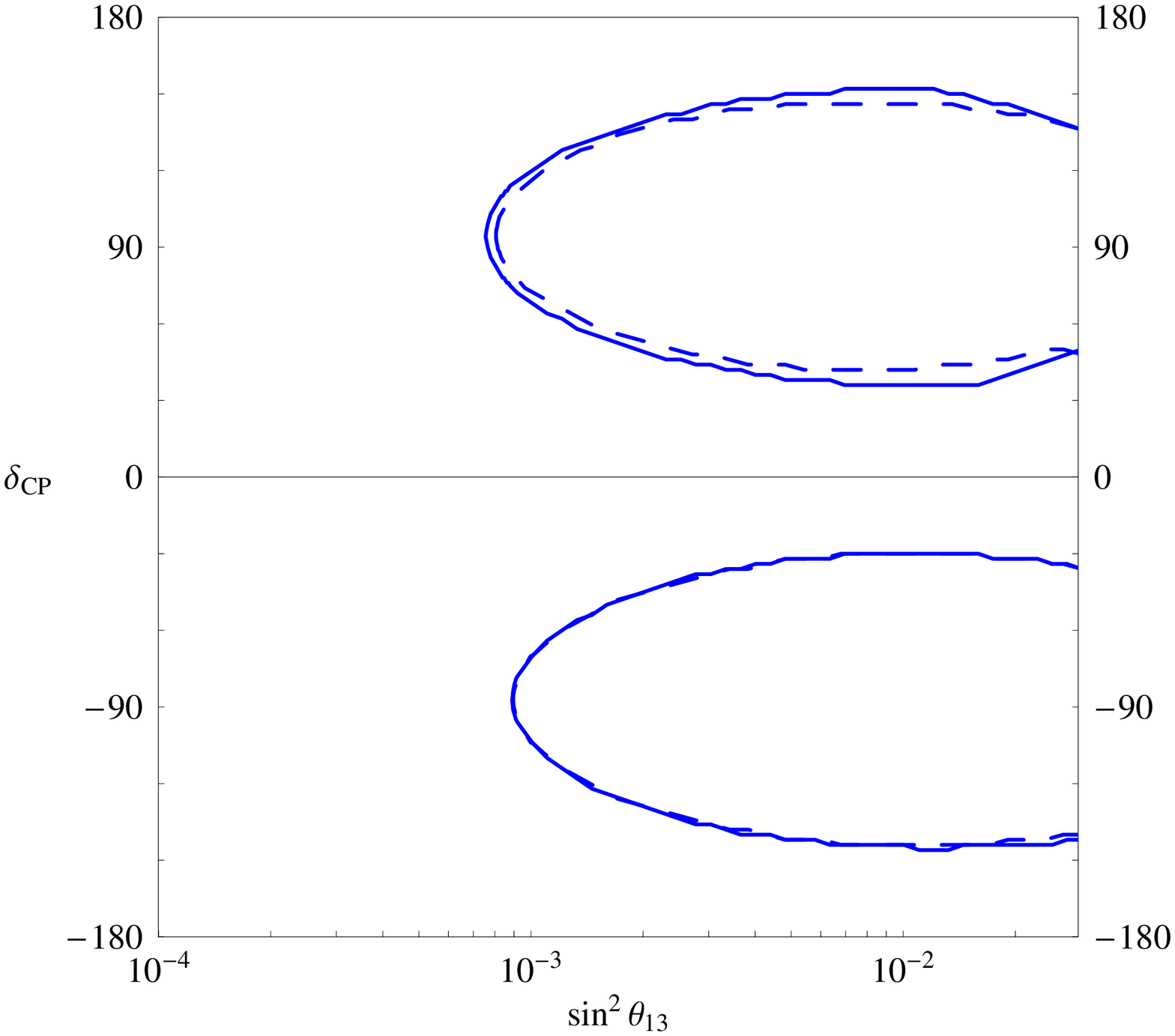} \\
\end{tabular}
\caption{CP discovery potential in the $(\theta_{13},\delta)$ plane at 99 \% C.L. for a super-beam 
(left) and a beta-beam (right), considering the 
appearance channel only (dashed line), and a  
combination of the appearance and the disappearance channels (solid line) at the 
same facility. A $2$\% systematic error in the disappearance channel and a $5$\% systematic 
error in the appearance channel have been considered for both facilities (Donini \etal 2005c).
}
\label{f:clone2}
\end{center}
\end{figure}
\begin{figure}
  \centering
   \includegraphics[width=0.5\textwidth]{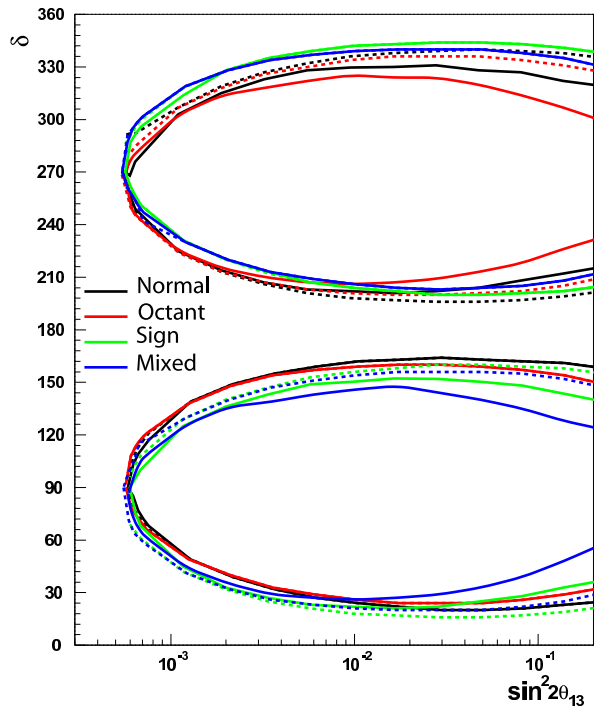}
   \protect\caption{Impact of the degeneracies on the $\cal{CP}$ discovery potential
   for the $\gamma=100$ beta-beam. We show the sensitivity at $3\sigma$
   $(\Delta\chi^2>9)$ computed for 4 different options about the
   true parameter values: Normal: $sign(\Delta m^2_{31}) = 1$, $\theta_{23} = 40^\circ$;
   Octant: $sign(\Delta m^2_{31}) = 1$, $\theta_{23} = 60^\circ$; Sign: $sign(\Delta m^2_{31}) = -1$,
   $\theta_{23} = 40^\circ$; Mixed: $ sign(\Delta m^2_{31})= -1$, $\theta_{23} = 60^\circ$.
   Dotted curves are computed neglecting
   degeneracies (Campagne \etal 2006).\label{f:100deg2}}
\end{figure}
Different strategies can be followed to resolve the degeneracy issue, such as the use of spectral 
analysis (Burguet-Castell \etal 2001), 
combination of various experiments (Minakata \etal 2003), and/of different channels (Donini \etal 2002). 
In (Donini \etal 2005c) the combination of the appearance and disappearance channels is explored both
for the original 
beta-beam and for a super-beam, showing that while there is no improvement for the 
former as soon as some level of systematic error is added, 
degeneracies are reduced for the latter. Figure \ref{f:clone2} presents 
the discovery potential for each of these options taken alone,
taking fully into account the degeneracies.
Note that the $\cal{CP}$ sensitivity contours have only an approximate symmetry for 
$|\delta| \ge \pi/2$ and $|\delta| \le \pi/2$, while there is no symmetry between positive an negative
values of $\delta$. 
The impact of the degeneracies on the $\cal{CP}$ discovery reach for the $\gamma=100$ 
option is also shown in Figure \ref{f:100deg2} (Campagne \etal 2006).
A discussion on the issue of degeneracies can be found in 
(Rigolin 2005 and 2006) as well. In (Donini and Fernandez-Martinez 2006), and (Fernandez-Martinez 2006),
the idea of alternating the ions, i.e. $^{6}$He and $^{18}$Ne with $^{8}$B and $^{8}$Li,
boosted at the same $\gamma$ of 100, is proposed to reduce the degeneracies. Indeed, high 
$^{8}$B and $^{8}$Li production rates could be possible thanks to a new production method called
"ionsation cooling" (Rubbia \etal 2006). Since the Q-values for the two sets of ions are very different,
the neutrino fluxes would be peaked at the second oscillation maximum, for the former, and at the first
oscillation maximum, for the latter, at a fixed baseline of 700 km. As a result, it is shown that
such a measurement, even though less sensitive to $\theta_{13}$, reduces the eightfold degeneracy to a
two-fold one, thanks to the different L/E dependence.  This option is particularly well suited if
$\theta_{13}$ is measurable at T2K-I. 

\begin{figure}[!t]
\centering
\includegraphics[width=0.95\textwidth]{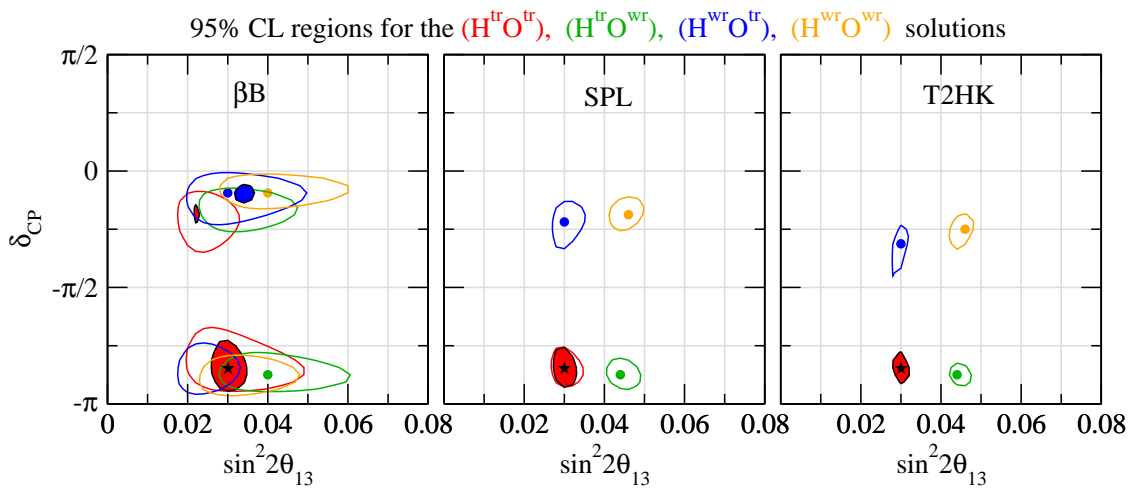}
  \protect\caption{Impact on the degeneracies 
  of a combined accelerator and atmospheric data analysis, in the case of
  a $\gamma=100$ beta-beam and a super-beam (SPL) at CERN on one hand, and
  of T2K phase II to Hyper-K (Itow \etal 2001) on the other hand. 
  The figure shows the
  allowed regions in $\sin^22\theta_{13}$ and
  $\delta_\mathrm{CP}$ for accelerator data alone (contour lines) and accelerator plus atmospheric
  data combined (colored regions). Solution for the true/wrong mass hierarchy (octant of
  $\theta_{23}$) are indicated with
  $\mathrm{H^{tr/wr} (O^{tr/wr})}$.
  The true parameter values are $\delta_\mathrm{CP} =
  -0.85 \pi$, $\sin^22\theta_{13} = 0.03$, $\sin^2\theta_{23} = 0.6$
  (Campagne \etal 2006).}
\label{f:100deg1}
\end{figure}

Another very interesting idea proposed recently to resolve the open problem of the degeneracies
is to combine accelerator experiments with data from atmospheric neutrinos provided by the same
detector (Huber \etal 2005b). In fact atmospheric neutrinos are sensitive to $\theta_{13}$
and the mass hierarchy due to earth matter effects in the $e$-like events. In addition effects from the solar parameters
and $\Delta m^2_{12}$ on $e$-like events in the sub-GeV energy range provide a sensitivity on $\theta_{23}$.
Due to these effects the sign and octant degeneracies can be broken. 
A first application is done in (Campagne \etal 2006)
which shows that 
the true solutions are identified by the combined accelerator and atmospheric data
both for a beta-beam and a super-beam (SPL) at CERN (Figure \ref{f:100deg1}).
In particular, the hierarchy can be identified at 2$\sigma$ C.L. from a combined analysis if 
$\sin^2 \theta_{13} \ge 0.03-0.05$ for the beta-beam and the super-beam while the
octant can be determined through the super-beam and the atmospheric data.
Let us recall that these experiments alone have no sensitivity at all on the octant and sign issues. 
This important synergy between accelerator and
atmospheric data is clearly very 
promising for the identification of the true value of the third mixing angle as well as of
the $\cal{CP}$ violating phase in a beta-beam experiment.

\section{Monochromatic neutrino beta-beams}
An exciting variant to the original beta-beam idea has been proposed, where
monochromatic neutrino beams are produced through the electron capture of
radioactive ion beams (Bernabeu {\it et al} 2005a, Sato 2005).
The baseline of such a facility would be the same as for the original
beta-beam facility (Zucchelli 2002) (Fig. \ref{fig:baseline}).
The production of such beams depends on the existence of ions decaying through
electron capture 
with a short enough lifetime for
electron capture to occur. In fact one needs to keep one electron bounded to the ion
and therefore to produce, accelerate and store partly charged ions. 
Since, even in a very good vacuum conditions, a few remaining atoms make the
ions quickly lose the electron, it is crucial that their lifetime is short
enough to have high neutrino intensities. 
Good candidates are offered by 
some recently discovered nuclei far from the stability line, which
have a fast decay through an
allowed transition to a Gamow-Teller resonance in the energetically
accessible region. 
Examples of such nuclei are given in Table
\ref{tab:ec-ions}.
The monochromatic neutrino flux produced
by electron capture and arriving in a detector at a given distance $L$ is 
\begin{equation}
{{d^2 N_{\nu}} \over {dSdE }} \approx {{\Gamma_{\nu}N_{ions} \gamma^2} \over
{\Gamma \pi L^2}} \delta(E-2\gamma E_0)
\end{equation}
with $\gamma >> 1$, $N_{ions}$ is the total number of decaying ions,
$\Gamma_{\nu}/\Gamma$ is the branching ratio for electron capture (Table
\ref{tab:ec-ions}). The quantities $E_0$ and $E=E_0/ [\gamma(1-\beta \cos\theta) ]$
represent the energy in the rest and in the laboratory frames respectively, 
with the
angle $\theta$ being the deviation of the neutrino with respect to the
prolongation of the long straight section of the decay ring.

\begin{table}
\caption{\label{tab:ec-ions} Decay properties of some rare-earth nuclei.
The half-life and the branching ratios for electron-capture relative to
beta-decay are given in the first columns, while the last column presents the
neutrino energy in the rest frame, defined as the Q$_{\beta}$-value 
minus the energy at which the
Gamow-Teller resonance is located (see text).
The given ions are example candidates for a beta-beam facility producing 
monochromatic beams through electron capture (Bernabeu \etal 2005a).}
\begin{tabular}{|c|ccc|} \hline
Decay & t$_{1/2}$ & EC/$\beta_+$ (\%) & $E_0 $(keV) \\ \hline
$^{148}$Dy $\rightarrow ^{148}$Tb & 3.1 m  & 96/4 & 2062 \\
$^{150}$Dy $\rightarrow ^{150}$Tb & 7.2 m  & 99.9/0.1 & 1397 \\
$^{152}$Tm $ 2^- \rightarrow ^{152}$Er & 8.0 s  & 45/55 & 4400 \\
$^{150}$Ho $ 2^- \rightarrow ^{150}$Dy & 72  s  & 77/33 & 3000 \\ \hline
\end{tabular}
\end{table}

\begin{figure}[t]
\begin{center}
\includegraphics[scale=0.4]{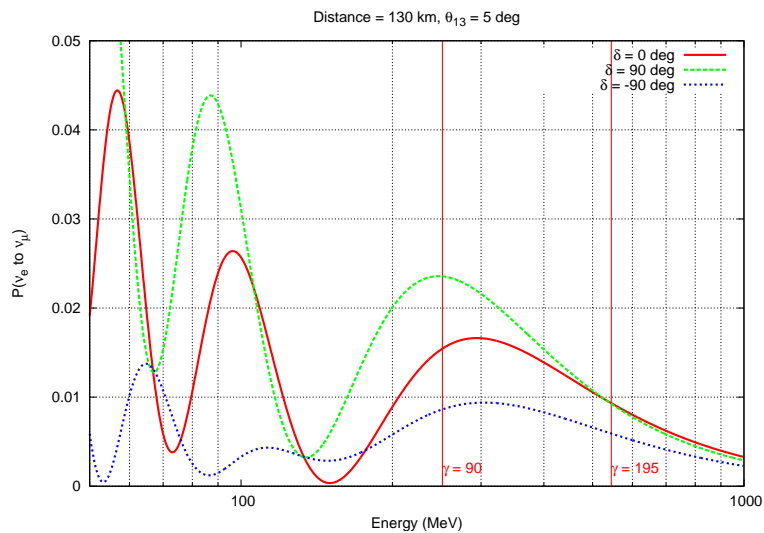}
\end{center}
\caption{Appearance probability for neutrino oscillations in the three
flavour formalism and a non-zero Dirac $\delta$ phase, as a function of
neutrino energy, with a distance between the source and the detector
of 130 km. The three curves correspond to three
different values of the phase $\delta$, while the vertical lines show the
two ion Lorentz boosts used in the simulation 
(Bernabeu \etal 2005a) \label{fig:prob-ec}.}
\end{figure}

There is a major difference in
the use of neutrino beams coming from electron capture and from beta-beams
in the search for ${\cal CP}$,
since only electron neutrino beams can be produced in the first case.
As a consequence, the study of ${\cal CP}$ violation cannot be made
through a comparison of neutrino versus anti-neutrino oscillations
Eq.(\ref{e:CPgeneral}), but by performing oscillation measurements at different ion
boosts. In fact, for a fixed distance between the source and the detector,
the appearance probability for $\nu_e \rightarrow \nu_{\mu}$ being energy
dependent, the effect of
the $\delta$ phase is to shift the oscillation interference pattern,
while the amplitude of the oscillation probability is determined by the
mixing angle $\theta_{13}$ (Figure \ref{fig:prob-ec}) (Bernabeu \etal 2005a).
As a consequence
the values of these unknown parameters can be extracted by
comparing the $\nu_e \rightarrow \nu_{\mu}$ oscillation probability at
different neutrino energies (ion boosts).
The main advantage of the electron capture case over beta-beams is
that in the former case the energy of all neutrinos can be peaked at the
energy(ies) of interest, to match the the oscillation maxima; while
in the latter case, because of the broad spectrum (Fig.\ref{fig:fluxes}), for
a given intensity, many neutrinos will be in an energy range which is less
sensitive to the phase and/or where the cross section is low.

\begin{figure}[t]
\begin{center}
\includegraphics[scale=0.8]{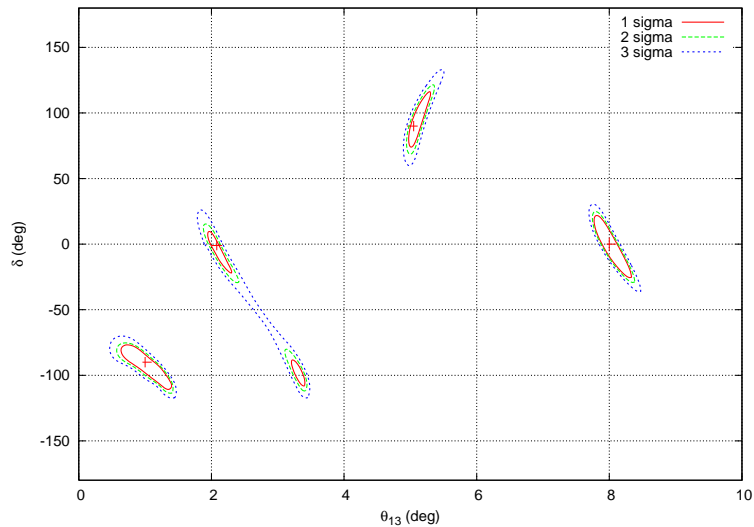}
\end{center}
\caption{Sensitivity to the ${\cal CP}$ violating phase $\delta$
and the still unknown mixing angle $\theta_{13}$. The results are
obtained considering neutrino beams from electron-capture of partly charged
$^{150}$Dy ions,
boosted at $\gamma=90,195$. The different curves
show the confidence levels for the
assumed values $(8^\circ,0),(5^\circ,90^\circ),(2^\circ,0)$ and 
$(1^\circ,-90^\circ)$.
The details on the used setup can be found in the text (Bernabeu \etal 2005a).
\label{fig:ec-CP}}
\end{figure}

The sensitivity to ${\cal CP}$ violation has been studied in
(Bernabeu \etal 2005a), also in combination with a $\beta^--$beam (Bernabeu \etal 2006).
Figure \ref{fig:ec-CP} shows the sensitivity in the
($\delta,\theta_{13}$) plane for four chosen values of these parameters.
The source is of $10^{18}~^{150}$Dy ions/year,
for a running time period of 10 years.
The ions run 5 years at the Lorentz boost $\gamma=195$ - the
maximum achievable at the CERN's SPS (Fig.\ref{fig:baseline}) - and 5 years
at $\gamma=90$, chosen to be as different as possible from the first one and
high enough to avoid backgrounds.
The water \v{C}erenkov detector has 440 kton fiducial mass and is located at 
a distance of 130 km (the CERN-Fr\'ejus distance).

The possibility of using monochromatic electron neutrino beams to perform
${\cal CP}$ violation searches appears very promising. A detailed study
concerning the ion production rates, the target and ion source design, 
the accumulation scheme as well as possible vacuum improvements is now
required, to definitely state the achievable flux.

\section{Conclusion and perspectives}
The recent discoveries in the field of neutrino physics
open important new developments. Some
require in particular pure, intense and well controlled neutrino beams,
such as those furnished by "beta-beams".
This novel method exploits the intense radioactive ion beams
 under study,
and offers important synergies with the nuclear physics community.
A new facility based on beta-beams has been proposed 
with the main goal of addressing the crucial issue of the existence of $\cal{CP}$ 
violation in the lepton sector.
In the original scenario information on the $\cal{CP}$ violating phase is
extracted from a comparison of neutrino versus anti-neutrino oscillations.
The neutrino beams are produced at CERN,
and sent to a gigantic Cerenkov detector
located in the Fr\'ejus Underground
Laboratory, 130 km from CERN. The first steps of the facility are
very similar to the EURISOL project, while the acceleration to GeV energies uses
already existing accelerator infrastructures (the PS and SPS) 
which implies important savings in the cost. The ions are then 
stored in a large storage ring that needs to be built.
This scenario has also the significant advantage that it uses reasonable 
extrapolation of existing technology. 

While a beta-beam alone already has a very interesting discovery potential,
it can be pushed even further if a very intense conventional beam is fired 
to the same detector as well. This combination also offers the possibility
to perform studies of the $\cal{T}$ and $\cal{CPT}$ symmetries. 
On the other hand, it has recently been pointed out that 
if the $\cal{CPT}$ symmetry is assumed, the neutrino
measurements with a beta-beam and a super-beam in the $\cal{T}$ channel
can be used to reduce the measurement time from 10 to 5 years.
The achieved sensitivity is very close to that of a single
experiment running for 10 years in both the neutrino and the 
anti-neutrino modes. 

The discovery reach of the original scenario
is being explored in great detail. Various options are considered,
such as running the neutrino and antineutrino emitters either together or
independently in the storage ring, and/or varying 
the Lorentz ion boost from 60 to the maximum acceleration achievable in the SPS.
(If at $\gamma$ of about 60 the experiment is a counting experiment only, already
at about 100 some information on the energy of the events can be exploited.)
This facility is well positioned to explore values between $10^{-3}$ to $10^{-4}$
for the third still unknown neutrino mixing angle $\theta_{13}$.
For the $\gamma=100$ option the maximal violation
corresponding to $\delta=\pi/2$ and $3 \pi/2$ can be discovered down to 
$\sin^2 2 \theta_{13}=6(8)~10^{-4}$ (99 \% C.L.) for the beta-beam (super-beam); while the best
sensitivity is obtained for $\sin^2 2 \theta_{13} > 10^{-2}$ since at this value 
$\cal{CP}$ violation can be established for 73 \% (75 \%) for all $\delta$ values with
a beta-beam (super-beam). 
The comparison with e.g. 
T2K phase II to Hyper-K in Japan -- a project of similar size and timescale -- 
shows a very similar discovery reach. 
Besides, the MEMPHYS detector has the same physics potential for proton decay and
for (extra-)galactic supernova neutrinos as UNO in the US and Hyper-K.

A variant of the original beta-beam idea consists in using 
ions decaying through electron-capture instead of beta-decay.
This option has the advantage that the neutrino beams are
mono-chromatic. (No anti-neutrino beams can be produced.)
Information on the $\delta$ phase is extracted in this case by performing experiments
at different ion boosts, since the phase introduces shifts in the appearance oscillation patterns.
The recent discovery of exotic nuclei that have fast decay through a 
strong Gamow-Teller transitions offer good potential candidates for the neutrino emitters.
The required baseline is here the same as in the original beta-beam scenario.
A specific feasibility study is now required as far as the achievable
intensities and vacuum issues are concerned for example, 
since  not fully stripped ions need to be accelerated and
stored.
Also, the sensitivity studies are to be pushed further.

Medium and high energy scenarios have been 
proposed as well, where the ion boosts are of several hundreds and thousands
respectively. Such energies require new baseline scenarios: 
new accelerator infrastructures,
such as a refurbished SPS (for the medium energy case), 
or the LHC (for the high energy case);
a much bigger storage ring to stack the ions;
a longer source to detector distance, to match the first neutrino 
oscillation maximum.
New sites for the far detector are to be identified 
like the Gran Sasso Underground Laboratory at
about 700 km (for a boost of 350), or the Canfranc Underground Laboratory
(for $\gamma$=150). 
These scenarios benefit of
increased statistics, of the energy dependence of the events and eventually of matter effects,
 with an
enlarged discovery potential: an increased sensitivity;
information on the still unknown neutrino mass hierarchy;
a reduction of the degeneracies with a better
identification of the true location of the unknown angle
and of the phase.
From the physics point of view these scenarios with $\gamma > 150$, and
particularly the medium energy ones, are very promising. However,
they keep, at least for the moment, at a more speculative level than
the original one, since they are not based on existing accelerator complex,
nor on a robust evaluation of the ion intensities.
Clearly these options deserve further studies.

The analysis of a $\cal{CP}$ violating experiment will depend on the knowledge of all other neutrino
oscillation parameters at the time it takes place. At the moment the unknown sign of 
$\Delta m_{23}^2$ and octant of $\theta_{23}$ introduce supplementary fake regions in the
multidimensional oscillation parameter space besides those that correspond to the true solutions of
$\theta_{13}$ and $\delta$. A significant reduction of the 
degeneracies can be obtained in the medium and
high energy options. On the other hand, 
the original beta-beam scenario has no sensitivity to
the octant and to the sign issues, and is affected by degeneracies, even
combined with a super-beam. 
The synergy of the accelerator measurements with atmospheric data in the same detector, 
as recently proposed, has been shown to be a very promising strategy to get rid of the 
clone solutions in this case as well. 
In particular, the hierarchy can be identified at 2$\sigma$ C.L. from a combined analysis if 
$\sin^2 \theta_{13} \ge 0.03-0.05$ for the beta-beam and the super-beam, while the
octant can be determined through the super-beam, thanks to the atmospheric data.

The establishment of a beta-beam facility is a natural site to have low energy beta-beams
with neutrinos in the 100 MeV region. Such options open completely new possibilities
besides $\cal{CP}$, $\cal{T}$, and $\cal{CPT}$ symmetries studies.
The physics potential is being investigated. It 
covers different topics from nuclear structure and neutrino-nucleus
to fundamental interaction studies as well as core-collapse Supernova physics.
Since the neutrino flux emittance is low, a devoted storage ring appears as more appropriate
for these applications.

In conclusion, a beta-beam facility has
a rich and broad physics potential. The feasibility study of the original scenario,
as well as of the small storage for low energy beta-beams, is now ongoing within
the EURISOL Design Study. Both its outcome and future physics studies 
will furnish the necessary elements to
assess the final discovery reach. The ongoing work has already shown that
a beta-beam facility can clearly offer a very competitive strategy for future neutrino
accelerator experiments, and much more.

\vspace{.3cm}

\noindent
{\sc Acknowledgments}
\vspace{.05cm}

\noindent
It has been a very lively and exciting experience to
contribute to the developments of beta-beams since the very beginning. 
I wish to thank A.B. Balantekin, M. Benedikt, J. Bernabeu, J. Burguet-Castell,
J. Bouchez, A. Chanc\'e, J.J. Gomez-Cadenas, P. Hernandez, N. Jachowitz,
J.H. De Jesus, M. Lindroos, G.C. McLaughlin, R. Lazauskas, M. Mezzetto, 
J. Payet, J. Serreau, T. Weick and P. Zucchelli, for the interesting and stimulating
discussions.
I am particularly indebted to B. Berthier for his precious help, at the very beginning
of low energy beta-beams. Thanks also to M. Mezzetto for careful reading of
this manuscript.
The author acknowledges the Beams for European Neutrino Experiments (BENE)
network, the EURISOL Design Study as well as the CNRS - Etats-Unis 2005 and
2006 fundings.

\vspace{.5cm}

\section*{References}
\begin{harvard}

\item[] Ahmad Q R [the SNO Collaboration] 2001 {\it Phys. Rev. Lett.} 
{\bf 87} 071301 [nucl-ex/0106015]
\item[] Ahn M H [the K2K Collaboration] 2003 {\it Phys. Rev. Lett.}
{\bf 90} 041801 [hep-ex/0212007]
\item[] Agarwalla SK, Raychaudhuri A, Samanta A
2005 {\it Phys. Lett.} B {\bf 629} 33
\item[] Akimune H \etal 1997 {\it Phys. Lett.} B {\bf 394} 23
\item[] Albright C \etal 004 "The APS Neutrino Study" [physics/0411123]
\item[] ALEPH, DELPHI, L3, OPAL, and SLD Collaborations 2005
[hep-ex/0509008]    
\item[] Angrik J \etal [KATRIN Collaboration] 2004 "The KATRIN Design Report"
\item[] Anthony P L \etal [SLAC E158 Collaboration] 2005
{\it Phys.\ Rev.\ Lett.}  {\bf 95} 081601 [hep-ex/0504049]
\item[] Apollonio M \etal [the CHOOZ Collaboration] 
1999 {\it Phys. Lett.} B {\bf 466} 415 [hep-ex/9907037]
\item[] Ardellier \etal 2004 [the Double CHOOZ Collaboration]
[hep-ex/0405032]
\item[] Athanassopoulos C \etal [LSND Collaboration] 1998
{\it Phys.\ Rev.\ Lett.}  {\bf 81} 1774 [nucl-ex/9709006]
\item[] Athanassopoulos C \etal [LSND Collaboration] 1996
{\it Phys.\ Rev.\ Lett.}  {\bf 77} 3082 [nucl-ex/9605003]
\item[] Aunola M and Suhonen J 1996 {\it Nucl. Phys.} A {\bf 602} 133
\item[] Autin B \etal 2003 {\it Jour. Phys.} G {\bf 29} 1785
 [physics/0306106] 
\item[] Avignone F T 2000 \etal {\it Phys. Atom. Nucl.} {\bf 63} 1007
 [physics/0306106]; see also http://www.phys.ornl.gov/orland 
\item[] Balantekin A B, De Jesus J H, Volpe C 2006a {\it Phys. Lett.} B {\bf 634}
[hep-ph/0512310]
\item[] Balantekin A B, De Jesus J H, Lazauskas R, Volpe C 2006b
{\it Phys. Rev.} D {\bf 73} 073011 [hep-ph/0603078] 
\item[] Balantekin A B and Fuller G M 2003 {\it J.\ Phys.} G {\bf 29} 2513
[astro-ph/0309519]
\item[] Barger V, Marfatia D and Whisnant K 2002 {\it Phys. Rev.} 
D {\bf 65} 073023 [hep-ph/0112119]
\item[] Beacom J F and Vagins M R 2004 {\it Phys. Rev. Lett.} 
{\bf 93} 171101 [hep-ph/0309300]
\item[] Beacom J F and Vogel P 1999
{\it Phys. Rev. Lett.}  {\bf 83} 5222  [hep-ph/9907383]
\item[] Beacom J F, Farr W M and Vogel P 2002
{\t Phys.\ Rev.} D {\bf 66} 033001 [hep-ph/0205220]       
\item[] Benedikt M, Hancock S, Lindroos M 2004 Proceedings to ``EPAC 2004'', 
Lucerne, Switzerland, 5-9 July
\item[] Benedikt M 2005 {\it Nucl. Phys. Proc. Suppl.} {\bf 149} 54
\item[] Benedikt M, Chanc\'e A, Lindroos M, Payet J 2006 Private communication 
\item[] Bennett S C and Wieman C E 1999
{\it Phys.\ Rev.\ Lett.}  {\bf 82} 2484 [hep-ex/9903022]	
\item[] Bernabeu J, Burguet-Castell J, Espinoza C, M. Lindroos M 2005a
JHEP 0512 {\bf 014}  [hep-ph/0505054]
\item[] Bernabeu J, Burguet-Castell J, Espinoza C 2005b
Talk given at ``EPS International Europhysics Conference on High Energy
Physics (HEP-EPS 2005)'', Lisbon, Portugal, 21-27 Jul [hep-ph/0512297] 
\item[] Bernabeu J, Burguet-Castell J, Espinoza C, M. Lindroos M 2005c
Proceedings to ``GUSTAVOFEST: Symposium in Honor of Gustavo C. Branco: CP
Violation and the Flavor Puzzle'', Lisbon, Portugal, 19-20 Jul
[hep-ph/0512299]
\item[] Bernabeu J, Burguet-Castell J, Espinoza C, M. Lindroos M 2006
{\it Nucl. Phys. Proc. Suppl.} {\bf 155} 222 [hep-ph/0510278]
\item[] Bernabeu J \etal 1988 IPNO/TH-88-58, FTUV-88/20
\item[] Bilenky S M, Giunti C, Grifols J A, Masso E 2003
{\it Phys. Rept.} {\bf 379} 69
\item[] Borzov I N and Goriely S 2000 {\it Phys. Rev. C} {\bf 62} 035501-1
\item[] Bouchez J, Lindroos M, Mezzetto M 2004 {\it AIP Conf. Proc.} {\bf 721}
37 [hep-ex/0310059]
\item[] Bouchez J 2005 {\it Nucl. Phys. Proc. Suppl.} {\bf 147} 93
\item[] Burguet-Castell J \etal 2004 {\it Nucl. Phys.} B {\bf 695} 217
[hep-ph/0312068]
\item[] Burguet-Castell J \etal 2005 {\it Nucl. Phys.} B {\bf 725} 306
[hep-ph/0503021]
\item[] Burguet-Castell J, Gavela M B, Gomez-Cadenas J J, Hernandez P ad Mena O
2001 {\it Nucl. Phys.} B {\bf 608} 301 [hep-ph/0103258]
\item[] Campagne J E, Maltoni M, Mezzetto M, Schwetz T 2006 [hep-ph/0603172]
\item[] Chanc\'e A and Payet J 2005 Private communication
\item[] Christenson \etal 1964 {\it Phys. Rev. Lett.} {\bf 13} L38
\item[] Daraktchieva Z \etal  [MUNU Collaboration] 2005
{\it Phys.\ Lett.} B {\bf 615} 153 [hep-ex/0502037]
\item[] Davidson S, Forte S, Gambino P, Rius N and Strumia A 2002
{\it JHEP} {\bf 0202} 037 [hep-ph/0112302]     
\item Davis R 1964 {\it Phys. Rev. Lett.} {\bf 12}
\item[] De Gouvea A and Jenkins J 2006 [hep-ph/0603036]
\item[] Derbin A I \etal 1993 {\it JETP Lett.}  {\bf 57} 768
\item[] Deutsch J P, Macq P C and Van Elmbt L 1977
{\it Phys.\ Rev.} C {\bf 15} 1587     
\item[] Diwan M V \etal 2003 {\it Phys. Rev.} D {\bf 68} 012002 [hep-ph/0303081]
\item[] Donini A and Fernandez-Martinez 2006 [hep-ph/0603261]
\item[] Donini A, Fernandez-Martinez E, Migliozzi P, Rigolin S,
Scotto-Lavina L, Tabarelli de Fatis, Terranova F 2005a {\it Nucl. Phys.} B {\bf 710} 402
[hep-ph/0511134]
\item[] Donini A, Fernandez-Martinez E, Migliozzi P, Rigolin S,
Scotto-Lavina L 2005b {\it Nucl. Phys.} B {\bf 710} 402
[hep-ph/0406132]
\item[] Donini A, Fernandez-Martinez E, Rigolin S 
2005c {\it Phys. Lett.} B {\bf 621} 276 [hep-ph/0411402]
\item[] Donini A, Meloni D, Rigolin S  2004 {\it JHEP} {\bf 0406} 011 [hep-ph/0312072]
\item[] Donini A, Meloni D and Migliozzi P 2002 
{\it Nucl. Phys.} B {\bf 646} 321 [hep-ph/0206034]
\item[] Elliott S R and Engel J 2004 {\it J.\ Phys.} G {\bf 30} R183 [hep-ph/9604415]
\item[] Efremenko Y Private communication
\item[] Eguchi K \etal [the KamLAND Collaboration] 2003 {\bf 90}
{\it Phys. Rev. Lett.} 021802 [hep-ex/0212021] 
\item[] Ereditato A and Rubbia A 2006 {\it Nucl. Phys. Proc. Suppl.} {\bf 155} 233
[hep-ph/0510131]  
\item[] Ereditato A and Rubbia A 2004 Proceedings to ``HIF04'', La Biondola, June [hep-ph/0412230]
\item[] The EURISOL Project, see http://ganinfo.in2p3.fr/eurisol/
\item[] Fernandez-Martinez 2006 [hep-ph/0605101]
\item[] Fogli G L and Lisi E 1996 {\it Phys.\ Rev.} D {\bf 54} 3667 [hep-ph/9604415]
\item[] Fogli G L, Lisi E, Mirizzi A, Montanino D 2005 {\it JCAP} {\bf 0504} 002
\item[] Fukuda \etal [the Super-Kamiokande Collaboration] 
1998 {\it Phys. Rev. Lett.} {\bf 81} 1562 [hep-ex/9807003] 
\item[] GSI see http://www.gsi.de 
\item[] Guglielmi A, Mezzetto M, Migliozzi P, Terranova F 2005 
``The High Energy Frontier'' [hep-ph/0508034] 
\item[] Gomez-Cadenas JJ 2005 {\it Nucl. Phys. Proc. Suppl.} {\bf 145} 161
\item[] Giunti C and Laveder M 2002
[hep-ph/0202152]    
\item[] Hardy J C and Towner I S 2005
{\t Phys.\ Rev.} C {\bf 71} 055501 [nucl-th/0412056]
\item[] Haxton W C 2004 Talk presented at "The r-process: The astrophysical origin of the heavy elements"
[nucl-th/0406012]
\item[] Haxton W C, Langanke K, Qian Y Z, Vogel P 1997 {\it Phys. Rev. Lett.} {\bf 78} 2694
[astro-ph/9612047]
\item[] Heger \etal 2005 {\it Phys. Lett.} B {\bf 606} 258 [astro-ph/0307546]
\item[] Horowitz C J 2002 {\it Phys. Rev. } D {\bf 65} 043001 [astro-ph/0109209]
\item[] Huber P, Lindner M, Rolinec M, Winter W 2006 {\it Phys. Rev.} D {\bf 73} 053002 [hep-ph/0506237]
\item[] Huber P, Lindner M, Winter W 2005a {\it Comput. Phys. Commun.} {\bf 167} 195 [hep-ph/0407333]
\item[] Huber P, Maltoni M and Schwetz T 2005b {\it Phys. Rev.} D {\bf 71} [hep-ph/0501037] 
\item[] Itow Y \etal 2001 [hep-ex/0106019]
\item[] Jachowicz N and McLaughlin G C 2005
Proceedings of the "Erice International School of Nuclear Physics", 
Erice, Italy 16-24 September [nucl-th/0511069] 
\item[] Jachowicz N and McLaughlin G C 2006 submittted [nucl-th/0604046]
\item[] Jung C K 2000 [hep-ex/0005046] 
\item[] Kneller J P, McLaughlin G C and Surman R 2004 [astro-ph/0410397]
\item[] Kolbe E \etal 2003 {\it J. Phys.} G {\bf 29} 2569 [nucl-th/0311022]
\item[] Kolbe E and Langanke K 1999 {\it Phys. Rev.} C {\bf 60} 052801
[nucl-th/9905001]
\item[] Kortelainen M and Suhonen J 2002 {\it Europhys. Lett.} {\bf 58} 666
\item[] Kortelainen M and Suhonen J 2004 {\it Phys. Atom. Nucl.} {\bf 67} 1202
\item[] Kubodera K and Nozawa S 1994 {\it Int. J. Mod. Phys.} E {\bf 3} 101, 
and references therein
\item[] Langanke K and Martinez-Pinedo G 2004 {\it Nucl. Phys.} A {\bf 731} 365 
\item[] Lee Y K, Mo L W and Wu C S 1963
{\it Phys.\ Rev.\ Lett.} {\bf 10} 253
\item[] Lee T D and Wu C S 1965 {\it Ann.\ Rev.\ Nucl.\ Sci.} {\bf 15} 381  
\item[] Li H B \etal [TEXONO Collaboration] 2003 {\it Phys. Rev. Lett.} 
{\bf 90} 131802  
\item[] Lindroos M 2003 CERN-AB-2004-006-OP [physics/0312042] 
\item[] Lindroos M and Volpe C 2004 {\it Nucl. Phys. News} {\bf 14} 15   
\item[] Lindroos M 2005 EURISOL DS/TASK12/TN-05-02
\item[] Loinaz W \etal 2004  {\it Phys.\ Rev.} D {\bf 70} 113004
[hep-ph/0403306] 
\item[] Magistris M and Silari M 2003 note CERN-TIS-2003-017-RP-TN
\item[] Maki Z, Nakagawa M, Sakata S 1962 
{\it Prog. Theor. Phys.} {\bf 28} 870
\item[] McLaughlin G C and Volpe C 2004
{\it Phys. Lett.} B {\bf 591} 229 [hep-ph/0312156]
\item[] McLaughlin G C 2004 {\it Phys. Rev.} C {\bf 70} 045804
[nucl-th/0404002]
\item[] McLaughlin G C and Fuller G M 1995 {\it Astrophys. J.} {\bf 455} 202
\item[]  Meyer B S, McLaughlin G C and Fuller G M 1998
{\it Phys.\ Rev.} C {\bf 58} 3696 [astro-ph/9809242]     
\item[] Mezzetto M 2003 {\it Jour. Phys.} G {\bf 29} 1771
[hep-ex/0302007]
\item[] Mezzetto M 2005 {\it Nucl. Phys. Proc. Suppl.} {\bf 149} 179
\item[] Mezzetto M 2006 {\it Nucl. Phys. Proc. Suppl.} {\bf 155} 214
[hep-ex/0511005]
\item[]  Miller G A and Thomas A W 2005
{\it Int.\ J.\ Mod.\ Phys.} A {\bf 20} 95 [hep-ex/0204007]	
\item[] Migliozzi P 2005 {\it Nucl. Phys. Proc. Suppl.} {\bf 145} 199
\item[] Migliozzi P and Terranova F 2003 {\it Phys. Lett.} B {\bf 563} 73
[hep-ph/0302274]
\item[] Mikheev S P and Smirnov A Y 1986 {\it Nuovo Cimento} C {\bf 9} 17
\item[] Minakata H and Nunokawa H 2001 {\it JHEP} {\bf 0110} 001
[hep-ph/0108085]
\item[] Minakata H, Nunokawa H, Parke S J 2003 {\it Phys. Rev.} D {\bf 68} 013010 [hep-ph/0301210]
\item[] Mosca L 2005a {\it Nucl. Phys. Proc. Suppl.} {\bf 138} 203
\item[] Mosca L 2005b {\it Nucl. Phys. Proc. Suppl.} {\bf 138} 421
\item[] Muto K, Bender E and Klapdor H V 1989 {\it Z. Phys.} A {\bf 334} 177
\item[] Osterfeld F 1992 {\it Rev. Mod. Phys.} {\bf 64} 491 
\item[] Pontecorvo B 1957 {\it Sov. Phys. JETP} {\bf 6} 429
\item[] Qian Y Z \etal 1997 {\it Phys. Rev. C} 1532
\item[] Raffelt G G 1996
{\it Stars As Laboratories For Fundamental Physics: 
The Astrophysics Of Neutrinos,
Axions, And Other Weakly Interacting Particles},
  Chicago, USA: Univ. Pr.; and references therein 
\item[] Reines F, Gurr H S and Sobel H W 1976
{\it Phys. Rev. Lett.} {\bf 37} 315 
\item[] Richter B 2000 SLAC-PUB-8587 [hep-ph/0008222]
\item[] Rigolin S 2005 {\it Nucl. Phys. Proc. Suppl.} {\bf 145} 203 [hep-ph/0411403]
\item[] Rigolin S 2006 {\it Nucl. Phys. Proc. Suppl.} {\bf 155} 33
[hep-ph/0509366]
\item[] Rodin \etal 2003 {\it Phys. Rev.} C {\bf 68} 044302, and references therein 
\item[] Rubbia C 1977 CERN-EP/77-08
\item[] Rubbia C, Ferrari A, Kadii Y and Vlachoudis V 2006 [hep-ph/0602032] 
\item[] Rubbia A 2004 Proceedings to ``HIF04'', La Biondola, June [hep-ph/0412230]
\item[] Ruffert M, Janka H T, Takahashi K and Schafer G 1997
{\it Astron.\ Astrophys.}  {\bf 319} 122 [astro-ph/9606181]	
\item[] Sato Joe 2005 {\it Phys. Rev. Lett.} {\bf  95} 131804 [hep-ph/0503144]
\item[] Serreau J and Volpe C 2004
{\it Phys. Rev.} C {\bf 70} 055502 [hep-ph/0403293]
\item[] Singh SK, Sajjad Athar M, Ahmad S 2006a [nucl-th/0603001]
\item[] Singh SK, Sajjad Athar M, Ahmad S 2006b {\it Nucl. Phys.} A {\bf 764} 551
[nucl-th/0506046]
\item[] Sortais P 2003 Talk given at the Moriond workshop ``Radioactive beams
for nuclear physics and neutrino physics'', Les Arcs 
\item[] Terasawa M \etal 2004 {\it Astrophys. J.} {\bf 608} 470 
\item[] Terranova F, Marotta A, Migliozzi P, Spinetti M 2004
{\it Eur. Phys. J.} C {\bf 38} 69 
\item[] Terranova F 2005 {\it Nucl. Phys. Proc. Suppl.} {\bf 149} 185,
Proceedings to ``NUFACT04'', Osaka, Japan, 26 July, 1
August 2004
\item[] Thomas A W 1996 Talk given at the "Int. Symposium on Non-Nucleonic 
Degrees of Freedom Detected in the Nucleus", Osaka, Japan, Sept. 2-5 [nucl-th/9609052]
\item[] Vidyakin G S \etal 1992 {\it JETP Lett.}  {\bf 55} 206
\item[] Vogel P and Beacom J F 1999
{\it Phys.\ Rev.} D {\bf 60} 053003 [hep-ph/9903554]	
\item[] Vogel P and Engel J 1989 {\it Phys. Rev. } D {\bf 39} 3378 
\item[] Volpe C 2004 {\it Jour. Phys.} G {\bf 30} L1 [hep-ph/0303222]
\item[] Volpe C 2005a {\it Jour. Phys.} G  {\bf 31} 903 [hep-ph/0501233]
\item[] Volpe C 2005b {\it Nucl. Phys.} A {\bf 752} 38 [hep-ph/0409357]
\item[] Volpe C 2005c {\it Nucl. Phys. Proc. Suppl.} {\bf 143} 43
[hep-ph/0409249], and references therein
\item[] Volpe C 2006 {\it Nucl. Phys. Proc. Suppl.} B {\bf 155} 97 [hep-ph/0510242]
\item[] Volpe \etal 2000 {\it Phys. Rev. C} {\bf 62} 015501 [nucl-th/0001050]
\item[] Volpe \etal 2002 {\it Phys. Rev. C} {\bf 65} 044603 [nucl-th/0103039]
\item[] Weinberg S 1958 {\it Phys. Rev.}  {\bf 112} 1375
\item[] Wilkinson D H 2000 {\it Eur. Phys. J.} A {\bf 7} 307
\item[] Winter W 2006 Talk given at the "3rd Interntional Workshop on NO-VE: Neutrino 
Oscillations in Venice", Venice, Italy, Feb 7-10 [hep-ph/0603012]
\item[] Wolfenstein L 1978 {\it Phys. Rev. D} {\bf 17} 2369
\item[] Woosley  S E \etal 1990 {\it Astrophys. J.} {\bf 356} 272
\item[] Wu C S 1964 {\it Rev.\ Mod.\ Phys.} {\bf 36} 618
\item[] Zeller G P \etal [NuTeV Collaboration] 2002 
{\it Phys.\ Rev.\ Lett.}  {\bf 88} 091802
[Erratum-ibid.\  {\bf 90} (2003) 239902] [hep-ex/0110059]	  
\item[] Zucchelli P 2002 {\it Phys. Lett. } B {\bf 532} 166

\end{harvard}

\end{document}